\documentclass{article}

\usepackage{arxiv}

\usepackage[utf8]{inputenc}     
\usepackage[T1]{fontenc}        
\usepackage{hyperref}           
\usepackage{url}                
\usepackage{array}
\usepackage{booktabs}
\usepackage{booktabs}           
\usepackage{amsmath, amsfonts}  
\usepackage{nicefrac}           
\usepackage{microtype}          
\usepackage{graphicx}
\usepackage{xcolor}
\usepackage{lineno}
\usepackage{caption}
\usepackage[numbers,sort&compress]{natbib}
\usepackage{doi}
\usepackage{placeins}

\title{Topological Data Analysis of Spatial Patterning in Heterogeneous Cell Populations: Clustering and Sorting with Varying Cell-Cell Adhesion}

\author{ 
\href{https://orcid.org/0000-0001-8068-3101}{\includegraphics[scale=0.06]{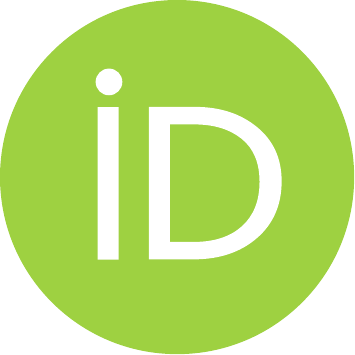}\hspace{1mm}Dhananjay Bhaskar\textsuperscript{1,2,3}}\thanks{Current Address: Department of Genetics, Yale School of Medicine} \\
	\texttt{dhananjay\_bhaskar@alumni.brown.edu}\\
	\And
	William Y. Zhang\textsuperscript{3,4}\thanks{Current Address: Operations Research Center, Massachusetts Institute of Technology} \\
	\texttt{william\_zhang1@alumni.brown.edu}\\
	\AND
	\href{https://orcid.org/0000-0003-3401-5094}{\includegraphics[scale=0.06]{orcid.pdf}\hspace{1mm}Alexandria Volkening\textsuperscript{5}}\\
	\texttt{avolkening@purdue.edu}
	\And
	\href{https://orcid.org/0000-0002-5432-1235}{\includegraphics[scale=0.06]{orcid.pdf}\hspace{1mm}Björn Sandstede\textsuperscript{3,4}}\\
	\texttt{bjorn\_sandstede@brown.edu}
	\And
	\href{https://orcid.org/0000-0002-9439-2548}{\includegraphics[scale=0.06]{orcid.pdf}\hspace{1mm}Ian Y. Wong\textsuperscript{1,2,3,6}\thanks{Corresponding author}}\\
	\texttt{ian\_wong@brown.edu}
}

\date{
\textsuperscript{1} School of Engineering, Brown University\\
\textsuperscript{2} Center for Biomedical Engineering, Brown University\\
\textsuperscript{3} Data Science Institute, Brown University\\
\textsuperscript{4} Division of Applied Mathematics, Brown University\\
\textsuperscript{5} Department of Mathematics, Purdue University\\
\textsuperscript{6} Legorreta Cancer Center, Brown University
}

\renewcommand{\shorttitle}{\textit{arXiv} Template}

\hypersetup{
pdftitle={Topological Data Analysis of Spatial Patterning in Heterogeneous Cell Populations},
pdfsubject={q-bio.QM, q-bio.CB, cs.LG},
pdfauthor={D.~Bhaskar, W.~Zhang, A.~Volkening, B.~Sandstede, I.~Wong},
pdfkeywords={Topological data analysis, Pattern formation, Cell sorting, Differential adhesion, Collective migration},
}

\begin{document}
\maketitle

\begin{abstract}
Different cell types aggregate and sort into hierarchical architectures during the formation of animal tissues. The resulting spatial organization depends (in part) on the strength of adhesion of one cell type to itself relative to other cell types. However, automated and unsupervised classification of these multicellular spatial patterns remains challenging, particularly given their structural diversity and biological variability. Recent developments based on topological data analysis are intriguing to reveal similarities in tissue architecture, but these methods remain computationally expensive. In this article, we show that multicellular patterns organized from two interacting cell types can be efficiently represented through persistence images. Our optimized combination of dimensionality reduction via autoencoders, combined with hierarchical clustering, achieved high classification accuracy for simulations with constant cell numbers. We further demonstrate that persistence images can be normalized to improve classification for simulations with varying cell numbers due to proliferation. Finally, we systematically consider the importance of incorporating different topological features as well as information about each cell type to improve classification accuracy. We envision that topological machine learning based on persistence images will enable versatile and robust classification of complex tissue architectures that occur in development and disease.
\end{abstract}



\section{Introduction}

Animal tissues are spatially organized into complex spatial patterns by varying adhesive interactions between different cell types \cite{Lecuit:2007cw,tsai_adhesion-based_2022}. For instance, mixtures of motile animal cells in a planar geometry can self-sort into their respective types, as well as aggregate into multicellular clusters \cite{Graner:1992kx,Rieu:1998bf,Belmonte:2008dv,Hogan:2009bc,Beatrici:2011fz,Mehes:2012fy,Kabla:2012gm,Strandkvist:2014ie,Nielsen:2015jl,GamboaCastro:2016da,carrillo_adhesion_2018,carrillo_population_2019,Leggett:2019hva,Li:2019bt,Krajnc:2020dd,Sahu:2020cc,Dey:2021ke,lucia2022cell,Skamrahl2022}. Historical work by Steinberg attempted to explain these behaviors using a physical analogy with surface tension, where cells exhibit ``differential adhesion'' with one another \cite{Steinberg:2007ih}, which was subsequently updated by Brodland to include contractility \cite{brodland_differential_2002}. In particular, two different cell types that each exhibit strong homotypic adhesion (to self) but weak heterotypic adhesion (to the other) would eventually segregate into separate clusters, each consisting of a single cell type \cite{Steinberg:1963kq}. On the other hand, two cell types with strong heterotypic adhesion would randomly intermix within a single cluster. Between these limiting cases, intermediate homotypic and heterotypic adhesion would result in a core-shell organization, where a first cell type would aggregate at the interior, and the second cell type would organize as a spread layer around the periphery. More recent work using \emph{Drosophila melanogaster} has addressed the role of cellular contractility \cite{kasza2011dynamics} and proliferation \cite{Gibson:2006gi} in tissue patterning, resulting in distinctive topologies including developmental compartment boundaries \cite{Major:2006ir}, hierarchical hexagonal patterning in the retina \cite{Hayashi:2004it, Hilgenfeldt2008}, and multicellular rosettes during germband extension \cite{Blankenship:2006jg}. Contractility-based sorting has also been characterized in germ-layer organization during zebrafish gastrulation \cite{Krieg2008}. ``Checkerboard'' cellular patterns of sensory hair cells and supporting cells have been observed in the auditory epithelium of the mouse cochlea \cite{Togashi2011}. More complicated finger-like or labyrinth-like tissue patterning have also been attributed to reaction-diffusion (Turing) mechanisms \cite{Kondo:2010bx}. Further, Lim and coworkers have demonstrated synthetic cell-cell adhesions that are fully modular and tunable, enabling rational design of multicellular architecture \cite{stevens_programming_2023}. Given this rich diversity of tissue architectures generated experimentally and \textit{in silico}, an emerging challenge is to achieve an automated and unbiased classification of distinct spatial patterns \cite{bishop_2016}. An intriguing possibility is to implement an interpretable and computationally-efficient machine learning framework that can be generalized to classify spatial patterns comprised of multiple interacting cell types.

Topological data analysis (TDA) is a promising approach for machine learning of high-dimensional architectures that extracts the ``shape'' of a dataset based on spatial connectivity \cite{carlsson2009topology}. TDA considers how discrete data points may be connected pairwise (dimension 0 homology) or linked into closed loops around an empty region (dimension 1 homology). Persistent homology then treats the stability of these connected structures at varying spatial scales \cite{herbertedelsbrunner2009}. Topological features can then be represented using a characteristic persistence ``barcode'' or ``diagram,'' and their relative ``similarity'' can be determined based on the ``cost'' of rearranging one diagram to resemble the other \cite{Amezquita:2020bi}.  Such topological approaches have recently been used to visualize the spatial organization of a single motile species \cite{Topaz:2015gd, Ulmer:2019hx, atienza2019persistent, Bhaskar:2019hj, McGuirl201917763, skinner2021topological, Nardini2021,Stolz2022} and particulate systems \cite{kramar_quantifying_2014}. These past investigations mostly utilized connected components (dimension 0 homology) to analyze populations of fixed size that were near confluency. However, a comparison of two populations with differing size will be biased since the number of connected components is not identical. Our recent work has shown that classification based on closed loops (dimension 1 homology) is a robust approach to classify spatial patterns with varying population size, particularly in the presence of empty regions \cite{bhaskar2020topological}.

Alternatively, persistence images represent topological features based on the weighted sum of Gaussian features, which yields a standardized finite vector representation that is amenable to machine learning \cite{Adams2017}. These persistence images can be compressed into a fixed-length numerical representation using dimensionality reduction and manifold learning techniques such as Principal Component Analysis (PCA), Uniform Manifold Approximation and Projection (UMAP) \cite{McInnes2018}, Potential of Heat Diffusion for Affinity-based Transition Embedding (PHATE) \cite{Moon2019} and autoencoder (AE) \cite{bank_autoencoders_2021,baldi_autoencoders_2012}. In principle, unsupervised hierarchical clustering can then be applied to determine similar spatial structures and compared to some ground truth. Nevertheless, the effectiveness of persistence images for classifying spatial patterns comprised of multiple species has not been previously addressed. 

In this article, we investigate the use of persistence images for unsupervised classification of multicellular spatial patterns that emerge from two interacting cell types. We systematically tune homotypic and heterotypic interactions between cell types to generate distinct architectures ranging from dispersed individuals to intermixed clusters to partially or wholly sorted clusters, as well as more complex striped phases, hierarchical hexagonal patterns, and rosettes. The spatial organization of these patterns is then represented by persistence images, persistence curves, or classical order parameters. Further, dimensionality reduction and hierarchical clustering was performed based on dimension 0 and/or dimension 1 homology for each cell type, as well as accounting for both cell types. As a case study, we first considered pattern formation in populations of fixed size. We then considered pattern formation in populations where cells can proliferate with contact inhibition. This work establishes the importance of topological features such as connected components (dimension 0 homology) and closed loops (dimension 1 homology) for classifying spatial patterns, as well as information about each cell type. Altogether, we envision this computational approach will enable new quantitative insights into the emergence of complex tissue architectures via spatiotemporal interactions between multiple cell types.

\section*{Materials and Methods}

\begin{figure}[ht]
\centering
\includegraphics[width=5.2in]{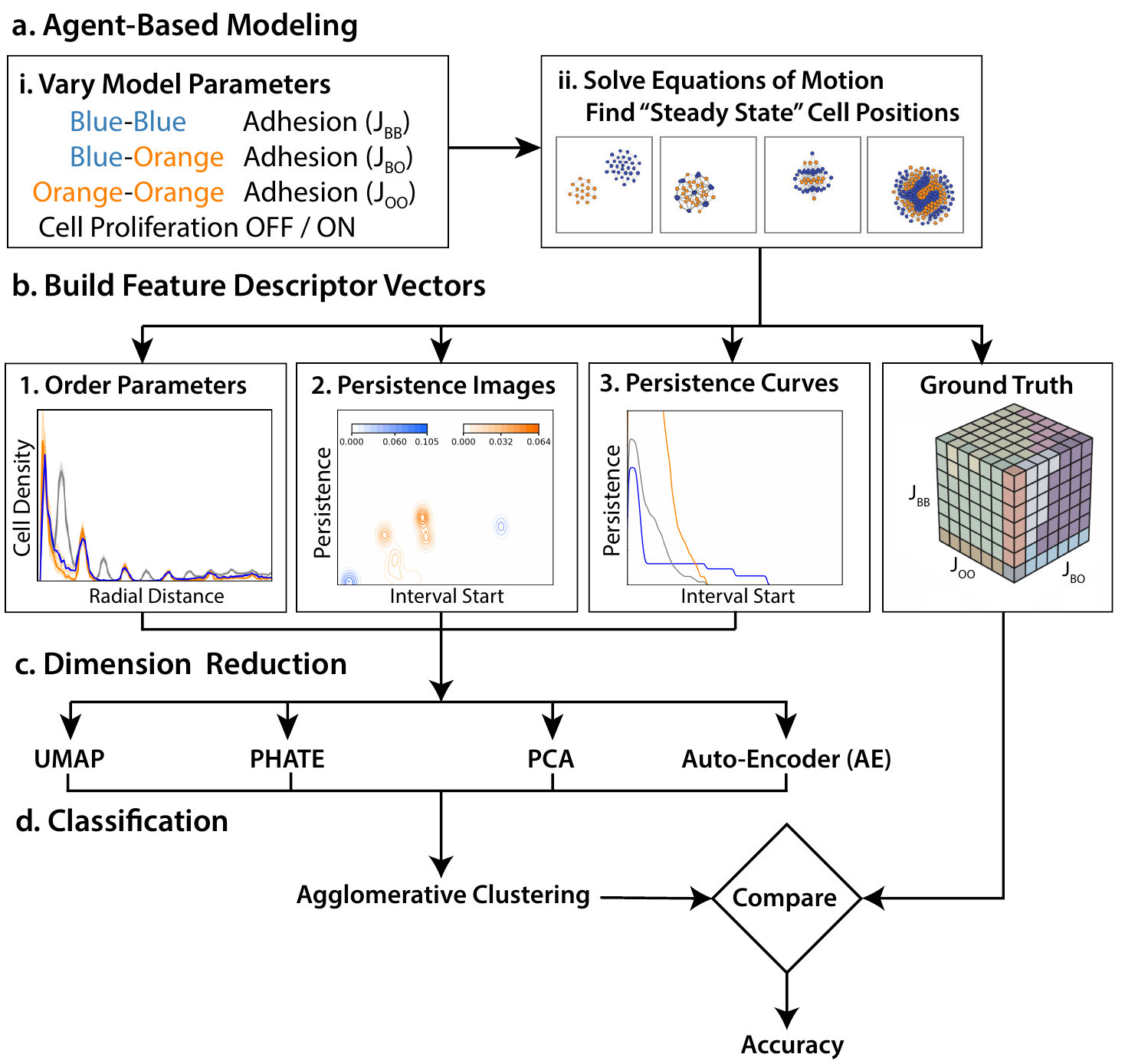}
\caption{{\bf Flow diagram of the proposed methodology for unsupervised classification of cellular patterns in the co-culture model.} (a) Adhesion parameter sweep simulations of an agent-based model at varying and constant population sizes. (b) Cell positions at steady state are featurized using persistent homology and order parameters. (c) Fixed length feature vectors are dimension reduced and classified using hierarchical clustering. Accuracy is computed by comparing to ground truth.}
\label{fig:persistence}
\end{figure}

We investigated how two different types of interacting cells (discrete agents) self-organize into multicellular patterns as cell-cell adhesion was varied (\textbf{Fig.~\ref{fig:persistence}a}). For simplicity, cells were defined as either ``blue'' or ``orange,'' and a total of $512$ different combinations with varying adhesion between blue-blue, orange-orange, or blue-orange cell types were simulated at constant population size (i.e. no proliferation) (\textbf{Fig.~\ref{fig:persistence}a,i}). These simulations were then implemented for another $343$ different conditions (3 replicates each) where cells were permitted to proliferate when there was unoccupied space nearby (i.e. contact inhibited proliferation). The equations of motion for each cell was solved for self-propulsion with random re-orientation, until cells reached some ``steady-state'' configuration after 5,000,000 time-steps (\textbf{Fig.~\ref{fig:persistence}a,ii}). These varying multicellular patterns were subsequently considered for classification.

Steady-state multicellular patterns were then manually classified to define a ``ground truth'' (\textbf{Fig.~\ref{fig:persistence}b,i}). These patterns were further converted to persistence images, which can also be represented as feature descriptor vectors (\textbf{Fig.~\ref{fig:persistence}b,ii}). For comparison, these patterns were also converted to feature vectors based on conventional order parameters that represent radial or angular symmetries of blue, orange, or both sets of cells (\textbf{Fig.~\ref{fig:persistence}b,iii}). Finally, these multicellular patterns were converted to normalized persistence curves. Unsupervised classification of these feature descriptor vectors was then implemented using UMAP \cite{McInnes2018}, PHATE \cite{Moon2019}, PCA and autoencoder (AE) \cite{bank_autoencoders_2021,baldi_autoencoders_2012} (\textbf{Fig.~\ref{fig:persistence}c}) for comparison with the ground truth classification (\textbf{Fig.~\ref{fig:persistence}b,i}).

\subsection*{Agent-Based Modeling of Two Interacting Cell Types}

\begin{figure}[!h]
\centering
\includegraphics[width=8cm]{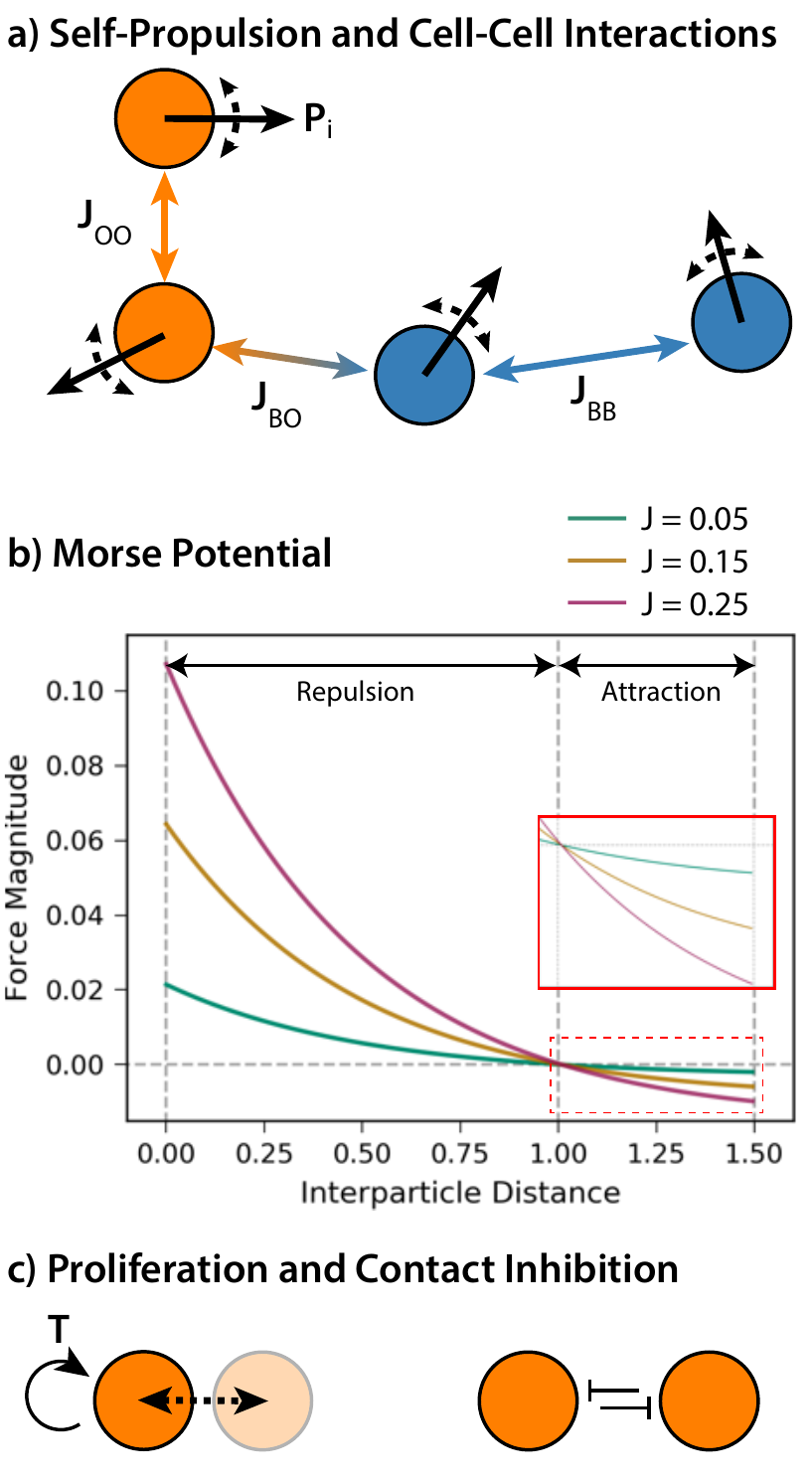}
\caption{{\bf Schematic of the agent-based model.}
Cells are modeled as rigid bodies undergoing non-inertial motion. (a) Homotypic and heterotypic cell-cell adhesion parameters. (b) Cell-cell adhesion varies with inter-particle distance based on the Morse potential. (c) Cell proliferation is implemented using an internal cell cycle timer. Additionally, cells with greater than four neighbors are unable to proliferate.}
\label{fig:agentmodel}
\end{figure}

Cells were represented as rigid-body agents that undergo non-inertial motion (Eq.~\ref{eqn:ABM_FE}) due to a random self-propulsion force, $\mathbf{P}_i^t$, and cell-cell interactions governed by the attraction-repulsion force, $\mathbf{F}_{ij}^t$ (\textbf{Fig.~\ref{fig:agentmodel}a}). Cell positions were initialized in a $[-20,20]\times[-20,20]$ simulation box with periodic boundaries on all sides. The equation of motion was given by:

\begin{equation}
\label{eqn:ABM_FE}
    \mathbf{x}_i^{t+\Delta t} = \mathbf{x}_i^t + \frac{\Delta t}{\eta}\Big( \mathbf{P}_i^t + \sum_{j \ne i}^{N(t)} \mathbf{F}_{ij}^t \Big)
\end{equation}
where $\mathbf{x}_i^t$ denotes the position of the $i$th cell at time $t$, $\Delta t$ denotes the time-step, and $\eta$ denotes a friction coefficient.

Two cells $i$ and $j$ in close proximity (within neighborhood distance $r_{\text{max}} = 1.5$) would attract or repel each other in accordance with an the attraction-repulsion force, $\mathbf{F_{ij}}$, which is the gradient of some potential $U(||\mathbf{x}_j-\mathbf{x}_i||)$ (Eq. \ref{eqn:ABM_Adh}) (\textbf{Fig.~\ref{fig:agentmodel}b}): 

\begin{equation}
\label{eqn:ABM_Adh}
    \mathbf{F}_{ij} = -\nabla U(||\mathbf{x}_j-\mathbf{x}_i||) \frac{\mathbf{x}_j - \mathbf{x}_i}{||\mathbf{x}_j - \mathbf{x}_i||} \mathbf{1}_{||\mathbf{x}_j - \mathbf{x}_i|| \le r_{\text{max}}}
\end{equation}

In particular, we define $U$ in terms of a Morse potential with a first term that represents long-range attraction and a second term that represents short-range repulsion (Eq. \ref{eqn:ABM_U}):

\begin{equation}
\label{eqn:ABM_U}
    U(r_{ij}) = -J_{ij}\Bigg[\text{exp}\Big(\frac{-r_{ij}}{l_A}\Big)-\frac{1}{4}\text{exp}\Big({-\frac{r_{ij}}{l_R}}\Big)\Bigg]
\end{equation}

where the adhesion parameter $J_{ij} = J(T(i),T(j))$ determines the magnitude of the attraction-repulsion potential, and depends on cell types $T(i)$ and $T(j)$ for cells $i$ and $j$ respectively, which we denote as blue or orange types (e.g. $\tau_B$ and $\tau_O$, respectively). We further define the characteristic length scale of the first long-range attraction term as $l_A=14.0$, and the second short-range repulsion term as $l_R=0.5$, based on our previous work with epithelial cells \cite{Leggett:2019hva, bhaskar2020topological}. Finally, the second short-range repulsion term was further scaled to be $1/4$ of the first long-range attraction term.\\

Cell positions were initialized at $t=0$ from a uniform distribution with rejection sampling to ensure a minimum separation of $1.0$ between all cells. The magnitude of the polarization force, $|\mathbf{P}|$, was fixed at $0.005$ and the adhesion parameter, $J$, was varied between $0.01$ and $0.25$. As a consequence, cells moved directionally at a constant speed for some duration, followed by random reorientation. This behavior has been experimentally validated for epithelial cells \cite{Potdar2010}, and is analogous to classical ``run and tumble'' models of bacteria.\\

The friction coefficient $\eta=1.0$, and time-step $\Delta t = 0.02$, were held constant throughout the simulation. Each simulation ran for $5,000,000$ time-steps. Repolarization takes place every $2,500$ time-steps, with an initial offset to prevent synchronization. In scenarios with proliferation, cell division was modeled with a cell cycle duration of $80,000$ time-steps, and permitted to occur as long as cells have less than 4 neighbors (i.e. contact inhibition of proliferation) (\textbf{Fig.~\ref{fig:agentmodel}c}). This scenario with both proliferation and sorting occur is reminiscent of tissue repair after damage, such as the regeneration of zebrafish skin patterns after laser ablation \cite{Yamaguchi2007}.

Altogether, for scenarios at constant cell number (without proliferation), a total of 512 combinations of adhesion parameters were simulated (8 different values of $J_{BB}$ $\times$ 8 different values of $J_{OO}$ $\times$ 8 different values of $J_{BO}$), with 3 independently initialized replicates. However, for scenarios with proliferation, simulations with low adhesion were not included since they were too computationally expensive and did not reach steady state over the timescales considered for the other simulations ($J \approx 0.00,\ 0.001$). Thus, only 216 combinations of adhesion parameters were simulated with proliferation (6 different values of $J_{BB}$ $\times$ 6 different values of $J_{OO}$ $\times$ 6 different values of $J_{BO}$), with 3 independently initialized replicates. 

\subsection*{Computation of Persistence Diagrams, Persistence Images, and Order Parameters}

\begin{figure}[h]
    \centering
    \includegraphics[width=14cm]{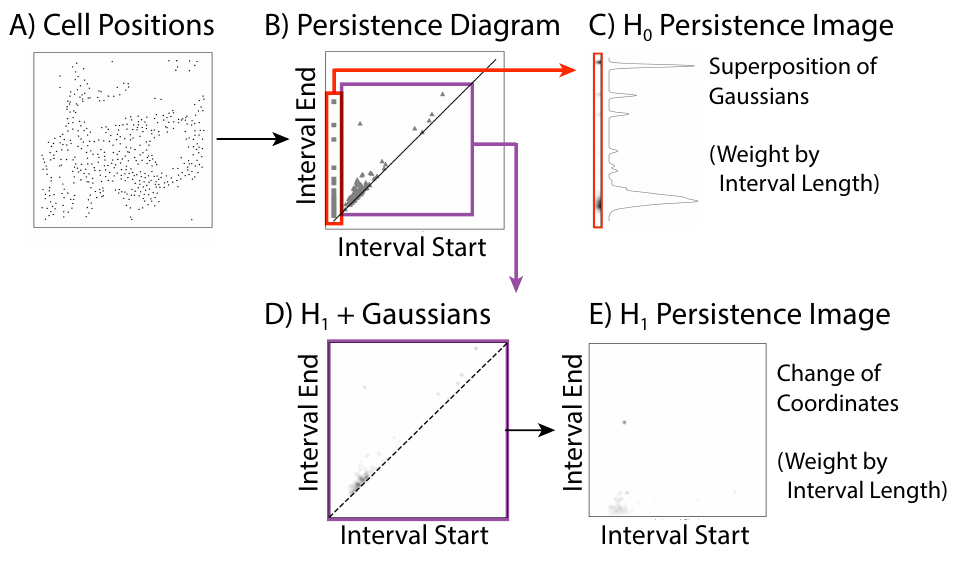}
    \caption{A Rips filtration is used on point cloud representing cell positions (A) to compute the persistence diagram (B). Persistence images (C, E) are generated by replacing interval start/end coordinates in the persistence diagram with a Gaussian weighted by distance from the diagonal. In dimension 0 homology (notated $H_{0}$, the interval start is always 0 (each cell position is a connected component), resulting in a 1D image (C). In dimension 1 homology (notated $H_{1}$), the intervals represent topological loops arising at non-zero values of the filtration radius, resulting in a 2D image.}
    \label{fig:pdtopersimg}
\end{figure}

Given a point cloud representing cell positions, we extracted topological features by constructing a chain of simplicial complexes based on a ``proximity parameter'' $\epsilon$ (separation distance between cell centroids) (\textbf{Fig.~\ref{fig:pdtopersimg}A}). Specifically, we used Euclidean distance between the cell positions to compute persistent homology via the Vietoris-Rips filtration. To avoid confusion, we denote these distances as ``interval start'' and ``interval end'' rather than ``birth'' and ``death,'' which have different meanings in cell biology and topology literature. We used TDA to obtain a list of (\emph{start, end}) pairs for topological features. These \emph{start} and \emph{end} pairs can be represented using a barcode diagram, or a persistence diagram for visualization (\textbf{Fig.~\ref{fig:pdtopersimg}B}). In the worst case scenario, this computation is  performed in $O(n^3)$ time, where $n$ is the number of generators of the filtered complex~\cite{zomorodian_computing_2005}, although this has been sped up using the standard reduction algorithm. In order to perform machine learning on simulation data, it was helpful to encode the topological information into vectors, so persistence images were introduced as a vectorized representation of topological features that are encoded in persistence diagrams.

Persistence images summarize the topological features using an intensity function over a measurement of length connected to the $\epsilon$ radius \cite{Adams2017}. Generally, in order to compute a persistence image, we take the points from the persistence diagram and place Gaussians centered around each point, additionally weighting the Gaussians so that more significant points have larger weight. Then we take the sum over these Gaussians to produce an intensity. However, depending on the dimensionality of the feature we are interested in, this process, as well as the shape of the output, varies. For each simulation at ``steady state'', persistence images were computed, using only orange cell positions, only blue cell positions, and both cell type positions together. In each case, we calculated the persistence images in dimensions 0 and 1. As an example, we now discuss further details for the computation of persistence images for dimension 0 and dimension 1.

In dimension 0 homology, we consider only the connected component features and place Gaussians at each point weighted by (\emph{end - start}) resulting in a 1-dimensional intensity function (\textbf{Fig.~\ref{fig:pdtopersimg}C}). Mathematically,

\begin{align*}
    \text{PI}(x) = \sum\limits_{(start,end) \in \mathcal{D}} (end - start) g_{(start),\sigma}(x)
\end{align*} 
where $\mathcal{D}$ is the persistence diagram and $g_{(start),\sigma}$ is a Gaussian with mean $start$ and variance $\sigma^{2}$\\

In dimension 1 homology, the presence/absence of topological loops is indicated by (\emph{start, end}) coordinates. We transform the coordinate system to measure persistence along the $y$-axis, (\emph{start, end - start}), and place Gaussians around these new points, weighted by the distance to the diagonal (\textbf{Fig.~\ref{fig:pdtopersimg}D}). In other words, weight each Gaussian by $\sqrt{(\frac{start - end}{2})^{2} + (\frac{end - start}{2})^{2}}$. The result is a 2-dimensional intensity function, with higher intensity for more persistent features (\textbf{Fig.~\ref{fig:pdtopersimg}E}). Mathematically,

\begin{align*}
\text{PI}(x,y) = \sum\limits_{(start,end) \in \mathcal{D}} (end - start) g_{(start,end - start),\Sigma}(x,y)
\end{align*} 
where $g_{(start,end-start),\Sigma}$ is a 2-dimensional Gaussian centered at (\emph{start, end-start}) with covariance matrix $\Sigma$.

Persistence curves are an alternative method of summarizing the persistence diagram by defining a function along its diagonal. Consider a point $(t,t)$ on the diagonal of the persistence diagram $\mathcal{D}$, which defines a  rectangular region where (\emph{start, end}) $\in [0,t] \times [t, \infty)$. We apply a function $\psi$ to all points (interval coordinates) within this region and compute a summary statistic $T$, resulting in a scalar value $\textsc{PC}(t)$. More formally, we define the persistence curve, $\textsc{PC}$ as a real-valued function over the diagonal, \cite{chung_smooth_2020}

\begin{align*}
    \textsc{PC}(\mathcal{D},\psi,T)(t) = T(\psi(\mathcal{D};b,d,t)|(b,d)\in \mathcal{D}_{t}), t\in \mathcal{R} 
\end{align*}

There are various options for choosing the functions $\psi$ and $T$. One example is the \emph{Betti curve}, where $\psi$ is the indicator function and $T$ is the summation operator. The Betti curve counts the number of points in the $[0, t] \times [t,\infty)$ rectangular region for all points $t$ along the diagonal. Another example is the \emph{life curve}, where $\psi$ is \emph{(end - start)} and $T$ is the summation operator. Here, we implement the Gaussian persistence curve \cite{chung_gaussian_2022}, where $\psi$ places a Gaussian of fixed bandwidth at each point in the $[0, t] \times [t,\infty)$ region, weighted by distance from the diagonal, and $T$ is the integral sum. One benefit of using Gaussian persistence curves is efficiency: they are easy to implement and fast to compute, with theoretical guarantees on stability and injectivity \cite{chung_gaussian_2022}. They have previously been applied to problems in image classification and neuroscience \cite{JMLR:v12:gonen11a,chung_topological_2018,chung_smooth_2020,barnes_comparative_2021}. Although we focus on persistence images in the main text, persistence curves are benchmarked in the supporting information and represent a promising avenue for future work.

\subsection*{Dimensionality Reduction and Classification}

Persistence images computed from cell positions were concatenated and compressed to a low-dimensional feature vector. Dimensionality reduction was then performed on the feature vectors using principal component analysis (PCA), Uniform Manifold Approximation and Projection (UMAP) \cite{McInnes2018}, Potential of Heat-diffusion for Affinity-based Trajectory Embedding (PHATE) \cite{Moon2019}, and an autoencoder (AE) \cite{bank_autoencoders_2021}, which facilitated visualization (in 2-D) and unsupervised classification using hierarchical clustering. Briefly, PCA projects data in low dimensions using a linear combination of features to maximize variance along each principal component. UMAP is a learned embedding based on ideas from manifold learning and TDA that constructs a high-dimensional graph representation of the data and optimizes a low-dimensional graph embedding to preserve structural similarities. PHATE computes an information distance based on a diffusion operator constructed from affinity between data points. It leverages multi-dimensional scaling (MDS) to perform dimensionality reduction using the information distance metric. An autoencoder learns the identity map by passing data, $x$, through an encoder network, $E$, to obtain a low-dimensional latent representation, $z = E(x)$, which can be decoded to reconstruct the original data via the decoder network, $\hat{x} = D(z) = D(E(x))$. The autoencoder is trained without supervision using the reconstruction loss penalty, $\mathcal{L} = \| x - \hat{x} \|$. The classification results were determined by comparing cluster labels obtained using hierarchical clustering to ground truth phase classification. Agglomerative hierarchical clustering (ward linkage) \cite{mullner_modern_2011} was performed on $20$-dimensional representations obtained by dimensionality reduction, with a stopping criterion (i.e. dendrogram cutoff) to prevent merging when the number of clusters reached the total number of distinct phases in the ground truth.

\section*{Results}

\subsection*{Sorting and Clustering of Two Non-Proliferating Cell Types with Varying Adhesion}

Differential adhesion simulations were performed using a self-propelled particle model with randomly initialized particle positions and periodic boundaries. Two cell types, labelled in blue ($\tau_B$) and orange ($\tau_O$), were simulated at constant population size ($N = 200$ total, $60\%$ blue and $40\%$ orange) with systematically varying parameters governing blue-blue adhesion ($J_{BB}$), orange-orange adhesion ($J_{OO}$), and orange-blue adhesion ($J_{OB}$), respectively. A random self-propulsion force of constant magnitude, $|\mathbf{P}| = 0.005$, was applied to each particle throughout the simulation. The total simulation time was chosen to exceed the time taken to reach stable steady-state configurations for each set of parameter values (\textbf{Fig.~\ref{fig:SI_snaps_noprolif_log}, ~\ref{fig:SI_snaps_noprolif_linear}}). 

\begin{figure}
    \centering
    \includegraphics[width=12cm]{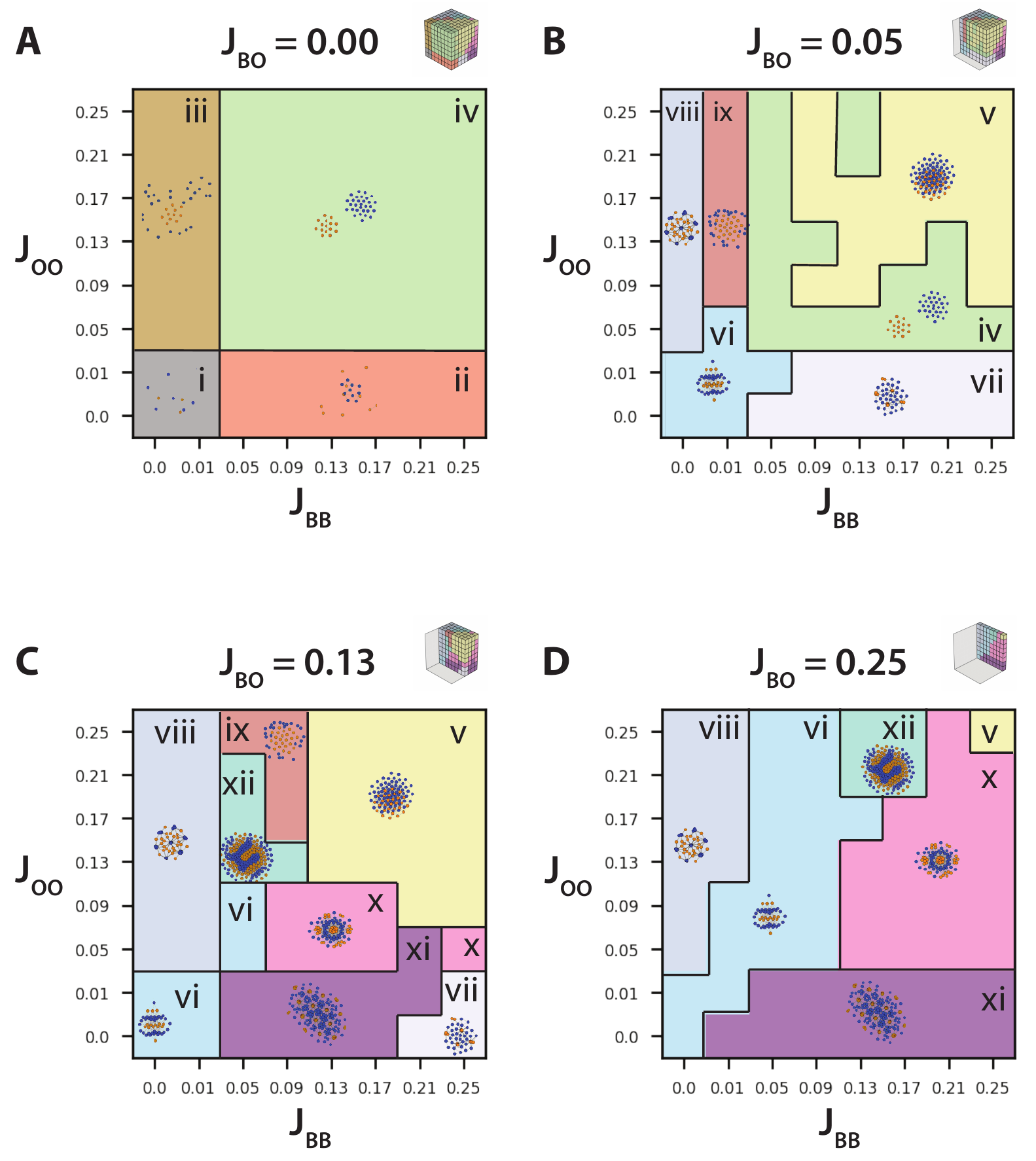}
    \caption{\textbf{Comparison of cluster and stripe patterning for two cell types (orange and blue) at constant population size with systematically varying blue-blue adhesion ($J_{BB}$), orange-orange adhesion ($J_{OO}$), and blue-orange adhesion ($J_{BO}$), respectively}. Representative slices with $J_{BB}$ vs $J_{OO}$ for $J_{BO} = 0.0$ (A), $J_{BO} = 0.05$ (B), $J_{BO} = 0.13$ (C), $J_{BO} = 0.25$ (D).}
    \label{fig:phasediagramnoprolif}
\end{figure}

In the limit of zero blue-orange adhesion ($J_{BO} = 0$) with weak blue-blue adhesion ($J_{BB} < 0.03$) and weak orange-orange adhesion ($J_{OO} < 0.03$), both orange and blue cells remained individually dispersed  (\textbf{Fig.~\ref{fig:phasediagramnoprolif}A,i, ~\ref{fig:SI_finalHT1}, ~\ref{fig:SI_finalHT2}}). At stronger blue-blue adhesion ($0.03 < J_{BB}$) and weak orange-orange adhesion ($J_{OO} < 0.03$), blue cells aggregated into clusters while orange cells remained individually dispersed (\textbf{Fig.~\ref{fig:phasediagramnoprolif}A,ii}). Conversely, at stronger orange-orange adhesion ($0.03 < J_{OO}$) and weak blue-blue adhesion  ($J_{BB} < 0.03$), orange cells aggregated into clusters while blue cells remained individually dispersed (\textbf{Fig.~\ref{fig:phasediagramnoprolif}A,iii}). Finally, at stronger blue-blue adhesion ($0.03 < J_{BB}$) and stronger orange-orange adhesion ($0.03 < J_{OO}$), blue cells aggregated into clusters and orange cells aggregated separately into clusters (\textbf{Fig.~\ref{fig:phasediagramnoprolif}A,iv}).

At slightly increased blue-orange adhesion ($J_{BO} = 0.05$), with comparable blue-blue ($J_{BB} = 0.05$) or orange-orange adhesion ($J_{OO} = 0.05$), sorting into separate blue and orange clusters was again observed (\textbf{Fig.~\ref{fig:phasediagramnoprolif}B,iv, ~\ref{fig:SI_finalHT3}, ~\ref{fig:SI_finalHT4}}). When blue-blue or orange-orange adhesions were stronger than blue-orange ($J_{BO} \le J_{OO}, J_{BO} \le J_{BB}$), a new type of cluster was observed with intermixed orange and blue cells (\textbf{Fig.~\ref{fig:phasediagramnoprolif}B,v}). This organization is reminiscent of ``checkerboard'' cellular patterns of sensory hair cells and supporting cells have been observed in mouse auditory epithelium of the cochlea \cite{Togashi2011}. Moreover, for weak blue-blue adhesion ($J_{BB} < 0.03$) and weak orange-orange adhesion ($J_{OO} < 0.05$), blue and orange cells organized into clusters with alternating stripes that were roughly 1-2 cells thick (\textbf{Fig.~\ref{fig:phasediagramnoprolif}B,vi}). These striped phases are somewhat reminiscent of those observed during skin patterning in zebrafish \cite{Volkening:2020jd}. At stronger blue-blue adhesion ($0.03 < J_{BB}$) and weak orange-orange adhesion ($J_{OO} < 0.03$), clusters consisted of orange cells dispersed in a hexagonal configuration, surrounded by blue cells.  (\textbf{Fig.~\ref{fig:phasediagramnoprolif}B,vii}). Conversely, at stronger orange-orange adhesion ($0.03 < J_{OO}$) and weak blue-blue adhesion  ($J_{BB} \approx 0$), clusters consisted of blue cells dispersed in a hexagonal configuration, surrounded by orange cells.  (\textbf{Fig.~\ref{fig:phasediagramnoprolif}B,viii}). Interestingly, this hierarchical patterning has a superficial resemblance to cone cells in ,the \emph{Drosophila} retina \cite{Hayashi:2004it}. At stronger orange-orange adhesion ($0.03 < J_{OO}$) and slightly stronger blue-blue adhesion  ($J_{BB} \approx 0.01$), clusters consisted of a core of orange cells surrounded by blue cells at the periphery (\textbf{Fig.~\ref{fig:phasediagramnoprolif}B,ix}).

At intermediate blue-orange adhesion ($J_{BO} = 0.13$), with stronger blue-blue or orange-orange adhesions  ($J_{BO} \le J_{OO}, J_{BO} \le J_{BB}$), clusters again were comprised of intermixed orange and blue cells (\textbf{Fig.~\ref{fig:phasediagramnoprolif}C,v, ~\ref{fig:SI_finalHT5}, ~\ref{fig:SI_finalHT6}}). For weaker blue-blue or orange-orange adhesions, ($J_{OO} \approx J_{BB} \le J_{BO}$), clusters with alternating orange and blue stripes were observed (\textbf{Fig.~\ref{fig:phasediagramnoprolif}C,vi}. At stronger blue-blue adhesion ($0.03 < J_{BB}$) and weak orange-orange adhesion ($J_{OO} < 0.03$), clusters consisted of orange cells dispersed in a hexagonal configuration, surrounded by tightly packed blue cells (\textbf{Fig.~\ref{fig:phasediagramnoprolif}C,xi}). When blue-blue adhesion and orange-orange adhesion were roughly comparable ($J_{BB} \approx J_{OO} \approx 0.09$), clusters consisted of tightly packed orange cells dispersed in a hexagonal configuration, surrounded by tightly packed blue cells (\textbf{Fig.~\ref{fig:phasediagramnoprolif}C,x}). At the strongest blue-blue adhesion ($J_{BB} \approx 0.20$) and weak orange-orange adhesion ($J_{OO} < 0.03$), spots of orange cells were again dispersed in a hexagonal configuration, surrounded by slightly sparser blue cells (\textbf{Fig.~\ref{fig:phasediagramnoprolif}C,vii}). At stronger orange-orange adhesion ($0.03 < J_{OO}$) and weak blue-blue adhesion  ($J_{BB} \approx 0$), clusters again consisted of blue cells dispersed in a hexagonal configuration, surrounded by orange cells.  (\textbf{Fig.~\ref{fig:phasediagramnoprolif}C,vii}). When orange-orange adhesion was comparable to blue-orange adhesion ($J_{OO} \approx J_{B0} = 0.13$) with weaker blue-blue adhesion ($J_{BB} = 0.07$), clusters were observed with finger-like or labyrinth-like patterns of orange and blue cells (\textbf{Fig.~\ref{fig:phasediagramnoprolif}C,xii}), analogous to those generated by reaction-diffusion (Turing) mechanisms \cite{Kondo:2010bx}. At strong orange-orange adhesion ($J_{OO} \approx 0.2$) with weak blue-blue adhesion ($J_{BB} \in [0.03, 0.11]$), clusters consisted of a core of orange cells surrounded by blue cells at the periphery (\textbf{Fig.~\ref{fig:phasediagramnoprolif}C,ix}). Notably, a number of cluster configurations observed at $J_{BO} = 0.13$  (\textbf{Fig.~\ref{fig:phasediagramnoprolif}C,v-ix}) were qualitatively similar to those previously observed at $J_{BO} = 0.05$ (\textbf{Fig.~\ref{fig:phasediagramnoprolif}B,v-ix}), but offset to higher values of $J_{BB}$ or $J_{OO}$.

At strong blue-orange adhesion ($J_{BO} = 0.25$) with comparable blue-blue and orange-orange adhesion ($J_{BB} \approx J_{OO}$), blue and orange cells organized into clusters with alternating stripes that were roughly 1-2 cells thick (\textbf{Fig.~\ref{fig:phasediagramnoprolif}D,vi, ~\ref{fig:SI_finalHT7}, ~\ref{fig:SI_finalHT8}}). At stronger blue-blue adhesion ($0.03 < J_{BB}$) and weak orange-orange adhesion ($J_{OO} < 0.03$), clusters again consisted of spots of orange cells dispersed in a hexagonal configuration, surrounded by tightly packed blue cells (\textbf{Fig.~\ref{fig:phasediagramnoprolif}C,xi}). Conversely, at stronger orange-orange adhesion ($0.03 < J_{OO}$) and weak blue-blue adhesion  ($J_{BB} \approx 0$), clusters again consisted of blue cells dispersed in a hexagonal configuration, surrounded by orange cells.  (\textbf{Fig.~\ref{fig:phasediagramnoprolif}D,viii}).  When orange-orange adhesion was comparable to blue-orange adhesion ($J_{OO} \approx J_{BO} =0.25$) with weaker blue-blue adhesion ($J_{BB} = 0.15$), clusters were observed with finger-like patterns of orange and blue cells (\textbf{Fig.~\ref{fig:phasediagramnoprolif}D,xii}).

When blue-blue adhesion and orange-orange adhesion were roughly comparable ($J_{BB} \approx J_{OO} \approx 0.13$), clusters consisted of tightly packed orange cells dispersed in a hexagonal configuration, surrounded by tightly packed blue cells (\textbf{Fig.~\ref{fig:phasediagramnoprolif}D,x}). When all three adhesions were comparable ($J_{BO} \approx J_{BB} \approx J_{OO} = 0.25$), clusters were observed with intermixed orange and blue cells (\textbf{Fig.~\ref{fig:phasediagramnoprolif}D,v}). Thus, we then performed unsupervised classification on the multicellular patterns for varying adhesion parameters (\textbf{Fig.~\ref{fig:phasediagramnoprolif}}) using persistence images (\textbf{Fig.~\ref{fig:SI_noprolif_PI}}), normalized persistence curves (\textbf{Fig.~\ref{fig:SI_noprolif_NPC}}), and order parameters (\textbf{Fig.~\ref{fig:SI_noprolif_OP}}).

\subsection*{Unsupervised Classification of Two Non-Proliferating Cell Types with Varying Adhesion}

\begin{figure}
    \centering
    \includegraphics[width=12cm]{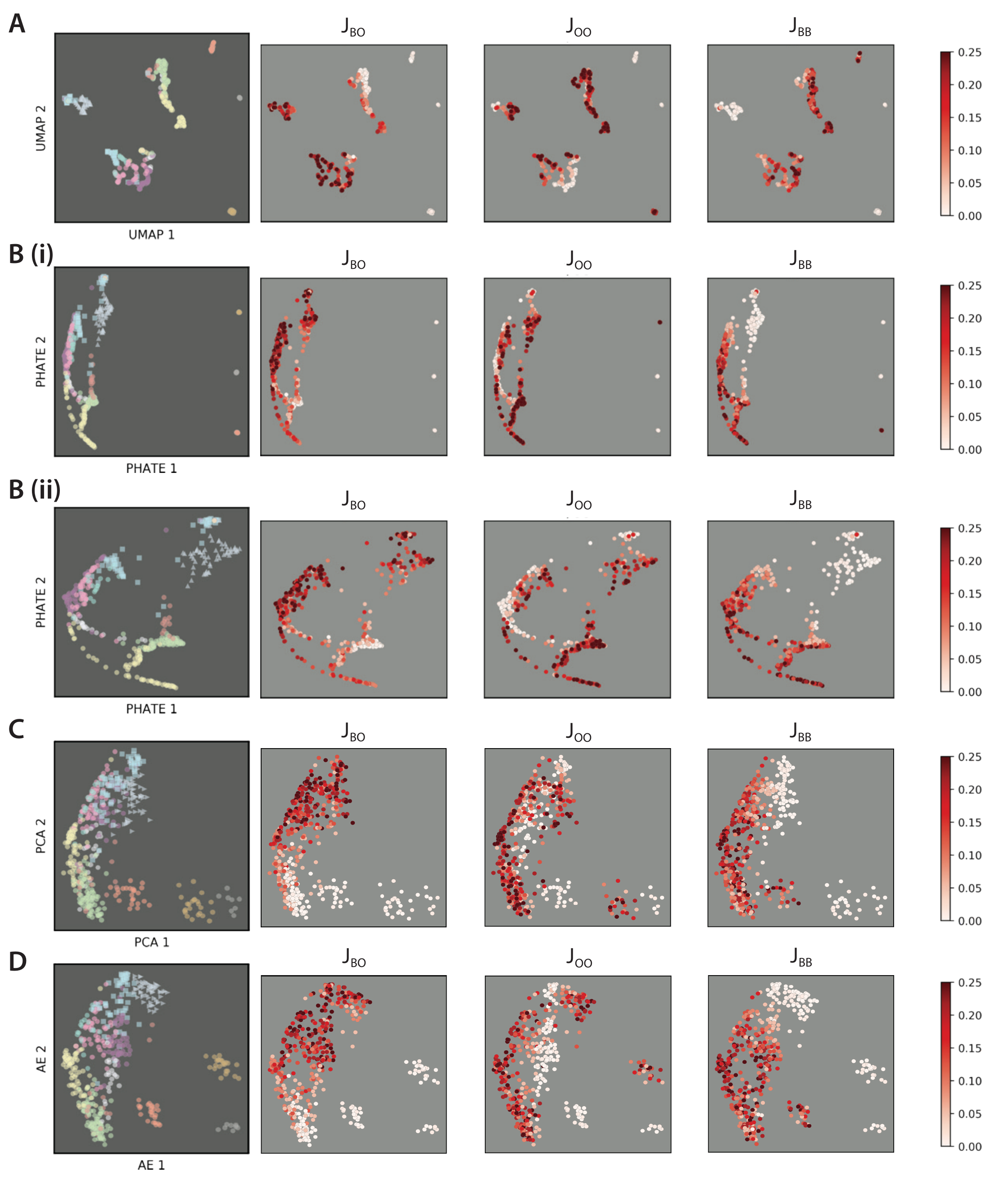}
    \caption{\textbf{2D embeddings of persistence images colored by ground truth and adhesion values for simulations with constant population size.} 2D embeddings obtained using UMAP (A), PHATE with all simulations (B,i) and zoomed in to only simulations with clusters (B,ii), PCA (C) and autoencoder (D).}
\label{fig:dimreductionnoprolif}
\end{figure}

Unsupervised classification of multicellular patterns using PCA, PHATE, AE, and UMAP was compared based on our ground truth labeling, and also colored by increasing values of $J_{BO}$, $J_{OO}$, and $J_{BB}$ (\textbf{Fig.~\ref{fig:dimreductionnoprolif}}). Ground truth labels ($12$ in total, denoted i - xii) were assigned based on manual inspection of particle configurations at the end of the simulation. For ease of visualization, axes were plotted so that the spatial configurations are more qualitatively consistent, as noted in the figure. In general, patterns with individually dispersed cells (e.g. i, ii, iii) were classified farther away from other patterns where both cell types were clustered (\textbf{Fig.~\ref{fig:dimreductionnoprolif}}). Further, for the crescent-like grouping of the remaining clustered cell patterns, $J_{BO}$ increased from right to left, $J_{OO}$ increased from the inside out, and $J_{BB}$ increased from left to right. Proceeding clockwise from the top right, the right arm of the crescent represented high $J_{BB}$ and $J_{OO}$ with low $J_{BO}$, and included sorted clusters (iv, green), partially sorted clusters (v, yellow), and orange clusters surrounded by blue cells (ix, brick red). Further, the center of the crescent represented high $J_{BB}$, low $J_{OO}$, and high $J_{BO}$, which included hexagonal configurations of orange cells (vii, gray; x, fuchsia; xi, purple). The left arm of the crescent represented low $J_{BB}$, low $J_{OO}$, and high $J_{BO}$, which included hexagonal configurations of blue cells (viii, gray blue). Finally, the upper left tip of the crescent represented low $J_{BB}$, high $J_{OO}$, and high $J_{BO}$, which included striped configurations (vi, sky blue; xii, turquoise). 

\begin{figure}
    \centering
    \includegraphics[width=12cm]{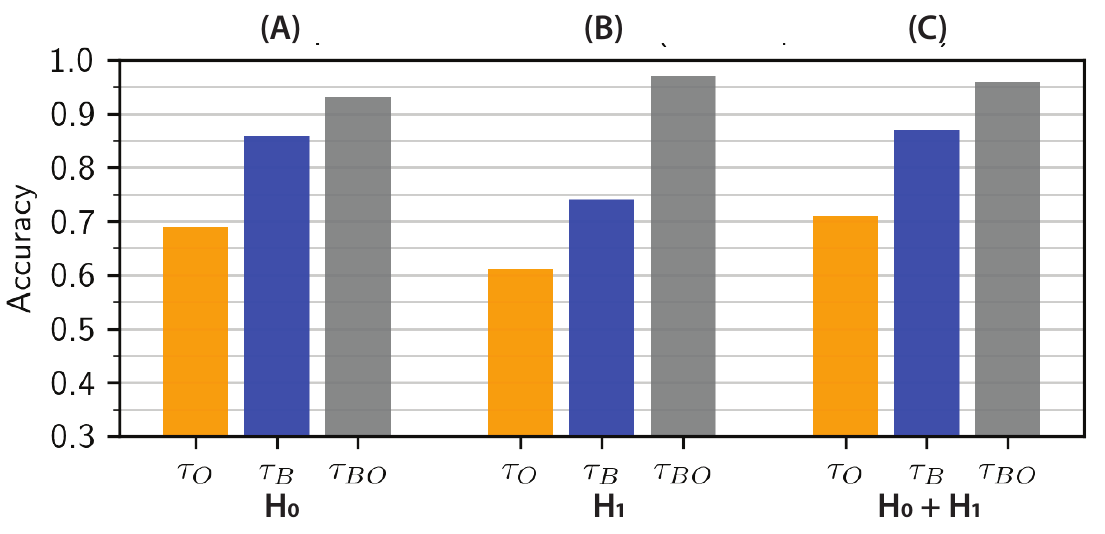}
    \caption[Unsupervised classification accuracy for persistence images at constant population size.]{\textbf{Unsupervised classification accuracy for persistence images at constant population size.} Unsupervised classification of simulations by hierarchical clustering of 20-dimensional autoencoder embeddings of persistence images. Classification accuracy is computed by comparing cluster labels to ground truth.}
\label{fig:persimg_classification_noprolif}
\end{figure}

The accuracy of unsupervised AE classification  (\textbf{Fig.~\ref{fig:dimreductionnoprolif}D}) relative to our manually annotated ground truth (\textbf{Fig.~\ref{fig:phasediagramnoprolif}}) was determined based on persistence images that considered only dimension 0 homology, only dimension 1 homology or both dimension 0 and dimension 1 (\textbf{Fig.~\ref{fig:persimg_classification_noprolif}, ~\ref{fig:SI_persimg_dim_noprolif}}). Further, unsupervised classification was performed using persistence images that included information on both blue and orange cells, blue cells only, and orange cells only (\textbf{Fig.~\ref{fig:persimg_classification_noprolif}}). 

Classification accuracy based on persistence images using only dimension 0 homology was moderately successful with only one cell type, ranging from 69\% for orange cells only, to 86\% for blue cells only, but increasing to 93\% for blue and orange cells combined (\textbf{Fig.~\ref{fig:persimg_classification_noprolif}A}). In comparison, classification accuracy using only dimension 1 homology was worse when using only orange cells (61\%) and blue cells (74\%), but considerably improved when considering both orange and blue cells (97\%) (\textbf{Fig.~\ref{fig:persimg_classification_noprolif}B}). Finally, classification accuracy using both dimension 0 and dimension 1 homology generally outperformed dimension 0 and dimension 1 only, ranging from 71\% for orange cells only, up to 87\% for blue cells only, and then 96\% for both orange and blue cells (\textbf{Fig.~\ref{fig:persimg_classification_noprolif}C}). Overall, classification accuracy was considerably better when considering blue cells only relative to orange cells only, which could be explained by the 60: 40 ratio of blue cells to orange cells in the simulations. Interestingly, combining information from both blue and orange cells incrementally improved classification accuracy for dimension 0 homology only (\textbf{Fig.~\ref{fig:persimg_classification_noprolif}A}), but resulted in much larger improvements for dimension 1 (\textbf{Fig.~\ref{fig:persimg_classification_noprolif}B}) only as well as dimension 0 and dimension 1 (\textbf{Fig.~\ref{fig:persimg_classification_noprolif}C}). For comparison, classification using persistence curves was slightly worse, ranging from 62\% to 83\% (\textbf{Fig. ~\ref{fig:SI_pc_classification}ABC, ~\ref{fig:SI_perscurve_dim_noprolif}}). Classification using radial order parameters was comparable to persistence images, ranging from 68\% to 94\% (\textbf{Fig. ~\ref{fig:SI_op_classification}ABC, ~\ref{fig:SI_op_dim_noprolif}}), but considerably worse for angular order parameters. Overall, the best classification occurred with persistence images and AE, considering both blue and orange cells for dimension 1 homology only (97\%), and comparable performance for dimension 0 and dimension 1 homology (96\%) (\textbf{Table ~\ref{tab:persimg_noprolif}, ~\ref{tab:perscurve_noprolif}, ~\ref{tab:op_noprolif}}). 

\subsection*{Sorting and Clustering of Two Proliferating Cell Types with Varying Adhesion}

Next, these differential adhesion simulations were then implemented with proliferation, so that a mother cell would divide into two daughter cells after some duration (e.g. 50,000 time-steps), randomly offset. Cell division events were implemented so that one daughter cell retained the velocity and direction of the mother cell, while the other daughter cell was placed nearby, but moving at equal velocity in the opposite direction. Moreover, a contact inhibition rule was included so that a cell was not permitted to divide if the local cell density was high (more than four nearest neighbors). Simulations were otherwise performed consistently with the previous scenario, starting with a 60:40 ratio of blue and orange cells while systematically varying blue-blue adhesion ($J_{BB}$), orange-orange adhesion ($J_{OO}$), and blue-orange adhesion ($J_{BO}$). Again, the total simulation time was chosen to exceed the time taken to reach stable steady-state configurations for each set of parameter values (\textbf{Fig.~\ref{fig:SI_snaps_prolif_log}, ~\ref{fig:SI_snaps_prolif_linear}}). For most of the representative simulations, the steady-state configuration was reached by 8-10 cell cycles  (15\% of the overall simulation duration), although certain scenarios involving the merging of clusters required up to 48 cell cycles (76\% of the overall simulation duration). Nevertheless, the multicellular stripe or spot organization within these clusters was typically evident by $\sim$10 cell cycles. This scenario with both proliferation and sorting was inspired by tissue repair, such as the regeneration of zebrafish skin patterns after laser ablation \cite{Yamaguchi2007}.

\begin{figure}
    \centering
    \includegraphics[width=12cm]{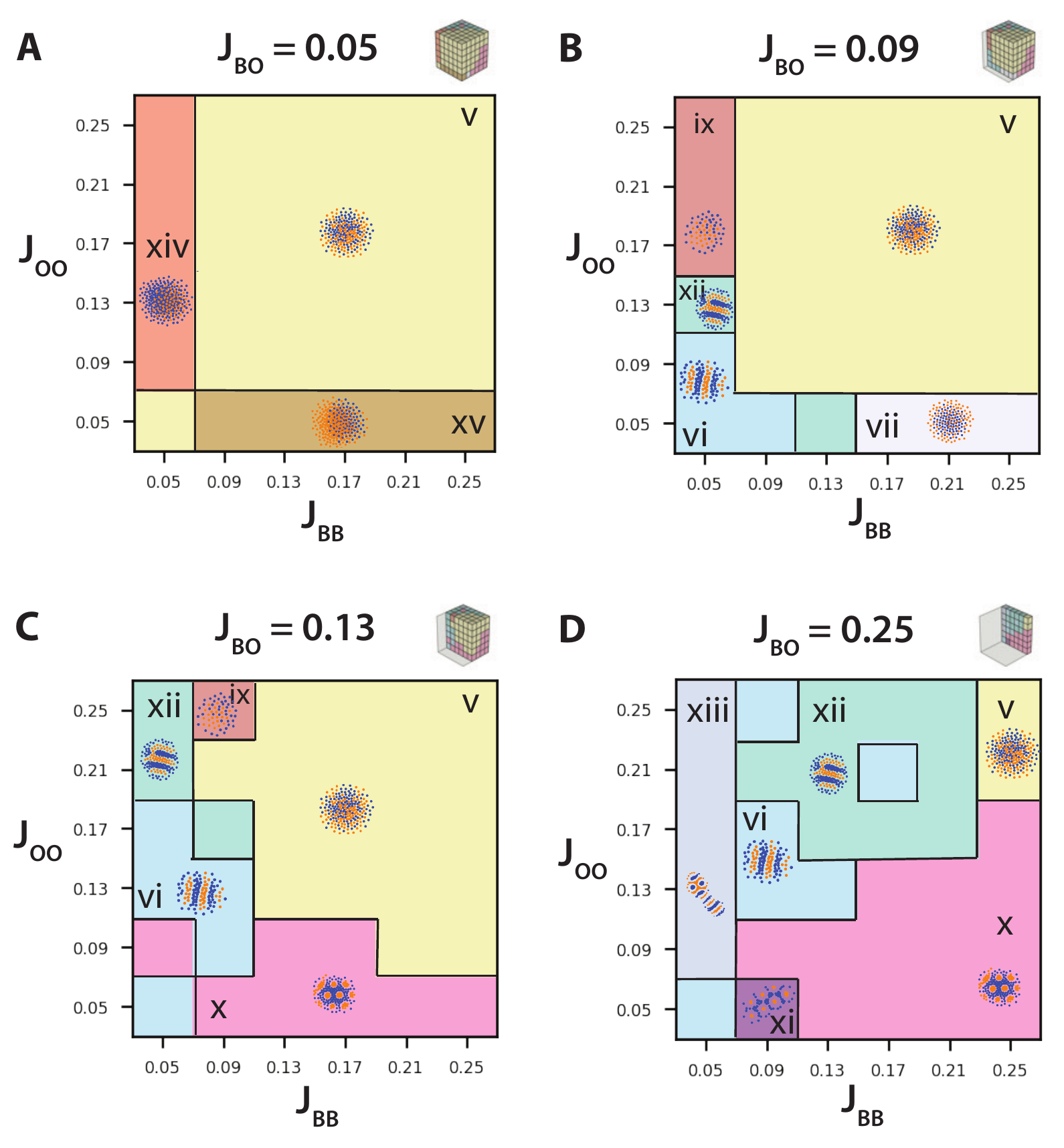}
    \caption{\textbf{Comparison of cluster and stripe patterning for two cell types (orange, blue) at varying population size with systematically varying blue-blue adhesion ($J_{BB}$), orange-orange adhesion ($J_{OO}$), and blue-orange adhesion ($J_{BO}$), respectively}. Representative slices with $J_{BB}$ vs $J_{OO}$ for $J_{BO} = 0.05$ (A), $J_{BO} = 0.09$ (B), $J_{BO} = 0.13$ (C), $J_{BO} = 0.25$ (D).}
    \label{fig:phasediagramprolif}
\end{figure}

For weak blue-orange adhesion $J_{BO} = 0.05$ with roughly comparable blue-blue and orange-orange adhesion ($J_{BB} \approx J_{OO}$), clusters appeared intermixed with irregular domain sizes (\textbf{Fig.~\ref{fig:phasediagramprolif}A,v, ~\ref{fig:SI_finalHT1p}}). We utilized the same numbering convention as in \textbf{Fig.~\ref{fig:phasediagramnoprolif}} for ease of comparison. Nevertheless, for weak blue-blue adhesions $J_{BB} = 0.05$ and varying orange-orange adhesions $0.07 \le J_{OO}$, clusters were partially sorted with more blue cells than orange cells (\textbf{Fig.~\ref{fig:phasediagramprolif}A,xiv}), which was not previously observed. By analogy, for weak orange-orange adhesions $J_{OO} = 0.05$ and varying blue-blue adhesions $0.07 \ge J_{BB}$, clusters were partially sorted with more orange cells than blue cells (\textbf{Fig.~\ref{fig:phasediagramprolif}A,xv}).

At slightly increased blue-orange adhesion $J_{BO} = 0.09$,  with roughly comparable blue-blue and orange-orange adhesion ($0.09 \le J_{BB} \approx J_{OO}$), clusters again appeared intermixed with irregular domain sizes (\textbf{Fig.~\ref{fig:phasediagramprolif}B,v, ~\ref{fig:SI_finalHT2p}}). In comparison, for weak blue-blue and orange-orange adhesion ($J_{BB} \approx J_{OO} \le 0.09 $), clusters with alternating orange and blue stripes were observed (\textbf{Fig.~\ref{fig:phasediagramprolif}B,vi}). For slightly increased blue-blue or orange-orange adhesions ($J_{OO} = 0.13, J_{BB} = 0.13$), clusters exhibited finger or labyrinth-like morphology (\textbf{Fig.~\ref{fig:phasediagramprolif}B,xii}). For strong blue-blue adhesions ($0.17 \le J_{BB}$) and weak orange-orange adhesions ($J_{OO} = 0.05$), clusters were sorted with  blue cells in the interior and orange cells at the periphery (\textbf{Fig.~\ref{fig:phasediagramprolif}B,vii}). In comparison, for strong orange-orange adhesions ($0.17 \le J_{OO}$) and weak blue-blue adhesions ($J_{BB} = 0.05$), clusters were sorted with  orange cells in the interior and blue cells at the periphery (\textbf{Fig.~\ref{fig:phasediagramprolif}B,ix}).

At intermediate blue-orange adhesion $J_{BO} = 0.13$, with roughly comparable blue-blue and orange-orange adhesion ($0.13 \le J_{BB} \approx J_{OO}$), clusters continued to be intermixed with irregular domain sizes (\textbf{Fig.~\ref{fig:phasediagramprolif}C,v, ~\ref{fig:SI_finalHT3p}, ~\ref{fig:SI_finalHT4p}}). For weak orange-orange adhesions ($J_{OO} < 0.13$), clusters exhibited hexagonally arrayed spots of packed orange cells surrounded by blue cells (\textbf{Fig.~\ref{fig:phasediagramprolif}C,x}). Instead, for weak blue-blue adhesions with comparable orange-orange and blue-orange adhesions ($J_{BB} < J_{OO} \approx J_{BO} = 0.13$), clusters exhibited a stripe phase (\textbf{Fig.~\ref{fig:phasediagramprolif}C,vi}). For weak blue-blue adhesions ($J_{BB} < 0.13$) with strong orange-orange adhesions ($J_{OO} \approx 0.21$), clusters exhibited a labyrinth or finger-like morphology (\textbf{Fig.~\ref{fig:phasediagramprolif}C,xii}). Finally, for slightly increased blue-blue adhesion ($J_{BB} = 0.09, J_{OO} = 0.25$), clusters exhibited a core of orange cells with peripheral blue cells (\textbf{Fig.~\ref{fig:phasediagramprolif}C,ix}).  

At the strongest blue-orange adhesion ($J_{BO} = 0.25$), intermixed clusters were only observed for comparably strong blue-blue and orange-orange adhesions ($J_{BB} \approx J_{OO} = 0.25$) (\textbf{Fig.~\ref{fig:phasediagramprolif}D,v, ~\ref{fig:SI_finalHT5p}, ~\ref{fig:SI_finalHT6p}}). When blue-blue and orange-orange adhesions were weaker but comparable ($J_{BB} \approx J_{OO} \le 0.21$), clusters with alternating orange and blue stripes were observed (\textbf{Fig.~\ref{fig:phasediagramprolif}D,vi}). At one location ($J_{BB} = 0.09, J_{OO} = 0.05$), clusters consisted of spots of orange cells dispersed in a hexagonal configuration, surrounded by tightly packed blue cells (\textbf{Fig.~\ref{fig:phasediagramprolif}D,xi}). For stronger blue-blue adhesions relative to orange-orange adhesions ($J_{OO} < J_{BB} \le 0.17$, clusters consisted of circular spots of orange cells surrounded by blue cells (\textbf{Fig.~\ref{fig:phasediagramprolif}D,x}). For weak blue-blue adhesions ($J_{BB} = 0.05$), a new cluster morphology was observed with mixed stripes and spots (\textbf{Fig.~\ref{fig:phasediagramprolif}D,xiii}). Finally, for stronger orange-orange adhesions relative to blue-blue adhesions ($J_{BB} < J_{OO} \le 0.25$), clusters exhibited a labyrinth or finger-like morphology (\textbf{Fig.~\ref{fig:phasediagramprolif}D,xii}). 

\subsection*{Unsupervised Classification of Two Proliferating Cell Types with Varying Adhesion}

\begin{figure}
    \centering
    \includegraphics[width=12cm]{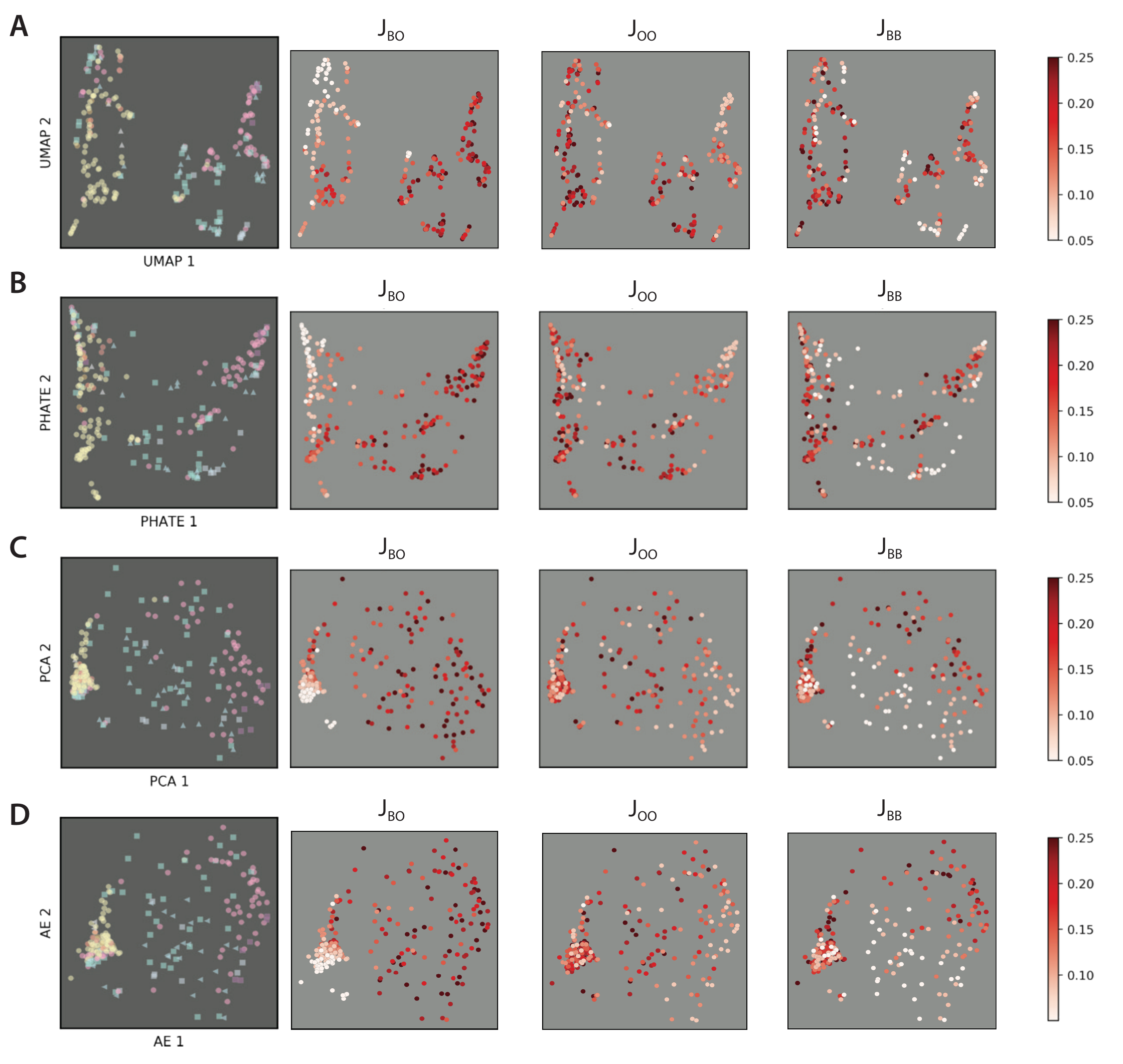}
    \caption{\textbf{2D embeddings of persistence images colored by ground truth and adhesion values for simulations with varying population size.} 2D embeddings obtained using UMAP (A), PHATE (B), PCA (C) and autoencoder (D).}
\label{fig:dimreductionprolif}
\end{figure}

Unsupervised classification of these multicellular patterns with proliferation was implemented using PCA, PHATE, AE, and UMAP for comparison with ground truth labeling, then plotted by increasing values of $J_{BB}$, $J_{OO}$, and $J_{BO}$ (\textbf{Fig.~\ref{fig:dimreductionprolif}}). Ground truth labels ($10$ in total, denoted v - vii, ix - xv) were assigned based on manual inspection of the final particle configurations. For ease of visualization, axes were again plotted so that the spatial configurations are more qualitatively consistent. In general, it was apparent that similarly classified conditions were more widely dispersed after dimensionality reduction for these simulations with proliferation relative to no proliferation (\textbf{Fig.~\ref{fig:dimreductionnoprolif}}), which can be attributed in part  to differences in total cell numbers. Since cells continued to proliferate until contact inhibited ``steady state,'' no simulations were observed with individually dispersed cells. 

After 2D embedding, UMAP and PHATE yielded roughly similar distributions (\textbf{Fig.~\ref{fig:dimreductionprolif}AB}), while PCA and AE also were comparable (\textbf{Fig.~\ref{fig:dimreductionprolif}CD}). Nevertheless,  conditions with different colors were poorly separated by UMAP and PHATE relative to PCA and AE (e.g. v, yellow and xv, tan). Roughly, $J_{BO}$ increased from top to bottom, $J_{OO}$ increased inward from the periphery, and $J_{BB}$ increased from left to right (\textbf{Fig.~\ref{fig:dimreductionprolif}}). Proceeding from the bottom upwards, the bottom grouping of cells represented high $J_{BB}$, low $J_{BB}$, and high $J_{BO}$, which included spots of orange cells in a hexagonal configuration (x, fuchsia and xi, purple) (\textbf{Fig.~\ref{fig:dimreductionnoprolif}}). Further, the middle grouping of cells represented low $J_{BB}$, high $J_{OO}$, and high $J_{BO}$, which included striped patterns (vi, sky blue; xii, turquoise; xiii, gray blue) (\textbf{Fig.~\ref{fig:dimreductionprolif}}). Finally, the top grouping of cells represented high $J_{BB}$, high $J_{OO}$, and low $J_{BO}$, which included intermixed clusters (v, yellow), partially sorted clusters (xiv, brick red; xv, tan), and clusters with orange cells at the interior and blue cells at the periphery (ix) (\textbf{Fig.~\ref{fig:dimreductionprolif}}). 

\begin{figure}
    \centering
    \includegraphics[width=14cm]{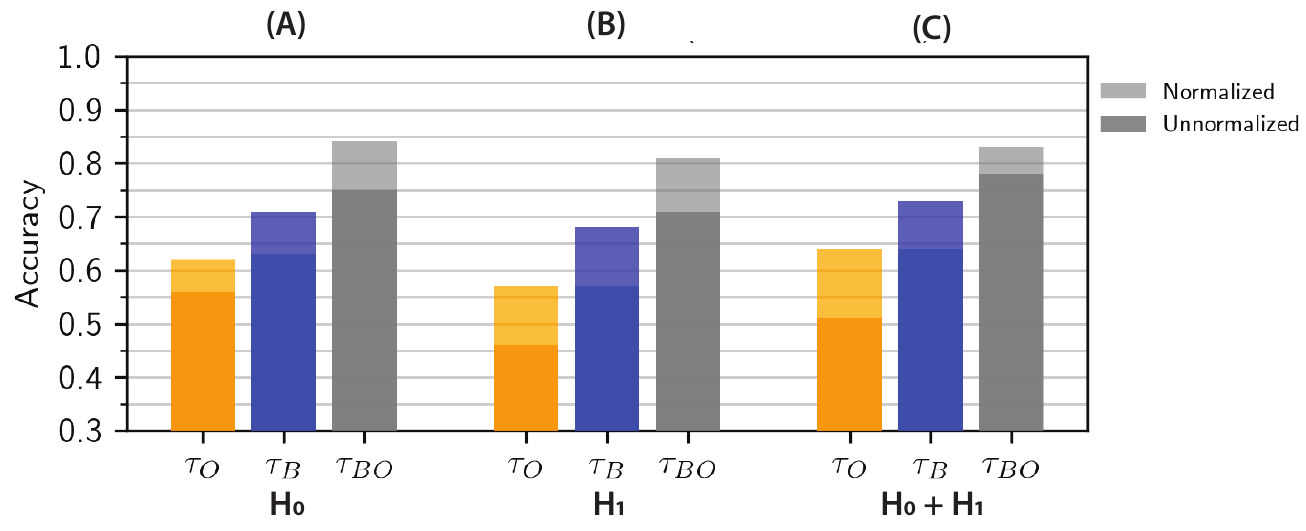}
    \caption[Unsupervised classification accuracy for persistence images at varying population size.]{\textbf{Unsupervised classification accuracy for persistence images at varying population size.} Unsupervised classification by hierarchical clustering of 20-dimensional autoencoder embedding of persistence images. Darker bars represent persistence images without normalization, whereas lighter bars represent the additional improvement after normalization by peak intensity per image. Classification accuracy of images with and without normalization is computed by comparing cluster labels to ground truth.}
\label{fig:persimg_classification_prolif}
\end{figure}

For these simulations with contact-inhibited proliferation, the accuracy of unsupervised AE classification (\textbf{Fig.~\ref{fig:dimreductionprolif}}) relative to our manually annotated ground truth (\textbf{Fig.~\ref{fig:phasediagramprolif}}) was again determined based on persistence images that considered only dimension 0 homology, only dimension 1 homology or both dimension 0 and dimension 1 (\textbf{Fig.~\ref{fig:SI_perscurve_dim_prolif}}). Further, unsupervised classification was performed using persistence images that included information on both blue and orange cells, blue cells only, and orange cells only. Notably, unsupervised AE classification for simulations with proliferation was considerably worse compared to simulations with constant cell number, particularly for dimension 1 homology or dimension 0 and 1 homology, which ranged from 37\% to 78\% (\textbf{Fig.~\ref{fig:persimg_classification_prolif}}). Qualitatively similar trends were also observed where classification accuracy tended to be worse when considering orange cells only, with some improvement for blue cells, as well as orange and blue cells, respectively. 

Persistence diagrams and images may be biased by different point cloud sizes, corresponding here to different numbers of cells, which likely affected the unsupervised classification. Thus, we  normalized the persistence images across all simulations by dividing each persistence image by its maximum intensity. As a consequence, classification accuracy for normalized images improved considerably by 5-10\%. For example, classification accuracy based on normalized persistence images improved to 61-84\% for dimension 0, to 56-81\% for dimension 1, and to 63-83\% for dimension 0 and 1, respectively (\textbf{Fig.~\ref{fig:persimg_classification_prolif}}). The general trend of improving classification from orange only to blue only to orange and blue remained consistent with normalization. For comparison, classification using persistence curves was comparable, ranging from 49\% to 84\% (\textbf{Fig. ~\ref{fig:SI_pc_classification}DEF, ~\ref{fig:SI_perscurve_dim_prolif}}). Classification using radial order parameters was also comparable to persistence images, ranging from 66\% to 88\% (\textbf{Fig. ~\ref{fig:SI_op_classification}DEF, ~\ref{fig:SI_op_dim_prolif}}), but considerably worse for angular order parameters. Further, classification of persistence curves and order parameters using PHATE was also associated with poorly separated groupings (\textbf{Fig. ~\ref{fig:SI_phate_repr_adh_vals}}).  Overall, dimension 0 only classification tended to outperform dimension 1 only classification, and dimension 0 and dimension 1 together tended to give the best classification. Moreover, accounting for both orange and blue cells tended to give comparable classification for persistence images, persistence curves, and order parameters (\textbf{Table ~\ref{tab:persimg_prolif}, ~\ref{tab:perscurve_prolif}, ~\ref{tab:op_prolif}}).

\section*{Discussion and Conclusion}

In this work, we investigated how two interacting and motile cell types (``blue'' and ``orange'') self-organize into multicellular patterns when the strength of blue-blue, blue-orange, or orange-orange adhesions are varied systematically.  Based on the relative positions of blue cells with respect to other blue cells and orange cells, as well as orange cells to other orange cells, these patterns could be represented using persistence images, persistence curves, and classical order parameters, then analyzed using dimensionality reduction and  hierarchical clustering. After optimization, unsupervised classification showed excellent agreement with our manually annotated ground truth (85-95\%) (\textbf{Fig.~\ref{fig:persimg_classification_noprolif}, ~\ref{fig:persimg_classification_prolif}}). Moreover, this classification is in good agreement with Steinberg's scaling arguments for differential adhesion \cite{Steinberg:1963kq}. For instance, the classifier reveals distinct regimes where both cell types are intermixed, which occurs when homotypic adhesions are weaker than heterotypic adhesion ($J_{BB}, J_{OO} < J_{OB}$, Fig.~\ref{fig:phasediagramnoprolif}v). Conversely, cells are sorted apart when heterotypic adhesions are much weaker than homotypic adhesion, ($J_{OB} << J_{BB}, J_{OO}$, Fig.~\ref{fig:phasediagramnoprolif}iv). We also classify a pattern where a core of orange cells surrounded by a shell of blue cells for mismatched homotypic adhesion with intermediate heterotypic adhesion ($J_{BB}<J_{OB}<J_{BB}$, ix). Our computational analysis is more granular, identifying distinct patterns driven by  more subtle differences in homotypic and heterotypic adhesion. For example, we observe discrete clusters of blue or orange cells of varying size arranged in checkerboard, stripe, or labyrinth patterns (Fig.~\ref{fig:phasediagramnoprolif}vi-xii).  These patterns are very robust for identical adhesion parameters with different initial conditions (Fig.~\ref{fig:SI_init_cond}). More recent conceptual models are based on interfacial tension, which address both cell-cell adhesion as well as cortical tension \cite{Lecuit:2007cw}. Our classified patterns are also in qualitative agreement with scaling arguments for interfacial tension (\cite{brodland_differential_2002, tsai_adhesion-based_2022}), noting that interfacial tension decreases with increasing adhesion. However, we note that Brodland's work uses a vertex-based model of cell shape, which yields certain numerical prefactors for the scaling argument. Thus, there are some quantitative differences with our agent-based model which does not consider interfacial tension and polygonal cell shape.


Based on manual inspection of the remaining discrepancies (5-15\%), we recognized that some conditions were located at the ``phase boundary'' between different regions, and could exhibit a mixture of different spatial patterns. Indeed, misclassified conditions were often far from the centroid of each grouping, which was also apparent from the 2D AE embedding (\textbf{Fig.~\ref{fig:SI_GMM_noprolif_H0}, ~\ref{fig:SI_GMM_noprolif_H1}, ~\ref{fig:SI_GMM_noprolif_H0p1}, ~\ref{fig:SI_GMM_prolif_H0}, ~\ref{fig:SI_GMM_prolif_H1}, ~\ref{fig:SI_GMM_prolif_H0p1}}). Although some caution is warranted when visually interpreting dimensionality reduced embeddings, we note that different colored groupings were relatively well separated by AE, particularly for the simulations with proliferation  (\textbf{Fig.~\ref{fig:dimreductionprolif},D}). In comparison, UMAP biased towards discrete clusters with some loss of global structure, while PHATE ``squeezed'' data points together into continuous branch-like structures \cite{Moon2019}. Thus, for simulations with proliferation, different colors are widely dispersed without clean separation in UMAP and PHATE embedding, including intermixed (v, yellow) and partially mixed particles (xv, brown), as well as stripes (vi, light blue) and spots (x, pink) (\textbf{Fig.~\ref{fig:dimreductionprolif},A,B}), which likely contributed to their markedly worse performance for classifying simulations with proliferation. For comparison, we have included supervised classification results obtained using a ``softer'' probabilistic algorithm where conditions can be classified by multiple adjacent regions. Briefly, we used a soft margin support vector machine (SVM), using the radial basis function for nonlinear transformation of the input, at various values of $C$ and computed the accuracy using $5$-fold cross-validation, which also exhibits excellent performance (\textbf{Table S7, S8}). 

Moreover, the classification was occasionally confounded by multicellular patterns with similar spatial organization but where the positions of the ``blue'' and ``orange'' cells were switched. For instance, consider a scenario where hexagonal arrays of orange clusters are surrounded by blue cells occur at high $J_{BB}$ and low $J_{OO}$,  (\textbf{Fig.~\ref{fig:phasediagramnoprolif}B,vii}). Switching the relative positions of blue and orange cells in this scenario results in hexagonal arrays of blue clusters surrounded by orange cells at low $J_{BB}$ and high $J_{OO}$ (\textbf{Fig.~\ref{fig:phasediagramnoprolif}B,viii}). The input to the classifier consists of feature vectors representing the persistence images of blue, orange, or blue and orange cells, which are weighted equally without explicitly specifying a ``cell type.'' Thus, from the perspective of the unsupervised classifier, these two configurations appeared indistinguishable, especially given that the effective blue and orange cell sizes were identical based on the interaction potential (Eq. \ref{eqn:ABM_U}). Analogous issues occurred for unsupervised classification using only one cell type (e.g. orange) without consideration of the other cell type (e.g. blue). For example, if only orange cells were considered, hexagonal arrays of blue clusters surrounded by orange cells appeared relatively similar to clusters with blue cells at the interior and concentric rings of orange and blue, particularly in dimension 1 homology that only considers topological loops (\textbf{Fig.~\ref{fig:SI_H1_misclassification}A}). Instead, if only blue cells were considered (and dimension 1 homology), hexagonal arrays of orange clusters would appear very similar to hexagonal arrays of orange individuals (\textbf{Fig.~\ref{fig:SI_H1_misclassification}B}). Similarly, if the spatial configuration of orange cells was comparable in scenarios with very different blue cell configurations, consideration of only orange cells in dimension 0 homology would also result in misclassification  (\textbf{Fig.~\ref{fig:SI_H0_misclassification}B}). The use of feature vectors representing spatial information about both blue and orange cells typically yielded improved classification. Nevertheless, classification using dimension 0 or dimension 1 homology usually resulted in comparable accuracy (using both colors), without significant improvement when both dimension 0 and dimension 1 were considered.

These simulations were initialized with a 60:40 ratio of blue to orange cells, which indirectly encodes a difference between cell types into the classifier. To check the robustness of these results, we repeated these simulations with varying domain sizes and verified that multicellular patterns occurred consistently, particularly the  organization of cells within clusters (\textbf{Fig.~\ref{fig:SI_domain_size},~\ref{fig:SI_domain_size_prolif}}). We note that classification accuracy remains consistently high across different domain sizes, both for constant population size as well as with proliferation (\textbf{Fig.~\ref{fig:SI_domain_size_classify}}). We also varied the cell type ratio from 90:10 to 10:90 over a comparable range of homotypic and heterotypic adhesion strengths (\textbf{Fig.~\ref{fig:SI_snaps_noprolif_ratios}}). Notably, increasing the  fraction of one cell type (e.g. blue) relative to the other (e.g. orange) was equivalent to increasing the relative homotypic adhesion (e.g. strengthening blue-blue adhesion relative to orange-orange). For example, a 90:10 ratio of blue:orange cells typically resulted in sparse orange cells surrounded by blue cells (\textbf{Fig.~\ref{fig:phasediagramnoprolif}D-H}). In comparison, a 10:90 ratio of blue:orange cells exhibited sparse blue cells surrounded by orange cells, also equivalent to switching the positions of blue and orange cells. Moreover, cells that could undergo contact-inhibited proliferation exhibited additional patterns with both stripes and spots (\textbf{Fig.~\ref{fig:phasediagramprolif}C,xiii}), as well as partially sorted patterns (\textbf{Fig.~\ref{fig:phasediagramprolif}C,xiv, xv}). We further verified that varying the proliferation rate with contact inhibition resulted in qualitatively similar spatial patterns when homotypic and heterotypic adhesions were held constant (\textbf{Fig.~\ref{fig:SI_snaps_prolif_rates}}). Overall, these artifacts of interchanging cells are unlikely to arise under more realistic conditions, since different cell types within a tissue exhibit appreciable differences in size or biomarker expression that would further inform unsupervised classification. We further considered simulations with three interacting cell types (50\% blue, 30\% orange, 20\% green), which resulted in analogous pattern formation as with two interacting cell types (\textbf{Fig.~\ref{fig:SI_multiple_cells_noprolif_100}, ~\ref{fig:SI_multiple_cells_noprolif}, ~\ref{fig:SI_multiple_cells_prolif}}). This algorithm also demonstrated excellent classification of these patterns based on the positions of two out of the three cell types, especially with blue and orange cells which comprise 80\% of the population (\textbf{Fig.~\ref{fig:SI_multicelltype_classify}}). We also classified blue-green and orange-green pairings, using the blue-orange ground truth phase classification as a reference. This resulted in a few misclassifications, primarily resulting from overlapping cell positions or blue and green cells (\textbf{Fig. ~\ref{fig:SI_multiple_cells_noprolif}D}), orange and green cells (\textbf{Fig. ~\ref{fig:SI_multiple_cells_noprolif}E}), as well as previously unseen spatial arrangements that emerged due to competition between two cell types for maximizing interaction with the third cell type (\textbf{Fig. ~\ref{fig:SI_multiple_cells_noprolif}F}).

Although self-organization into multicellular patterns occurs relatively robustly in development, the motility and interactions of individual cells are stochastic. Thus, an unsupervised classifier must be stable against variability that arises from biological ``noise.'' The transformation of coordinates to persistence diagrams in topological data analysis is provably stable with respect to bottleneck distance and Wasserstein distance~\cite{herbertedelsbrunner2009}. However, direct comparison of cell positions via bottleneck distance is undesirable, since the $L_\infty$ norm is entirely determined by a single topological feature. Comparisons based on the Wasserstein metric are computationally expensive, requiring the calculation of an optimal transport plan between pairs of persistence diagrams. Here, we have focused on persistence images as an underexplored approach for topological data analysis of multicellular pattern formation, relative to classical order parameters and persistence curves. Persistence images enable excellent unsupervised classification at intermediate computational cost with theoretical guarantees on stability. The Euclidean distance between persistence images based on the 2-D Gaussian kernel is bounded by the 1-Wasserstein distance between the corresponding persistence diagrams~\cite{Adams2017}. The kernel bandwidth and image resolution can be adjusted to achieve the desired trade-off between computational efficiency and stability guarantees. In comparison, persistence curves can be computed more efficiently but lead to worse classification. The loss of information by summing over the Gaussian kernel along the diagonal is offset by the greater computational efficiency of using a lookup table of cumulative distribution values in the calculation. In comparison, classical order parameters perform very well for classification, but are quite expensive computationally. One explanation for the relative success of classical order parameters, especially for simulations with proliferation, is that they have been normalized to account for different numbers of (identical) particles. In comparison, the optimum normalization for persistence images remains unresolved, and will be considered more thoroughly in a follow-up manuscript. 

In conclusion, we show that combining persistence images with autoencoders enables the unsupervised, computationally efficient classification of spatial patterns associated with two interacting cell types. This approach represents topological features of the multicellular architecture as the weighted sum of Gaussian features, yielding a standardized finite vector representation. We show that optimized dimensionality reduction using AE and hierarchical clustering can reveal topologically similar simulation conditions, in excellent agreement with our manually annotated ground truth, particularly for populations of fixed size. However, persistence images of simulations with varying population size required normalization for unbiased comparison, and performed slightly worse. In a follow-up manuscript, we will apply this approach to analyze stripe and spot patterning in zebrafish development, as well as to gain deeper mechanistic insight into persistence images. Overall, we envision that topology-based machine learning represents a powerful and human-interpretable framework to explore the diversity of complex tissue shapes and patterns that emerge from self-sorting and collective cell migration.

\section*{Acknowledgments}
This work was supported by the National Institute of General Medical Sciences (R01GM140108), a Brown University Data Science Initiative Seed Grant (D.B., W.Y.Z., and I.Y.W.), and the National Science Foundation (CCF-1740741, DMS-2038039, DMS-2106566) (B.S.). This research was conducted using computational resources and services at the Center for Computation and Visualization, Brown University.

\section*{Authorship}
\textbf{D.B.}: Conceptualization, Methodology, Software, Formal Analysis, Investigation, Visualization, Writing - Original Draft Preparation. \textbf{W.Y.Z.}: Methodology, Software, Formal Analysis,  Investigation, Visualization, Writing – Review \& Editing. \textbf{A.V.}: Methodology, Formal Analysis,  Writing – Review \& Editing. \textbf{B.S.}: Methodology, Formal Analysis,  Writing – Review \& Editing. \textbf{I.Y.W.}: Conceptualization, Methodology, Formal Analysis, Visualization, Writing - Original Draft Preparation, Supervision.

\section*{Competing Interests}
D.B. is currently supported by a Boehringer Ingelheim Fellowship at Yale University. All other authors declare no competing financial or non-financial interests.

\section*{Data Availability}
All simulation data generated for this study are available via an Open Science Foundation repository at \url{https://osf.io/md86n/?view_only=66ac8655cf1842e4bfe52ca0b8b59a04}\cite{bhaskar_code_2022} under an MIT License.

\section*{Code Availability}
All code used to perform the simulations, compute order parameters and persistent homology, and generate classification results is available on GitHub at \url{https://github.com/dbhaskar92/Coculture-ABM-Model} under an MIT license.


\bibliographystyle{unsrt}
\bibliography{references}

\begin{thebibliography}{10}

\bibitem{Lecuit:2007cw}
Thomas Lecuit and Pierre-Fran{\c c}ois Lenne.
\newblock {Cell surface mechanics and the control of cell shape, tissue
  patterns and morphogenesis.}
\newblock {\em Nat Rev Mol Cell Biol}, 8(8):633--644, August 2007.

\bibitem{tsai_adhesion-based_2022}
Tony Y.-C. Tsai, Rikki~M. Garner, and Sean~G. Megason.
\newblock Adhesion-{Based} {Self}-{Organization} in {Tissue} {Patterning}.
\newblock {\em Annu Rev Cell Dev Biol}, 38(1):349--374, 2022.

\bibitem{Graner:1992kx}
F~Graner and JA~Glazier.
\newblock {Simulation of biological cell sorting using a two-dimensional
  extended Potts model.}
\newblock {\em Phys. Rev. Lett.}, 69(13):2013--2016, September 1992.

\bibitem{Rieu:1998bf}
Jean-Paul Rieu, Naoki Kataoka, and Yasuji Sawada.
\newblock {Quantitative analysis of cell motion during sorting in
  two-dimensional aggregates of dissociated hydra cells}.
\newblock {\em Phys. Rev. E}, 57(1):924--931, January 1998.

\bibitem{Belmonte:2008dv}
Julio~M Belmonte, Gilberto~L Thomas, Leonardo~G Brunnet, Rita M~C de~Almeida,
  and Hugues Chat{\'e}.
\newblock {Self-Propelled Particle Model for Cell-Sorting Phenomena}.
\newblock {\em Phys. Rev. Lett.}, 100(24):220--4, June 2008.

\bibitem{Hogan:2009bc}
Catherine Hogan, Sophie Dupr{\'e}-Crochet, Mark Norman, Mihoko Kajita, Carola
  Zimmermann, Andrew~E Pelling, Eugenia Piddini, Luis~Alberto Baena-L{\'o}pez,
  Jean-Paul Vincent, Yoshifumi Itoh, Hiroshi Hosoya, Franck Pichaud, and
  Yasuyuki Fujita.
\newblock {Characterization of the interface between normal and transformed
  epithelial cells.}
\newblock {\em Nat Cell Biol}, 11(4):460--467, April 2009.

\bibitem{Beatrici:2011fz}
Carine~P Beatrici and Leonardo~G Brunnet.
\newblock {Cell sorting based on motility differences.}
\newblock {\em Phys Rev E Stat Nonlin Soft Matter Phys}, 84(3 Pt 1):031927,
  September 2011.

\bibitem{Mehes:2012fy}
El{\H o}d M{\'e}hes, Enys Mones, Val{\'e}ria N{\'e}meth, and Tam{\'a}s Vicsek.
\newblock {Collective motion of cells mediates segregation and pattern
  formation in co-cultures.}
\newblock {\em PLoS ONE}, 7(2):e31711, 2012.

\bibitem{Kabla:2012gm}
Alexandre~J Kabla.
\newblock {Collective cell migration: leadership, invasion and segregation}.
\newblock {\em J. R. Soc. Interface}, 9(77):3268--3278, July 2012.

\bibitem{Strandkvist:2014ie}
Charlotte Strandkvist, Jeppe Juul, Buzz Baum, Alexandre~J Kabla, and Tom Duke.
\newblock {A kinetic mechanism for cell sorting based on local variations in
  cell motility.}
\newblock {\em Interface Focus}, 4(6):20140013, December 2014.

\bibitem{Nielsen:2015jl}
Alexander~Valentin Nielsen, Annika~Lund Gade, Jeppe Juul, and Charlotte
  Strandkvist.
\newblock {Schelling model of cell segregation based only on local
  information}.
\newblock {\em Phys. Rev. E}, 92(5):488--4, November 2015.

\bibitem{GamboaCastro:2016da}
Marielena Gamboa~Castro, Susan~E Leggett, and Ian~Y Wong.
\newblock {Clustering and jamming in epithelial-mesenchymal co-cultures.}
\newblock {\em Soft Matter}, 12(40):8327--8337, October 2016.

\bibitem{carrillo_adhesion_2018}
José~Antonio Carrillo, Annachiara Colombi, and Marco Scianna.
\newblock Adhesion and volume constraints via nonlocal interactions determine
  cell organisation and migration profiles.
\newblock {\em Journal of Theoretical Biology}, 445:75--91, May 2018.

\bibitem{carrillo_population_2019}
Jose~A. Carrillo, Hideki Murakawa, Makoto Sato, Hideru Togashi, and Olena
  Trush.
\newblock A population dynamics model of cell-cell adhesion incorporating
  population pressure and density saturation.
\newblock {\em Journal of Theoretical Biology}, 474:14--24, August 2019.

\bibitem{Leggett:2019hva}
Susan~E Leggett, Zachary~J Neronha, Dhananjay Bhaskar, Jea~Yun Sim,
  Theodora~Myrto Perdikari, and Ian~Y Wong.
\newblock {Motility-limited aggregation of mammary epithelial cells into
  fractal-like clusters.}
\newblock {\em Proc. Natl. Acad. Sci. U.S.A.}, 116(35):17298--17306, August
  2019.

\bibitem{Li:2019bt}
Xinzhi Li, Amit Das, and Dapeng Bi.
\newblock {Mechanical Heterogeneity in Tissues Promotes Rigidity and Controls
  Cellular Invasion}.
\newblock {\em Phys. Rev. Lett.}, 123(5):058101, July 2019.

\bibitem{Krajnc:2020dd}
Matej Krajnc.
\newblock {Solid{\textendash}fluid transition and cell sorting in epithelia
  with junctional tension fluctuations}.
\newblock {\em Soft Matter}, 16(13):3209--3215, 2020.

\bibitem{Sahu:2020cc}
Preeti Sahu, Daniel~M Sussman, Matthias R{\"u}bsam, Aaron~F Mertz, Valerie
  Horsley, Eric~R Dufresne, Carien~M Niessen, M~Cristina Marchetti, M~Lisa
  Manning, and J~M Schwarz.
\newblock {Small-scale demixing in confluent biological tissues}.
\newblock {\em Soft Matter}, 16(13):3325--3337, 2020.

\bibitem{Dey:2021ke}
Supravat Dey and Moumita Das.
\newblock {Differences in mechanical properties lead to anomalous phase
  separation in a model cell co-culture.}
\newblock {\em Soft Matter}, 17(7):1842--1849, February 2021.

\bibitem{lucia2022cell}
Stephani~Edwina Lucia, Hyuntae Jeong, and Jennifer~H Shin.
\newblock Cell segregation via differential collision modes between heterotypic
  cell populations.
\newblock {\em Mol Biol Cell}, 33(13):ar129, 2022.

\bibitem{Skamrahl2022}
M.~Skamrahl, J.~Schünemann, M.~Mukenhirn, H.~Pang, J.~Gottwald, M.~Jipp,
  M.~Ferle, A.~Rübeling, T.~A. Oswald, A.~Honigmann, and A.~Janshoff.
\newblock Cellular segregation in cocultures is driven by differential adhesion
  and contractility on distinct timescales.
\newblock {\em Proc Natl Acad Sci U S A}, 120(15):e2213186120, 2023.

\bibitem{Steinberg:2007ih}
Malcolm~S Steinberg.
\newblock {Differential adhesion in morphogenesis: a modern view.}
\newblock {\em Curr. Opin. Genet. Dev.}, 17(4):281--286, August 2007.

\bibitem{brodland_differential_2002}
G.~Wayne Brodland.
\newblock The {Differential} {Interfacial} {Tension} {Hypothesis} ({DITH}): a
  comprehensive theory for the self-rearrangement of embryonic cells and
  tissues.
\newblock {\em J Biomech Eng}, 124(2):188--197, April 2002.

\bibitem{Steinberg:1963kq}
M~S Steinberg.
\newblock {Reconstruction of tissues by dissociated cells. Some morphogenetic
  tissue movements and the sorting out of embryonic cells may have a common
  explanation.}
\newblock {\em Science}, 141(3579):401--408, August 1963.

\bibitem{kasza2011dynamics}
Karen~E Kasza and Jennifer~A Zallen.
\newblock {Dynamics and regulation of contractile actin-myosin networks in
  morphogenesis.}
\newblock {\em Curr Opin Cell Biol}, 23(1):30--38, February 2011.

\bibitem{Gibson:2006gi}
Matthew~C Gibson, Ankit~B Patel, Radhika Nagpal, and Norbert Perrimon.
\newblock {The emergence of geometric order in proliferating metazoan
  epithelia.}
\newblock {\em Nature}, 442(7106):1038--1041, August 2006.

\bibitem{Major:2006ir}
Robert~J Major and Kenneth~D Irvine.
\newblock {Localization and requirement for Myosin II at the dorsal-ventral
  compartment boundary of the Drosophila wing.}
\newblock {\em Dev. Dyn.}, 235(11):3051--3058, November 2006.

\bibitem{Hayashi:2004it}
Takashi Hayashi and Richard~W Carthew.
\newblock {Surface mechanics mediate pattern formation in the developing
  retina.}
\newblock {\em Nature}, 431(7009):647--652, October 2004.

\bibitem{Hilgenfeldt2008}
S.~Hilgenfeldt, S.~Erisken, and R.~W. Carthew.
\newblock Physical modeling of cell geometric order in an epithelial tissue.
\newblock {\em Proc Natl Acad Sci U S A}, 105(3):907--911, 2008.

\bibitem{Blankenship:2006jg}
J~Todd Blankenship, Stephanie~T Backovic, Justina S~P Sanny, Ori Weitz, and
  Jennifer~A Zallen.
\newblock {Multicellular rosette formation links planar cell polarity to tissue
  morphogenesis.}
\newblock {\em Dev Cell}, 11(4):459--470, October 2006.

\bibitem{Krieg2008}
M.~Krieg, Y.~Arboleda-Estudillo, P.~H. Puech, J.~Käfer, F.~Graner, D.~J.
  Müller, and C.~P. Heisenberg.
\newblock Tensile forces govern germ-layer organization in zebrafish.
\newblock {\em Nat Cell Biol}, 10(4):429--436, 2008.

\bibitem{Togashi2011}
H.~Togashi, K.~Kominami, M.~Waseda, H.~Komura, J.~Miyoshi, M.~Takeichi, and
  Y.~Takai.
\newblock Nectins establish a checkerboard-like cellular pattern in the
  auditory epithelium.
\newblock {\em Science}, 333(6046):1144--1147, 2011.

\bibitem{Kondo:2010bx}
Shigeru Kondo and Takashi Miura.
\newblock {Reaction-diffusion model as a framework for understanding biological
  pattern formation.}
\newblock {\em Science}, 329(5999):1616--1620, September 2010.

\bibitem{stevens_programming_2023}
Adam~J. Stevens, Andrew~R. Harris, Josiah Gerdts, Ki~H. Kim, Coralie
  Trentesaux, Jonathan~T. Ramirez, Wesley~L. McKeithan, Faranak Fattahi,
  Ophir~D. Klein, Daniel~A. Fletcher, and Wendell~A. Lim.
\newblock Programming multicellular assembly with synthetic cell adhesion
  molecules.
\newblock {\em Nature}, 614(7946):144--152, February 2023.

\bibitem{bishop_2016}
Christopher~M. Bishop.
\newblock {\em Pattern Recognition and Machine Learning}.
\newblock Springer, 2016.

\bibitem{carlsson2009topology}
Gunnar Carlsson.
\newblock Topology and data.
\newblock {\em Bull. Amer. Math. Soc.}, 46(2):255--308, 2009.

\bibitem{herbertedelsbrunner2009}
Herbert Edelsbrunner.
\newblock {\em Computational Topology: An Introduction}.
\newblock American Mathematical Society, December 2009.

\bibitem{Amezquita:2020bi}
Erik~J Am{\'e}zquita, Michelle~Y Quigley, Tim Ophelders, Elizabeth Munch, and
  Daniel~H Chitwood.
\newblock {The shape of things to come: Topological data analysis and biology,
  from molecules to organisms.}
\newblock {\em Dev. Dyn.}, 249(7):816--833, July 2020.

\bibitem{Topaz:2015gd}
Chad~M Topaz, Lori Ziegelmeier, and Tom Halverson.
\newblock {Topological data analysis of biological aggregation models.}
\newblock {\em PLOS ONE}, 10(5):e0126383, 2015.

\bibitem{Ulmer:2019hx}
M~Ulmer, Lori Ziegelmeier, and Chad~M Topaz.
\newblock {A topological approach to selecting models of biological
  experiments.}
\newblock {\em PLOS ONE}, 14(3):e0213679, 2019.

\bibitem{atienza2019persistent}
Nieves Atienza, Luis~M Escudero, Maria~Jose Jimenez, and M~Soriano-Trigueros.
\newblock Persistent entropy: a scale-invariant topological statistic for
  analyzing cell arrangements.
\newblock {\em arXiv:1902.06467}, 2019.

\bibitem{Bhaskar:2019hj}
Dhananjay Bhaskar, Angelika Manhart, Jesse Milzman, John~T Nardini, Kathleen~M
  Storey, Chad~M Topaz, and Lori Ziegelmeier.
\newblock {Analyzing collective motion with machine learning and topology.}
\newblock {\em Chaos}, 29(12):123125, December 2019.

\bibitem{McGuirl201917763}
Melissa~R. McGuirl, Alexandria Volkening, and Björn Sandstede.
\newblock Topological data analysis of zebrafish patterns.
\newblock {\em Proc Natl Acad Sci USA}, 117(10):5113--5124, March 2020.

\bibitem{skinner2021topological}
Dominic~J Skinner, Boya Song, Hannah Jeckel, Eric Jelli, Knut Drescher, and
  J{\"o}rn Dunkel.
\newblock Topological metric detects hidden order in disordered media.
\newblock {\em Phys Rev Lett}, 126(4):048101, 2021.

\bibitem{Nardini2021}
J.~T. Nardini, B.~J. Stolz, K.~B. Flores, H.~A. Harrington, and H.~M. Byrne.
\newblock Topological data analysis distinguishes parameter regimes in the
  anderson-chaplain model of angiogenesis.
\newblock {\em PLoS Comput Biol}, 17(6):e1009094, 2021.

\bibitem{Stolz2022}
B.~J. Stolz, J.~Kaeppler, B.~Markelc, F.~Braun, F.~Lipsmeier, R.~J. Muschel,
  H.~M. Byrne, and H.~A. Harrington.
\newblock Multiscale topology characterizes dynamic tumor vascular networks.
\newblock {\em Science Advances}, 8(23), 2022.

\bibitem{kramar_quantifying_2014}
Miroslav Kramár, Arnaud Goullet, Lou Kondic, and Konstantin Mischaikow.
\newblock Quantifying force networks in particulate systems.
\newblock {\em Physica D: Nonlinear Phenomena}, 283:37--55, August 2014.

\bibitem{bhaskar2020topological}
Dhananjay Bhaskar, William~Y Zhang, and Ian~Y Wong.
\newblock {Topological data analysis of collective and individual epithelial
  cells using persistent homology of loops.}
\newblock {\em Soft Matter}, 17(17):4653--4664, May 2021.

\bibitem{Adams2017}
Henry Adams, Tegan Emerson, Michael Kirby, Rachel Neville, Chris Peterson,
  Patrick Shipman, Sofya Chepushtanova, Eric Hanson, Francis Motta, and Lori
  Ziegelmeier.
\newblock Persistence images: A stable vector representation of persistent
  homology.
\newblock {\em J Mach Learn Res}, 18(8):1--35, 2017.

\bibitem{McInnes2018}
L.~McInnes, J.~Healy, and J.~Melville.
\newblock Umap: Uniform manifold approximation and projection for dimension
  reduction.
\newblock {\em arXiv preprint arXiv:1802.03426}, 2018.

\bibitem{Moon2019}
K.~R. Moon, D.~van Dijk, Z.~Wang, S.~Gigante, D.~B. Burkhardt, W.~S. Chen,
  K.~Yim, A.~V.~D. Elzen, M.~J. Hirn, R.~R. Coifman, N.~B. Ivanova, G.~Wolf,
  and S.~Krishnaswamy.
\newblock Visualizing structure and transitions in high-dimensional biological
  data.
\newblock {\em Nat Biotechnol}, 37(12):1482--1492, 2019.

\bibitem{bank_autoencoders_2021}
Dor Bank, Noam Koenigstein, and Raja Giryes.
\newblock Autoencoders.
\newblock {\em arXiv:2003.05991 [cs, stat]}, April 2021.
\newblock arXiv: 2003.05991.

\bibitem{baldi_autoencoders_2012}
Pierre Baldi.
\newblock Autoencoders, {Unsupervised} {Learning}, and {Deep} {Architectures}.
\newblock In {\em Proceedings of {ICML} {Workshop} on {Unsupervised} and
  {Transfer} {Learning}}, pages 37--49. JMLR Workshop and Conference
  Proceedings, June 2012.
\newblock ISSN: 1938-7228.

\bibitem{Potdar2010}
A.~A. Potdar, J.~Jeon, A.~M. Weaver, V.~Quaranta, and P.~T. Cummings.
\newblock Human mammary epithelial cells exhibit a bimodal correlated random
  walk pattern.
\newblock {\em PLoS One}, 5(3):e9636, 2010.

\bibitem{Yamaguchi2007}
M.~Yamaguchi, E.~Yoshimoto, and S.~Kondo.
\newblock Pattern regulation in the stripe of zebrafish suggests an underlying
  dynamic and autonomous mechanism.
\newblock {\em Proc Natl Acad Sci U S A}, 104(12):4790--4793, 2007.

\bibitem{zomorodian_computing_2005}
Afra Zomorodian and Gunnar Carlsson.
\newblock Computing {Persistent} {Homology}.
\newblock {\em Discrete \& Computational Geometry}, 33(2):249--274, February
  2005.

\bibitem{chung_smooth_2020}
Yu-Min Chung, Michael Hull, and Austin Lawson.
\newblock Smooth {Summaries} of {Persistence} {Diagrams} and {Texture}
  {Classification}.
\newblock In {\em 2020 {IEEE}/{CVF} {Conference} on {Computer} {Vision} and
  {Pattern} {Recognition} {Workshops} ({CVPRW})}, pages 3667--3675, Seattle,
  WA, USA, June 2020. IEEE.

\bibitem{chung_gaussian_2022}
Yu-Min Chung, Michael Hull, Austin Lawson, and Neil Pritchard.
\newblock Gaussian {Persistence} {Curves}, May 2022.
\newblock arXiv:2205.11353 [cs, math].

\bibitem{JMLR:v12:gonen11a}
Mehmet G{{\"o}}nen and Ethem Alpaydin.
\newblock Multiple kernel learning algorithms.
\newblock {\em Journal of Machine Learning Research}, 12(64):2211--2268, 2011.

\bibitem{chung_topological_2018}
Yu-Min Chung, Chuan-Shen Hu, Austin Lawson, and Clifford Smyth.
\newblock Topological approaches to skin disease image analysis.
\newblock In {\em 2018 {IEEE} {International} {Conference} on {Big} {Data}
  ({Big} {Data})}, pages 100--105, December 2018.

\bibitem{barnes_comparative_2021}
Danielle Barnes, Luis Polanco, and Jose~A. Perea.
\newblock A {Comparative} {Study} of {Machine} {Learning} {Methods} for
  {Persistence} {Diagrams}.
\newblock {\em Frontiers in Artificial Intelligence}, 4, 2021.

\bibitem{mullner_modern_2011}
Daniel Müllner.
\newblock Modern hierarchical, agglomerative clustering algorithms, September
  2011.
\newblock arXiv:1109.2378 [cs, stat].

\bibitem{Volkening:2020jd}
Alexandria Volkening.
\newblock {Linking genotype, cell behavior, and phenotype: multidisciplinary
  perspectives with a basis in zebrafish patterns.}
\newblock {\em Curr. Opin. Genet. Dev.}, 63:78--85, August 2020.

\bibitem{bhaskar_code_2022}
Dhananjay Bhaskar.
\newblock Code and data for "{TDA} of {Spatial} {Patterning} in {Heterogeneous}
  {Cell} {Populations}", August 2023.

\end{thebibliography}

\clearpage

\newcommand{\SupportingPrefix}{S}
\renewcommand{\thefigure}{\SupportingPrefix\arabic{figure}}
\renewcommand{\thetable}{\SupportingPrefix\arabic{table}}
\setcounter{figure}{0}    
\setcounter{table}{0}
\setcounter{page}{1}

\vspace*{\fill}
\begin{center}
{\huge \textbf{Supplementary Information}}
\end{center}
\vspace*{\fill}

\begin{figure}[h]
    \centering
    \includegraphics[scale=0.27]{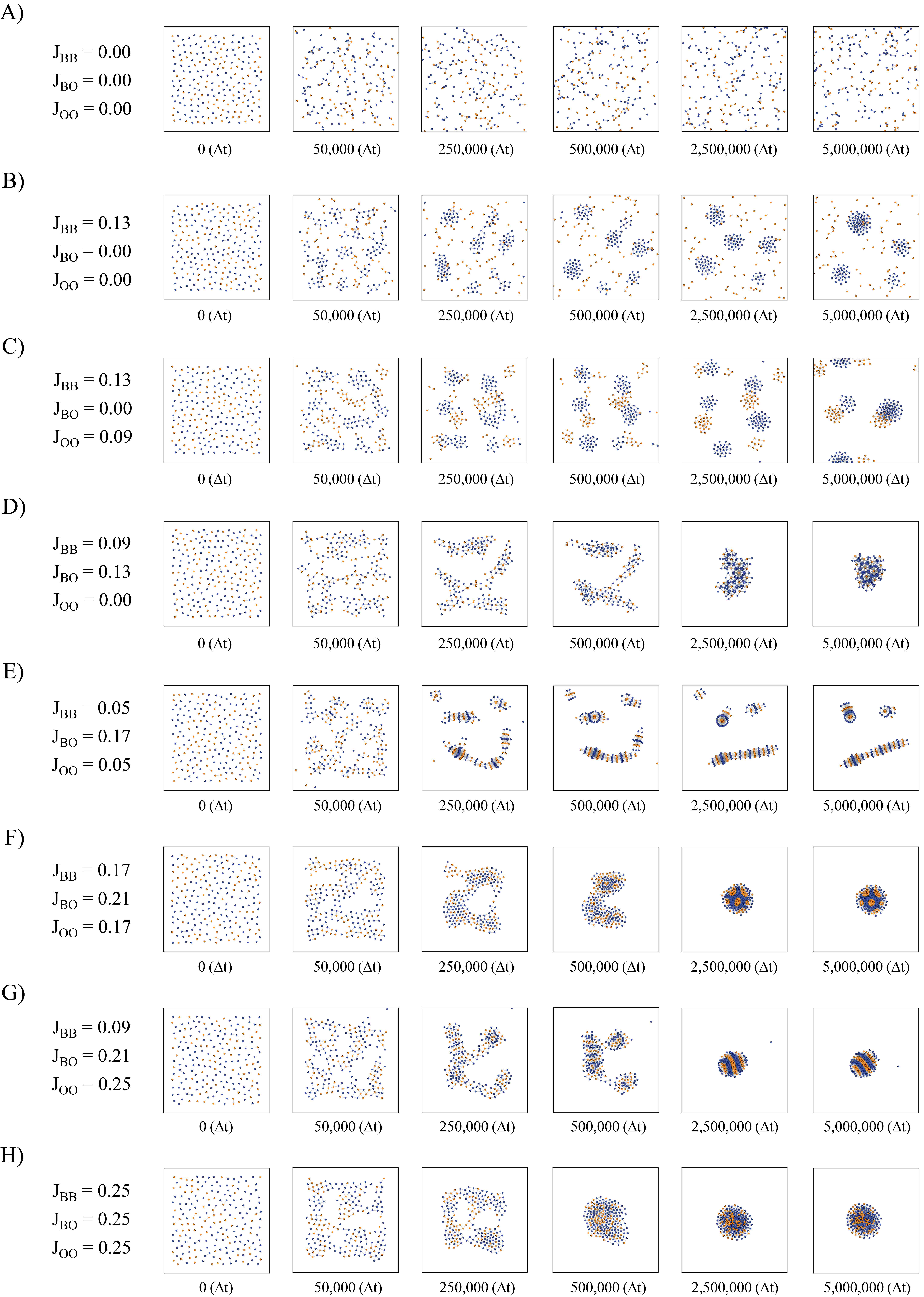}
    \caption{\footnotesize \textbf{Simulation snapshots showing self-organization in a heterogeneous population over time (log scale) at varying differential adhesion parameters.} (A) Particles remain individually dispersed at low adhesion values. (B) Blue particles aggregate into clusters due to high blue-blue adhesion. (C) Complete sorting simulation where both blue and orange particles form separate clusters due to high homotypic adhesion. (D-G) High heterotypic adhesion results in configurations that maximize the interaction between the two cell types, forming hexagonal, striped and spotted patterns. (H) Well-mixed clusters are obtained when homotypic and heterotypic adhesion values are greater than zero and (approximately) equal.}
    \label{fig:SI_snaps_noprolif_log}
\end{figure}

\begin{figure}[h]
    \centering
    \includegraphics[scale=0.27]{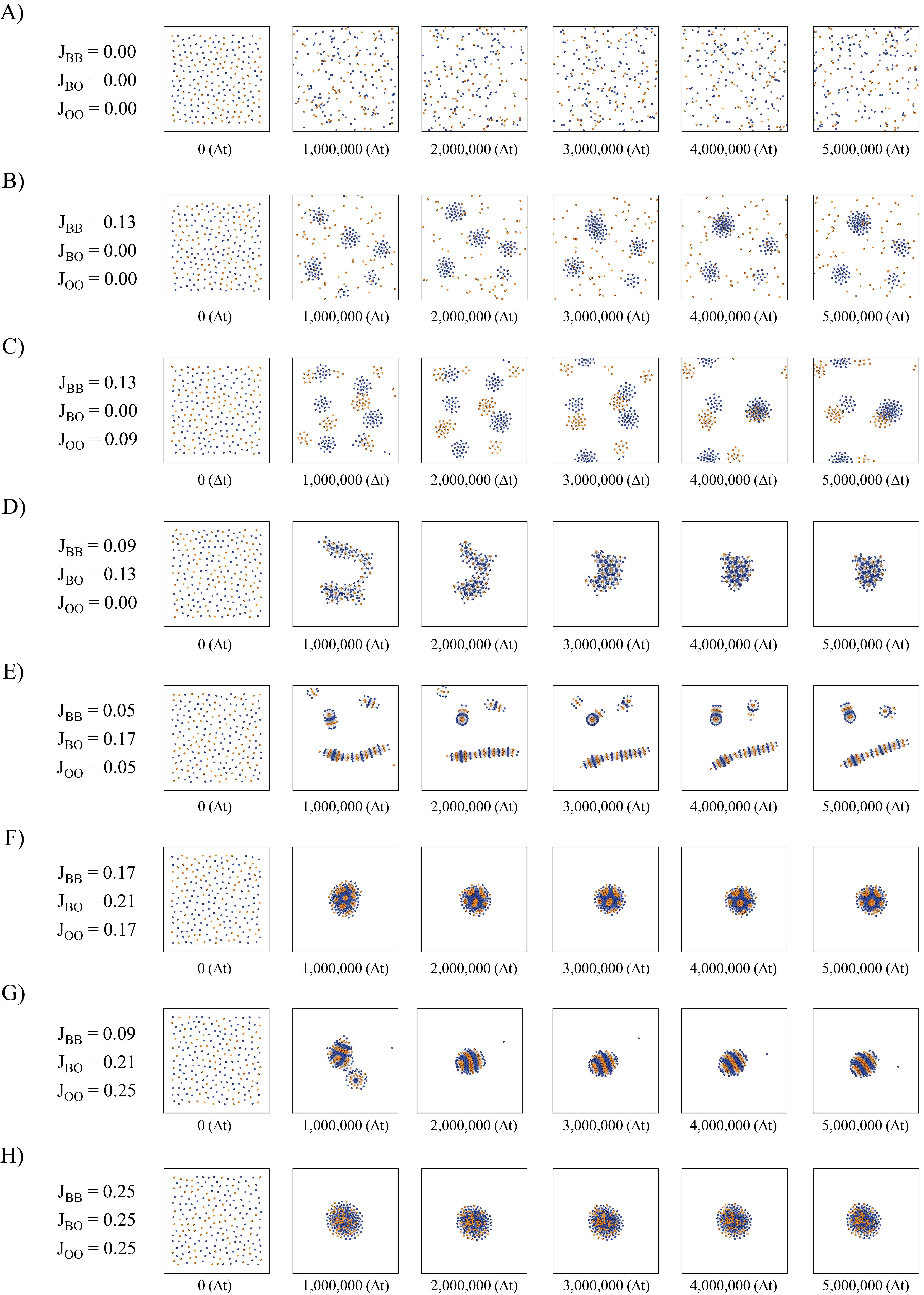}
    \caption{\footnotesize \textbf{Simulation snapshots showing self-organization in a heterogeneous population over time (linear scale) at varying differential adhesion parameters.} (A) Particles remain individually dispersed at low adhesion values. (B) Blue particles aggregate into clusters due to high blue-blue adhesion. (C) Complete sorting simulation where both blue and orange particles form separate clusters due to high homotypic adhesion. (D-G) High heterotypic adhesion results in configurations that maximize the interaction between the two cell types, forming hexagonal, striped and spotted patterns. (H) Well-mixed clusters are obtained when homotypic and heterotypic adhesion values are greater than zero and (approximately) equal.}
    \label{fig:SI_snaps_noprolif_linear}
\end{figure}

\begin{figure}[h]
    \centering
    \includegraphics[width=\linewidth]{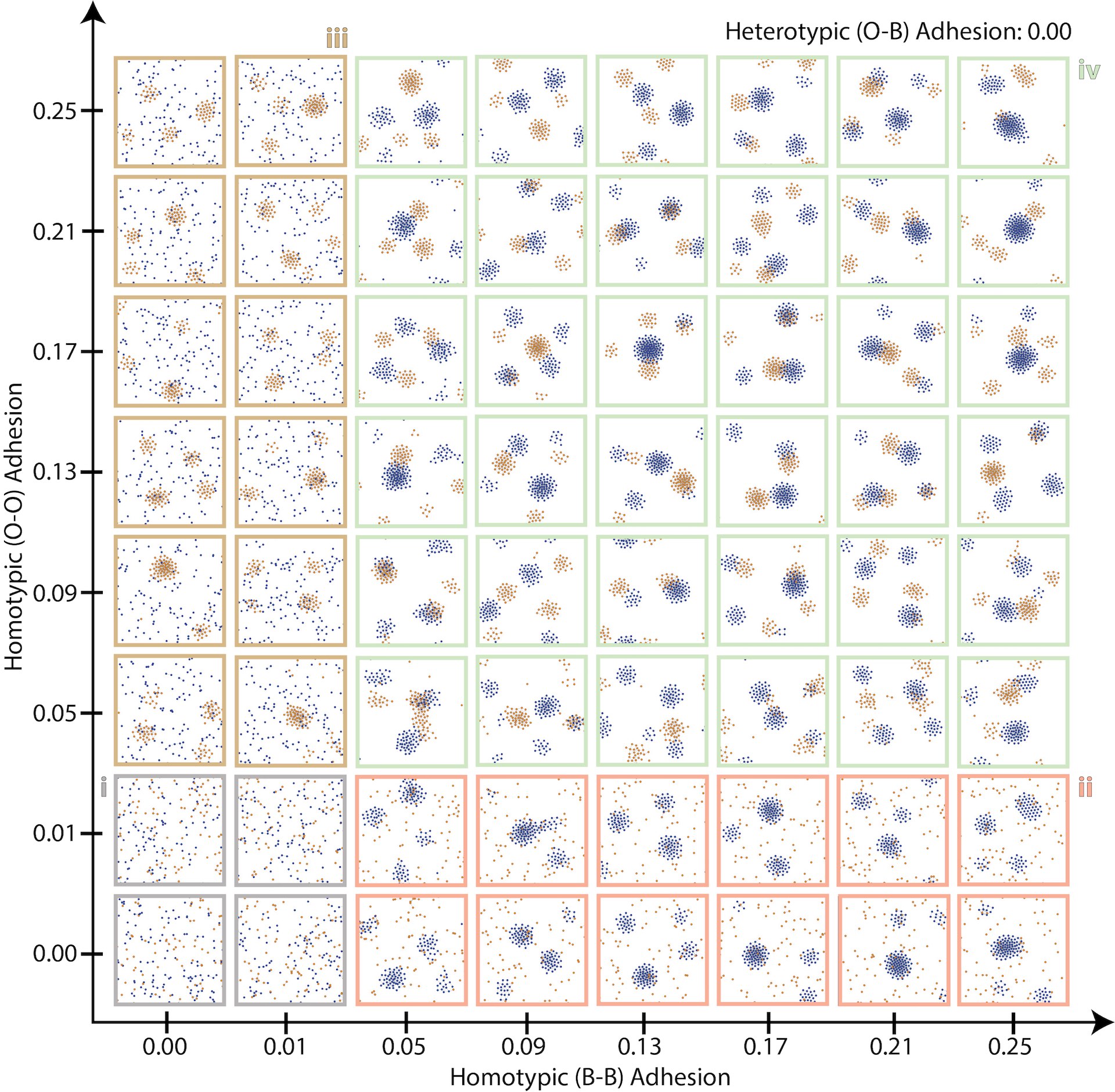}
    \caption{\footnotesize \textbf{Simulation snapshots of multicellular patterning of a heterogeneous, non-proliferating population at steady state with $\mathbf{J_{BO} = 0.00}$ and varying $\mathbf{J_{BB}, J_{OO}}$. }i. denotes individually dispersed blue and orange cells. ii denotes clusters of blue cells with individually dispersed orange cells. iii denotes individually dispersed blue cells with clusters of orange cells. iv. denotes separate clusters of orange and blue cells, respectively.}
    \label{fig:SI_finalHT1}
\end{figure}

\begin{figure}[h]
    \centering
    \includegraphics[width=\linewidth]{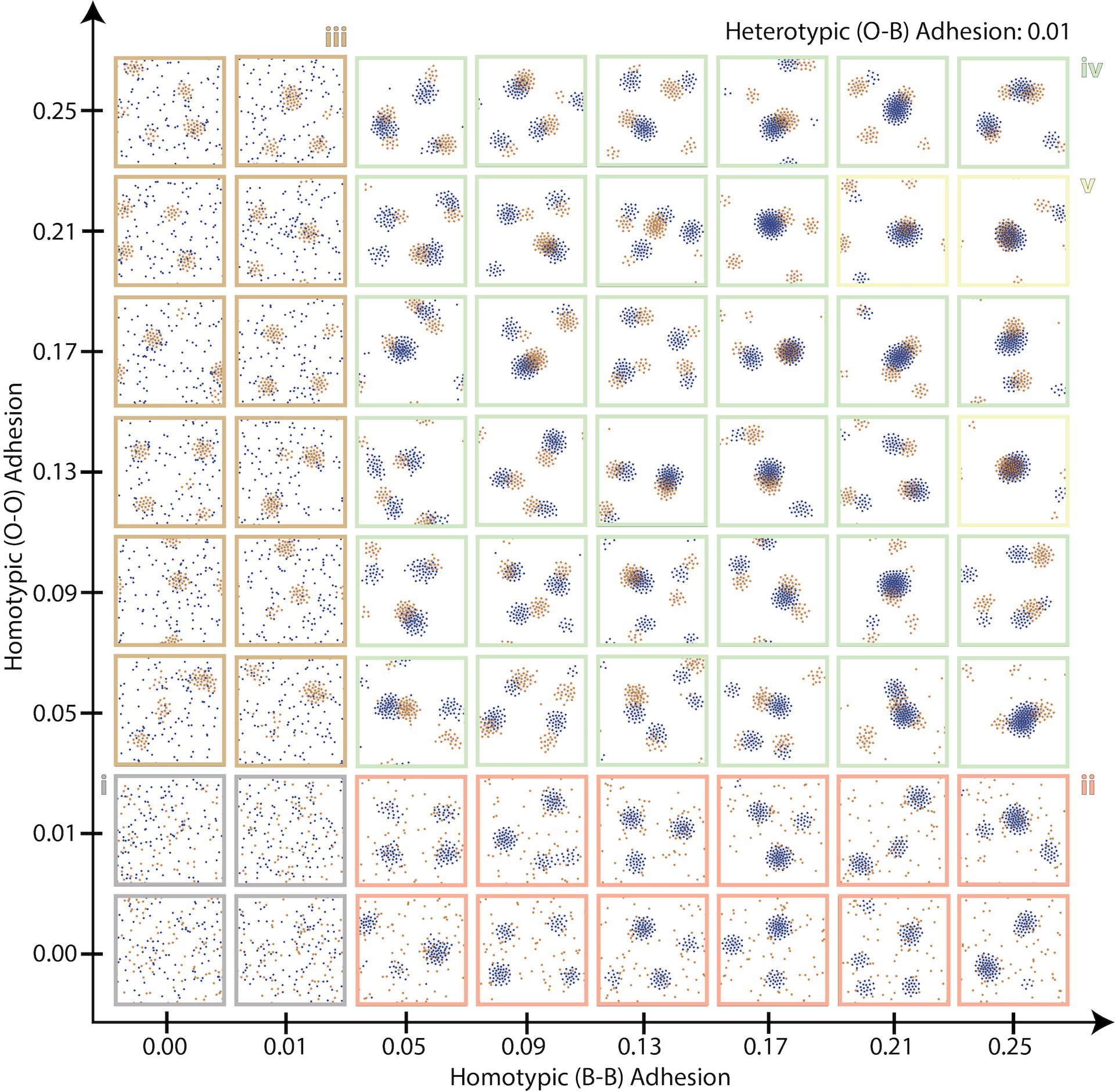}
    \caption{\footnotesize \textbf{Simulation snapshots of multicellular patterning of a heterogeneous, non-proliferating population at steady state with $\mathbf{J_{BO} = 0.01}$ and varying $\mathbf{J_{BB}, J_{OO}}$.} i. denotes individually dispersed blue and orange cells. ii denotes clusters of blue cells with individually dispersed orange cells. iii denotes individually dispersed blue cells with clusters of orange cells. iv. denotes separate clusters of orange and blue cells, respectively. v denotes clusters with intermixed blue and orange cells.}
    \label{fig:SI_finalHT2}
\end{figure}

\begin{figure}[h]
    \centering
    \includegraphics[width=\linewidth]{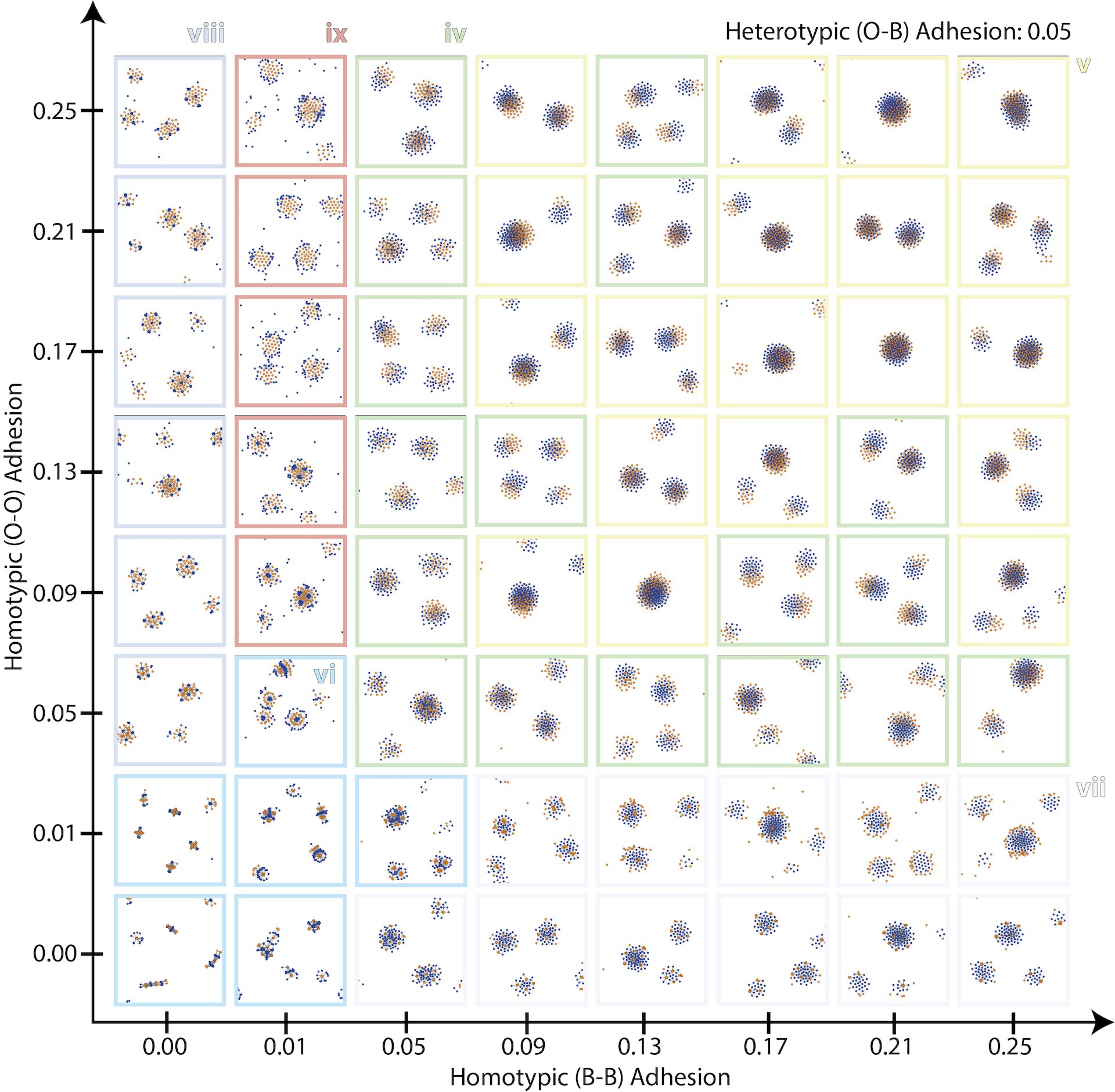}
    \caption{\footnotesize \textbf{Simulation snapshots of multicellular patterning of a heterogeneous, non-proliferating population at steady state with $\mathbf{J_{BO} = 0.05}$ and varying $\mathbf{J_{BB}, J_{OO}}$.} iv. denotes separate clusters of orange and blue cells, respectively. v denotes clusters with intermixed blue and orange cells. vi denotes clusters with alternating stripes of blue and orange cells. vii denotes hexagonal arrays of orange cells surrounded by blue cells. viii denotes hexagonal arrays of blue cells surrounded by orange cells. ix denotes a cluster with a core of orange cells surrounded by blue cells.}
    \label{fig:SI_finalHT3}
\end{figure}

\begin{figure}[h]
    \centering
    \includegraphics[width=\linewidth]{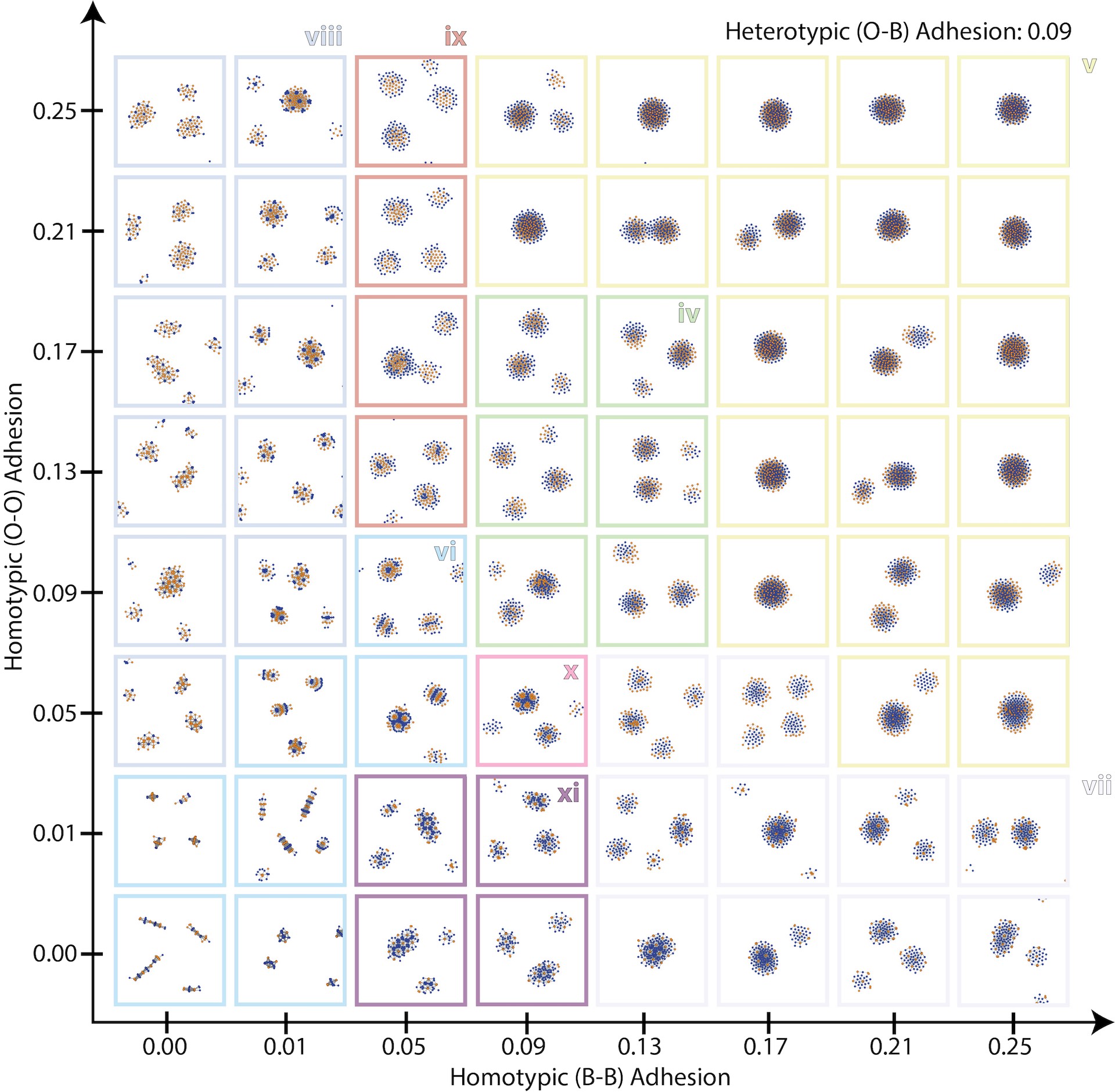}
    \caption{\footnotesize \textbf{Simulation snapshots of multicellular patterning of a heterogeneous, non-proliferating population at steady state with $\mathbf{J_{BO} = 0.09}$ and varying $\mathbf{J_{BB}, J_{OO}}$.} iv. denotes separate clusters of orange and blue cells, respectively. v denotes clusters with intermixed blue and orange cells. vi denotes clusters with alternating stripes of blue and orange cells. vii denotes hexagonal arrays of orange cells surrounded by blue cells. viii denotes hexagonal arrays of blue cells surrounded by orange cells. ix denotes a cluster with a core of orange cells surrounded by blue cells.}
    \label{fig:SI_finalHT4}
\end{figure}

\begin{figure}[h]
    \centering
    \includegraphics[width=\linewidth]{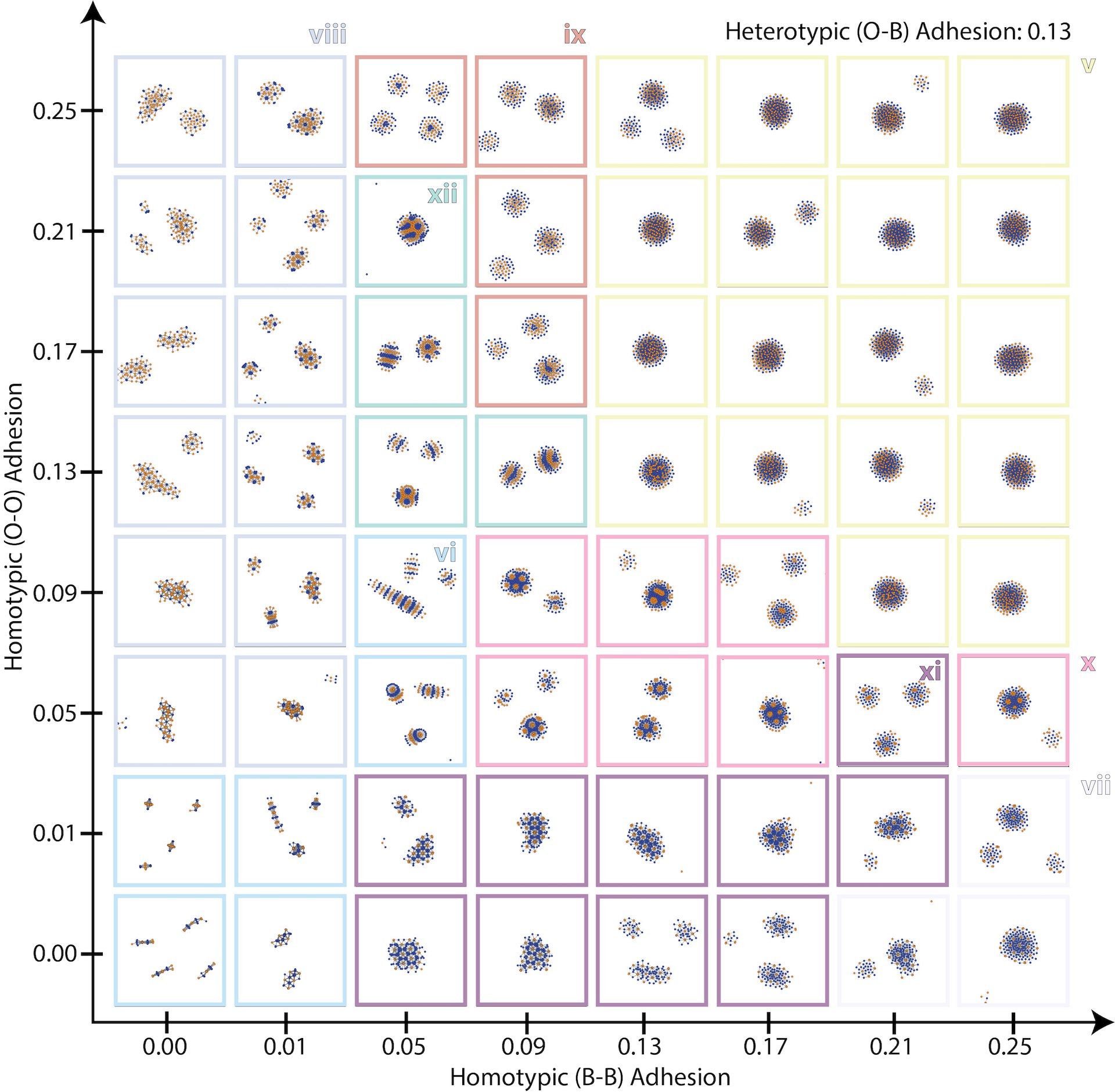}
    \caption{\footnotesize \textbf{Simulation snapshots of multicellular patterning of a heterogeneous, non-proliferating population at steady state with $\mathbf{J_{BO} = 0.13}$ and varying $\mathbf{J_{BB}, J_{OO}}$.} v denotes clusters with intermixed blue and orange cells. vi denotes clusters with alternating stripes of blue and orange cells. vii denotes hexagonal arrays of orange cells surrounded by blue cells. viii denotes hexagonal arrays of blue cells surrounded by orange cells. ix denotes a cluster with a core of orange cells surrounded by blue cells. x denotes tightly packed hexagonal arrays of orange cells surrounded by tightly packed blue cells. xi denotes hexagonal arrays of orange cells surrounded by tightly packed blue cells. xii denotes labyrinth patterns of blue and orange cells.}
    \label{fig:SI_finalHT5}
\end{figure}

\begin{figure}[h]
    \centering
    \includegraphics[width=\linewidth]{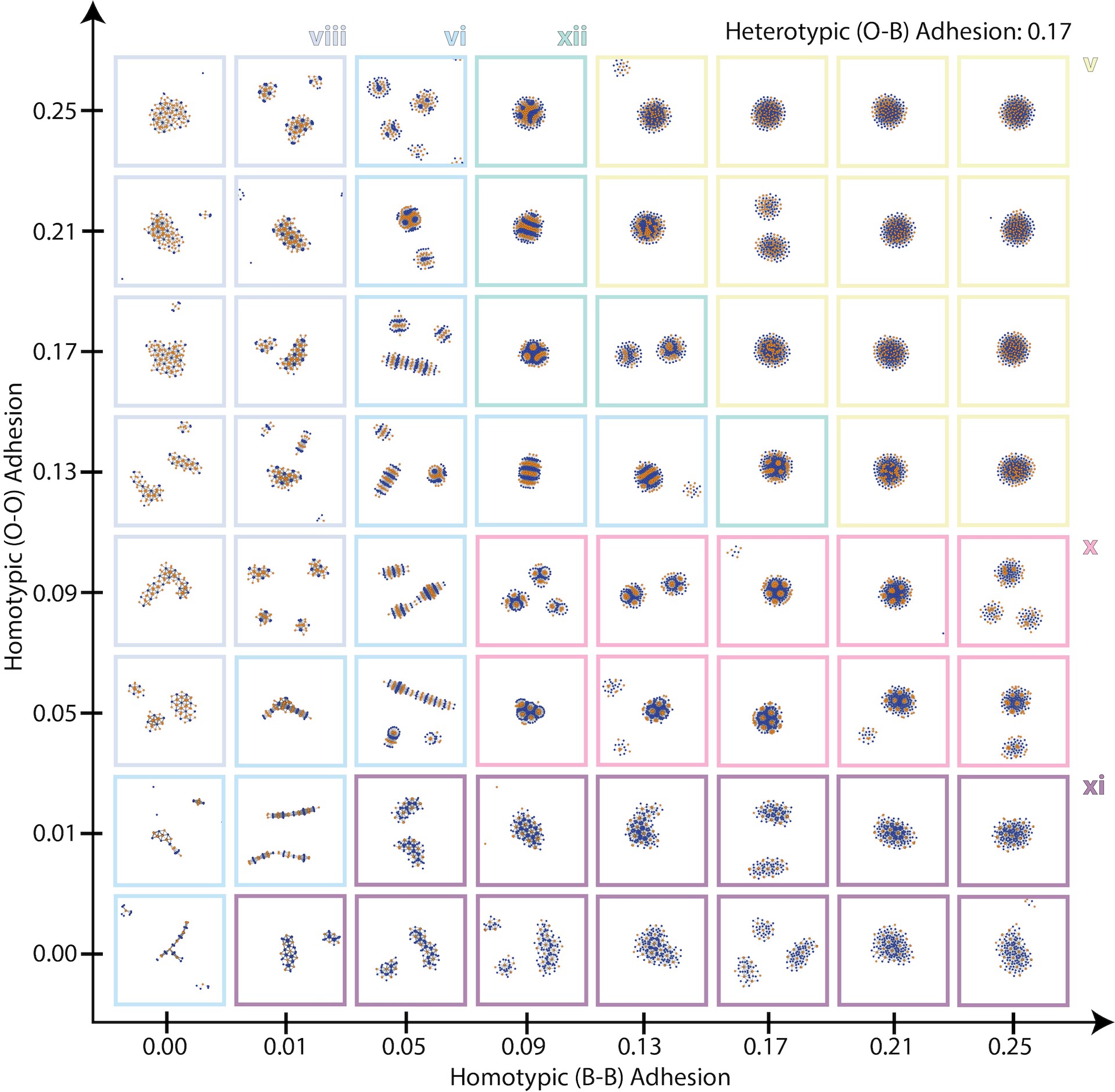}
    \caption{\footnotesize \textbf{Simulation snapshots of multicellular patterning of a heterogeneous, non-proliferating population at steady state with $\mathbf{J_{BO} = 0.17}$ and varying $\mathbf{J_{BB}, J_{OO}}$.} v denotes clusters with intermixed blue and orange cells. vi denotes clusters with alternating stripes of blue and orange cells. vii denotes hexagonal arrays of orange cells surrounded by blue cells. viii denotes hexagonal arrays of blue cells surrounded by orange cells. ix denotes a cluster with a core of orange cells surrounded by blue cells. x denotes tightly packed hexagonal arrays of orange cells surrounded by tightly packed blue cells. xi denotes hexagonal arrays of orange cells surrounded by tightly packed blue cells. xii denotes labyrinth patterns of blue and orange cells.}
    \label{fig:SI_finalHT6}
\end{figure}

\begin{figure}[h]
    \centering
    \includegraphics[width=\linewidth]{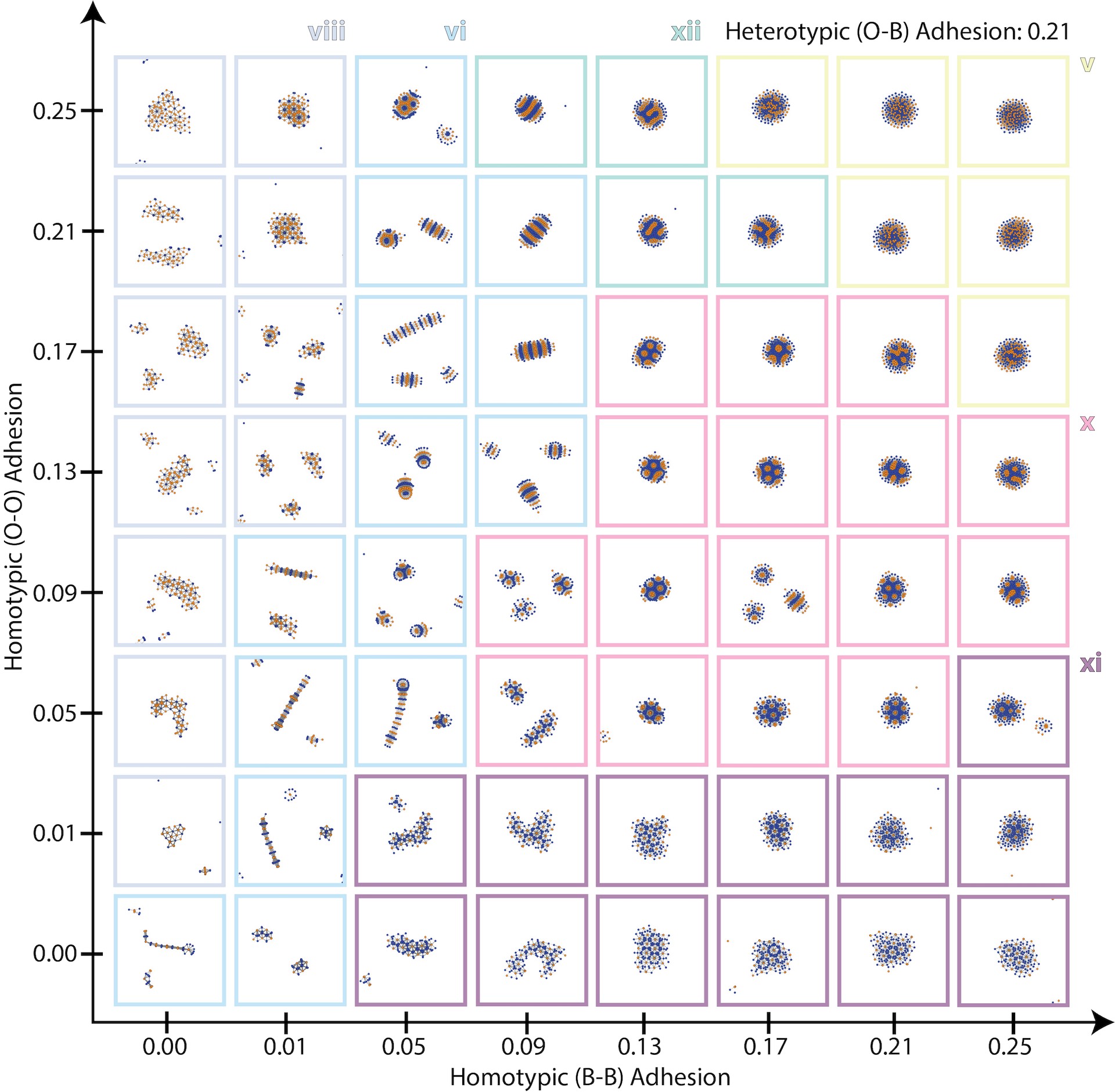}
    \caption{\footnotesize \textbf{Simulation snapshots of multicellular patterning of a heterogeneous, non-proliferating population at steady state with $\mathbf{J_{BO} = 0.21}$ and varying $\mathbf{J_{BB}, J_{OO}}$.} v denotes clusters with intermixed blue and orange cells. vi denotes clusters with alternating stripes of blue and orange cells. vii denotes hexagonal arrays of orange cells surrounded by blue cells. viii denotes hexagonal arrays of blue cells surrounded by orange cells. ix denotes a cluster with a core of orange cells surrounded by blue cells. x denotes tightly packed hexagonal arrays of orange cells surrounded by tightly packed blue cells. xi denotes hexagonal arrays of orange cells surrounded by tightly packed blue cells. xii denotes labyrinth patterns of blue and orange cells.}
    \label{fig:SI_finalHT7}
\end{figure}

\begin{figure}[h]
    \centering
    \includegraphics[width=\linewidth]{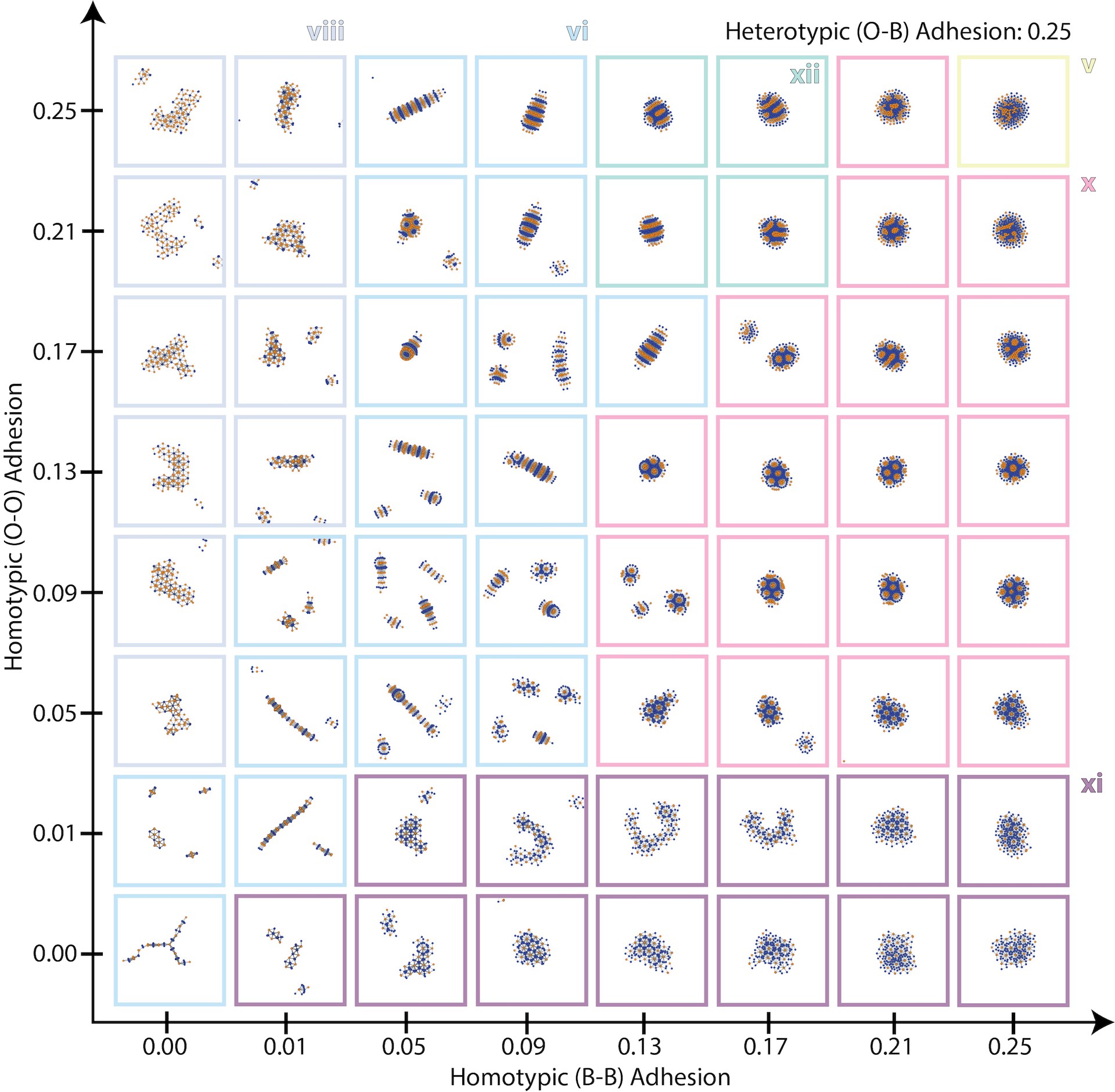}
    \caption{\footnotesize \textbf{Simulation snapshots of multicellular patterning of a heterogeneous, non-proliferating population at steady state with $\mathbf{J_{BO} = 0.25}$ and varying $\mathbf{J_{BB}, J_{OO}}$.} v denotes clusters with intermixed blue and orange cells. vi denotes clusters with alternating stripes of blue and orange cells. viii denotes hexagonal arrays of blue cells surrounded by orange cells. x denotes tightly packed hexagonal arrays of orange cells surrounded by tightly packed blue cells. xi denotes hexagonal arrays of orange cells surrounded by tightly packed blue cells. xii denotes labyrinth patterns of blue and orange cells.}
    \label{fig:SI_finalHT8}
\end{figure}

\begin{figure}[h]
    \centering
    \includegraphics[scale=0.6]{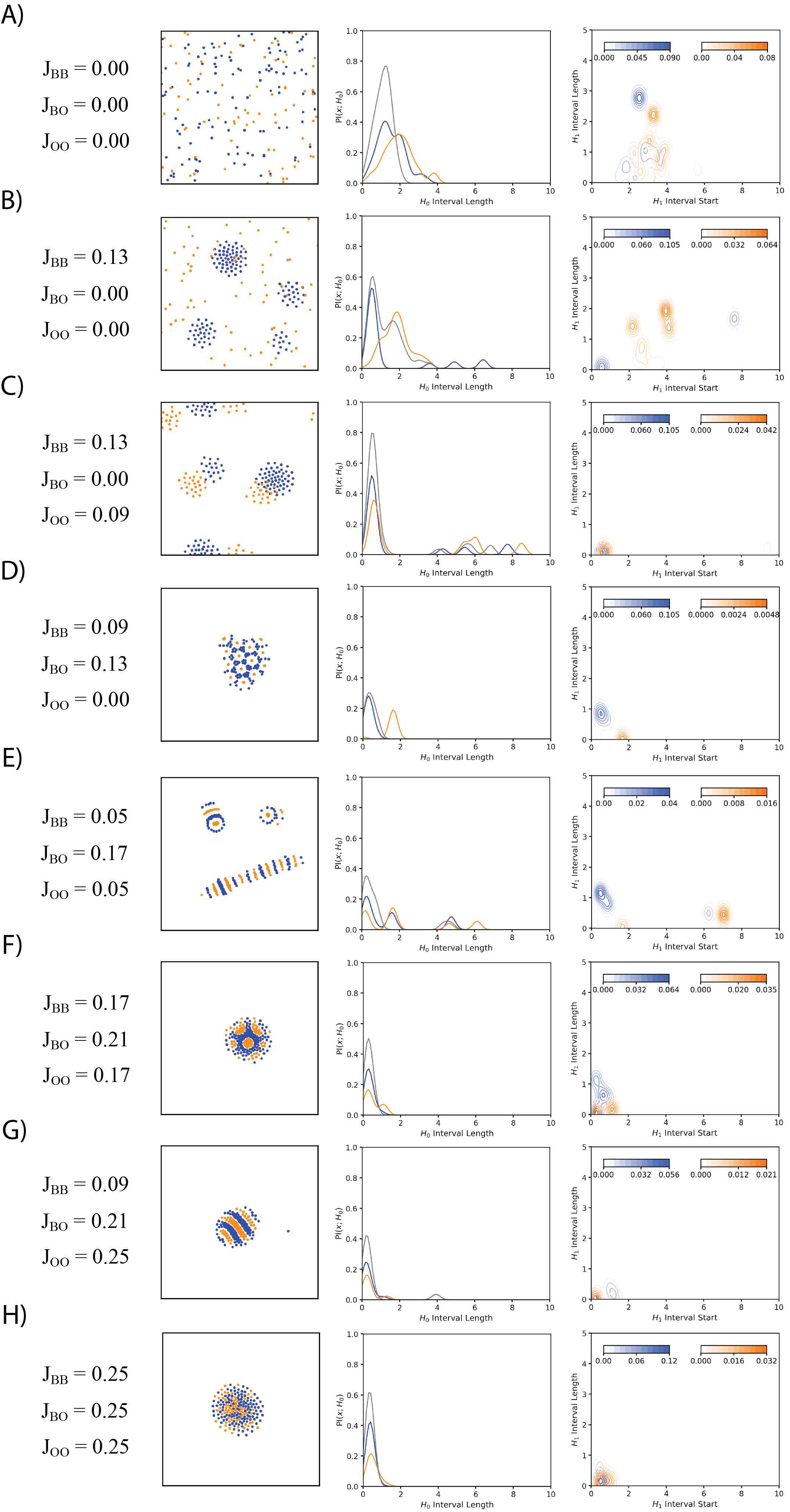}
    \caption{\footnotesize \textbf{Persistence images for representative particle configurations.} Adhesion parameter values are provided in the first column. Particle positions for the two cell types are plotted in blue and orange at the final time-step of the simulation in the second column. Intensity values for the dimension 0 persistence image is shown in the third column. Contour plots for dimension 1 persistence images are shown in the fourth column.}
    \label{fig:SI_noprolif_PI}
\end{figure}

\begin{figure}[h]
    \centering
    \includegraphics[scale=0.6]{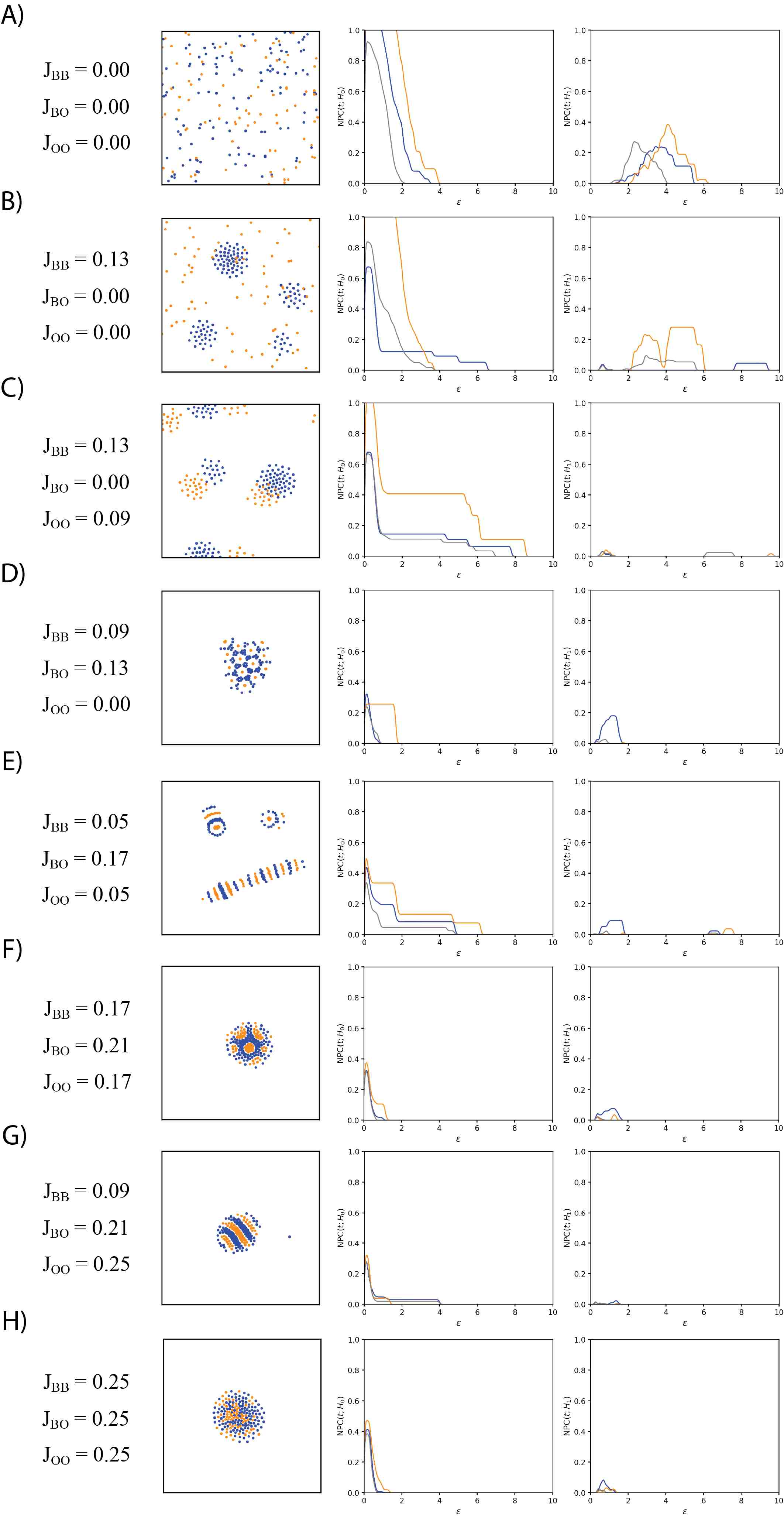}
    \caption{\footnotesize \textbf{Normalized persistence curves for representative particle configurations.} Adhesion parameter values are provided in the first column. Particle positions for the two cell types are plotted in blue and orange at the final time-step of the simulation in the second column. Persistence curves corresponding to dimension 0 and 1 persistent homology are shown in the third and fourth columns respectively.}
    \label{fig:SI_noprolif_NPC}
\end{figure}

\begin{figure}[h]
    \centering
    \includegraphics[scale=0.6]{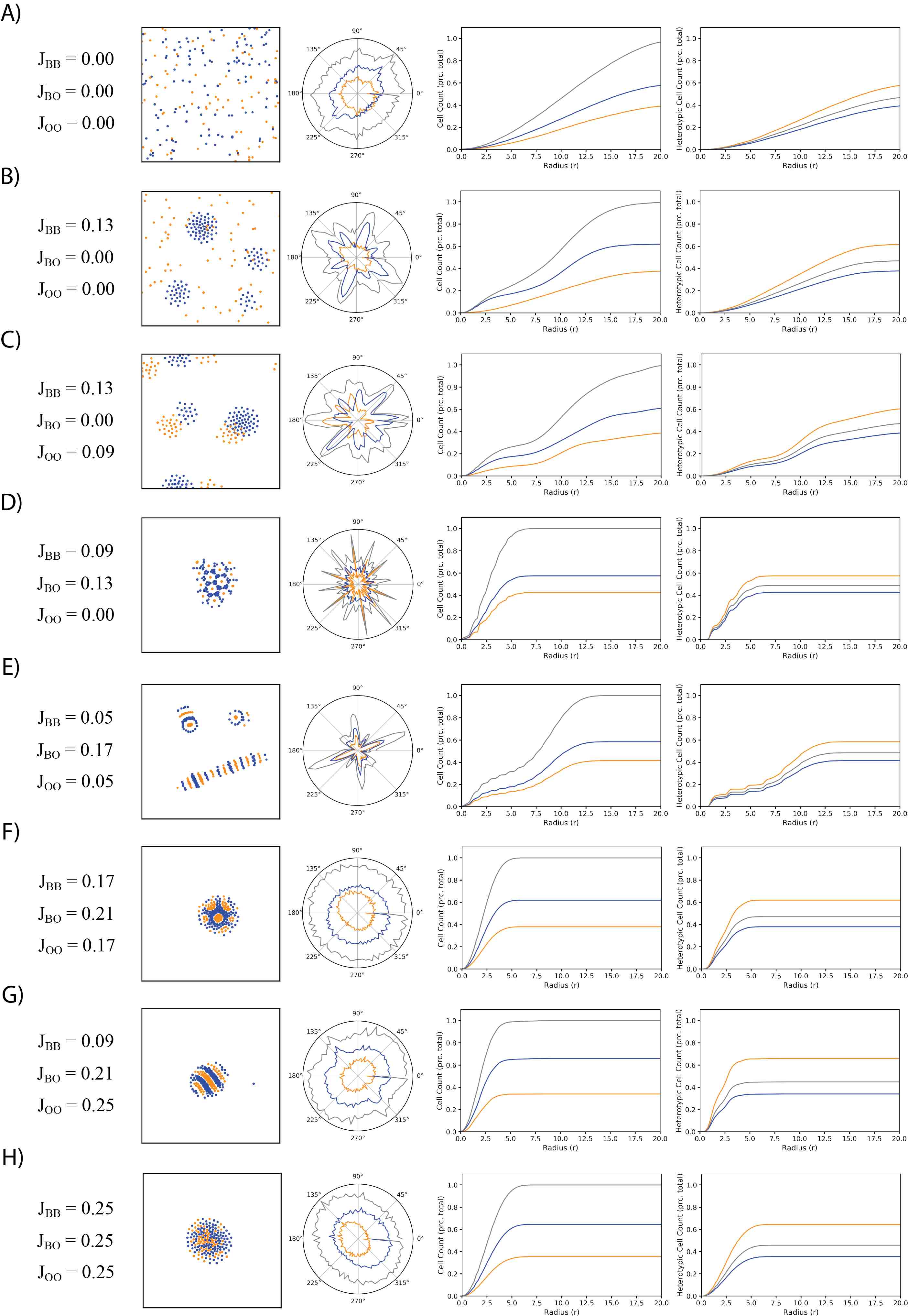}
    \caption{\footnotesize \textbf{Order parameters for representative particle configurations.} Adhesion parameter values are provided in the first column. Particle positions for the two cell types are plotted in blue and orange at the final time-step of the simulation in the second column. Ensemble-averaged angular distribution, radial distribution for homotypic neighbors and radial distribution for heterotypic neighbors are shown in the third, fourth and fifth columns respectively.}
    \label{fig:SI_noprolif_OP}
\end{figure}

\begin{figure}[h]
    \centering
    \includegraphics[scale=0.6]{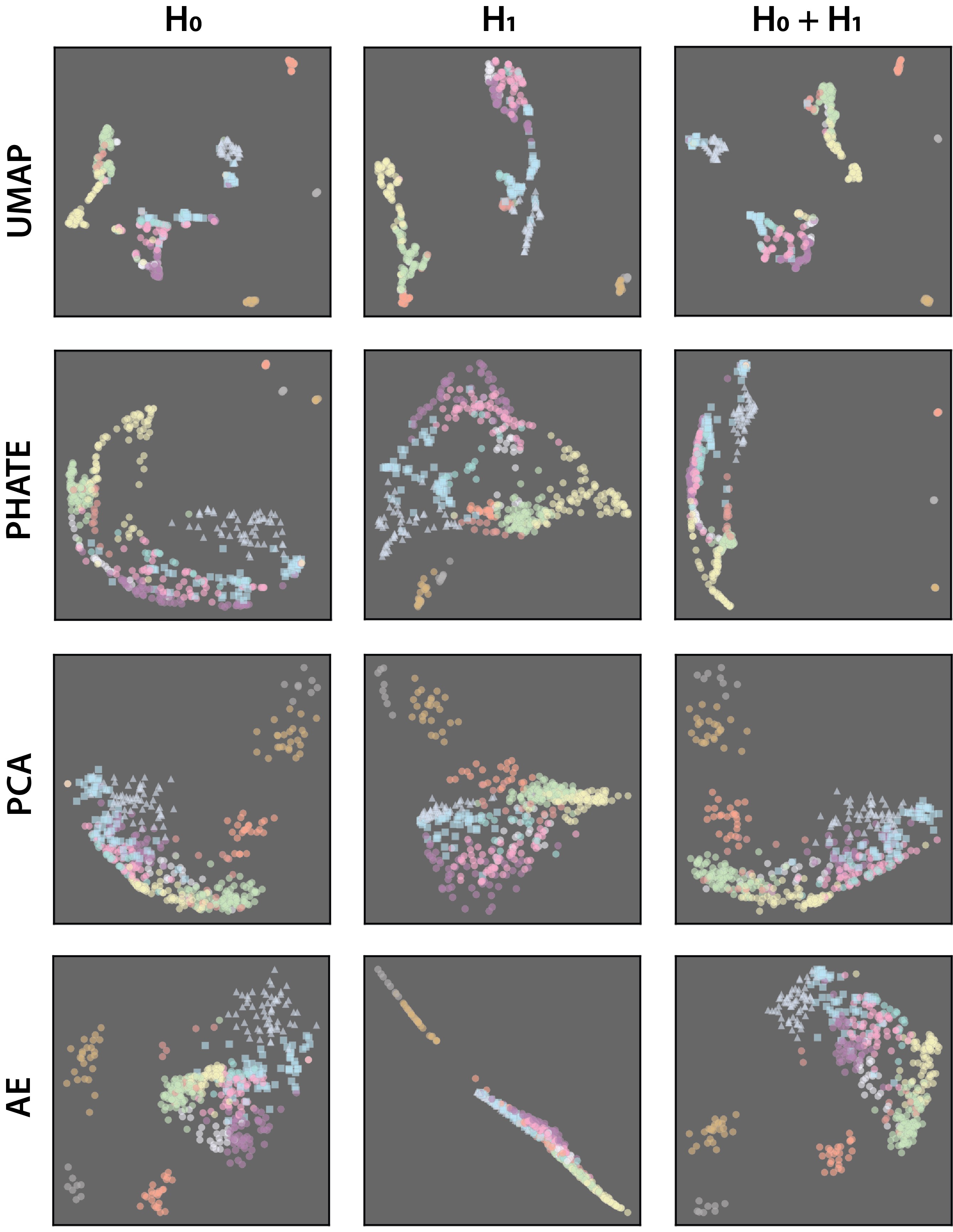}
    \caption{\footnotesize \textbf{2D embeddings of persistence images of non-proliferating particle configurations.} Low dimension representation of simulated particle configurations obtained using UMAP, PHATE, and PCA. Embeddings computed using dimension 0, dimension 1, and concatenated persistence images for simulations with proliferation disabled.}
    \label{fig:SI_persimg_dim_noprolif}
\end{figure}

\begin{figure}[h]
    \centering
    \includegraphics[scale=0.8]{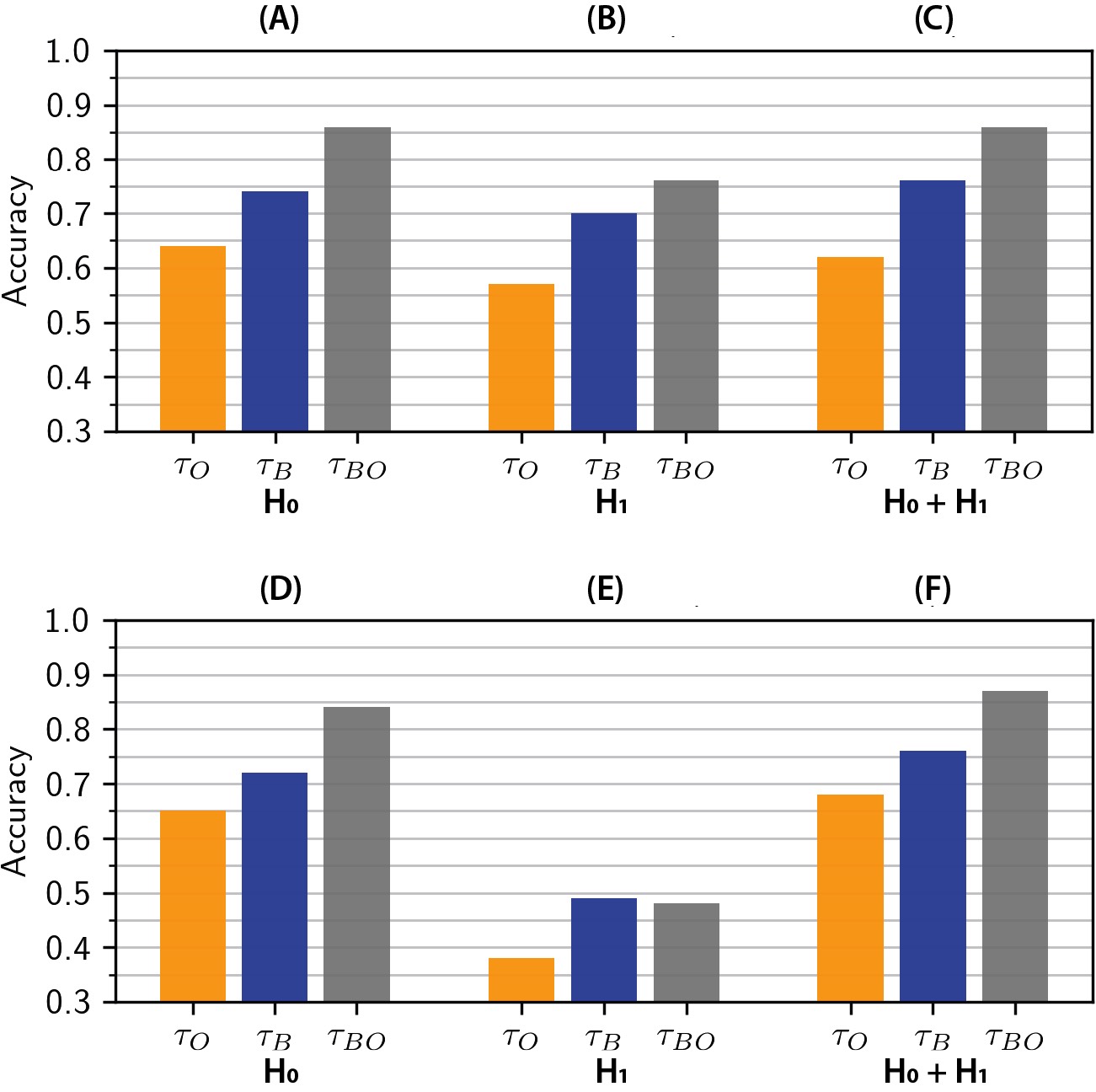}
    \caption{\footnotesize \textbf{Classification accuracy of persistence curves.} (A-C) at constant population size. (D-F) at varying population size.}
    \label{fig:SI_pc_classification}
\end{figure}

\begin{figure}[h]
    \centering
    \includegraphics[scale=0.6]{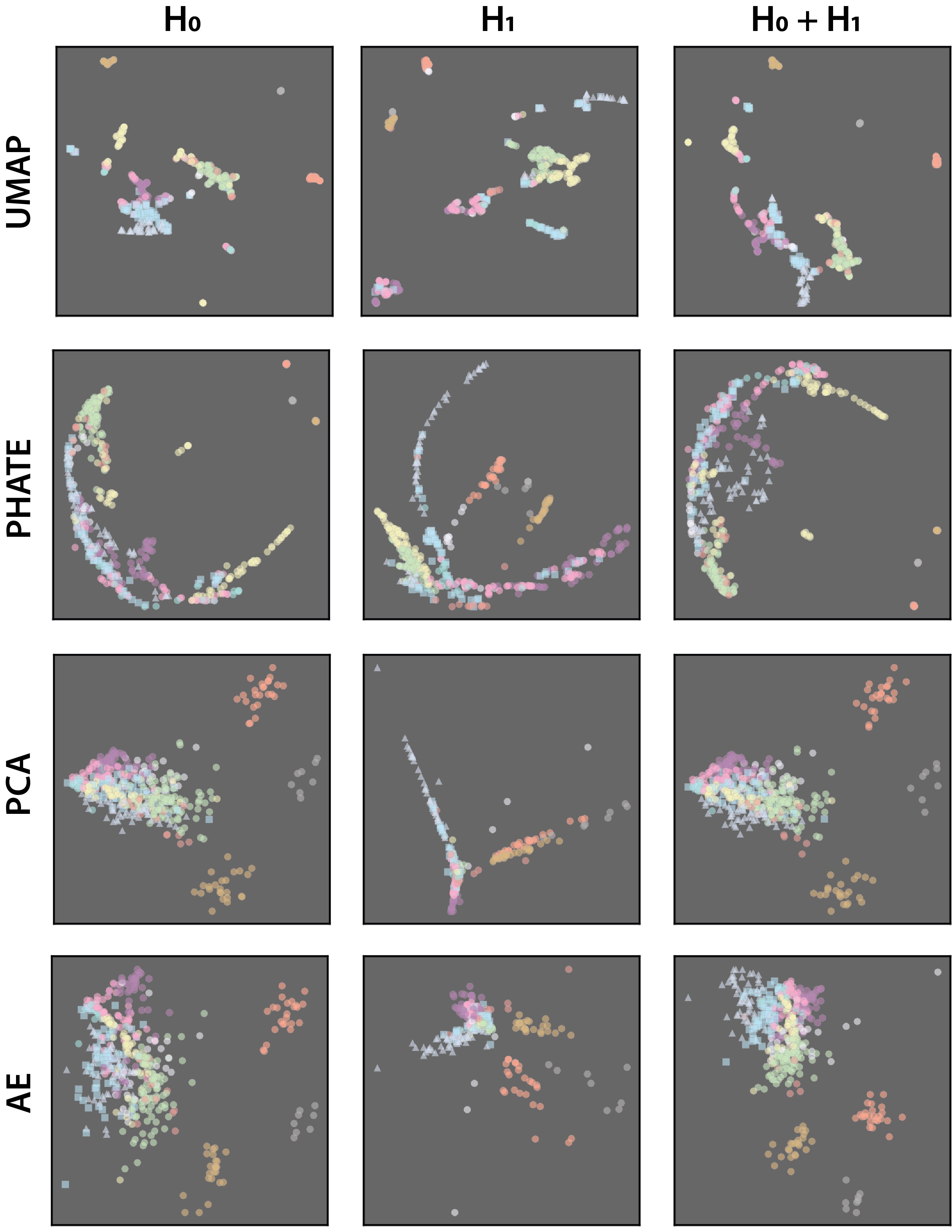}
    \caption{\footnotesize \textbf{2D embeddings of persistence curves of non-proliferating particle configurations.} Low dimension representation of simulated particle configurations obtained using UMAP, PHATE, and PCA. Embeddings computed using dimension 0, dimension 1, and concatenated persistence curves for simulations with proliferation disabled.}
    \label{fig:SI_perscurve_dim_noprolif}
\end{figure}

\begin{figure}[h]
    \centering
    \includegraphics[scale=0.8]{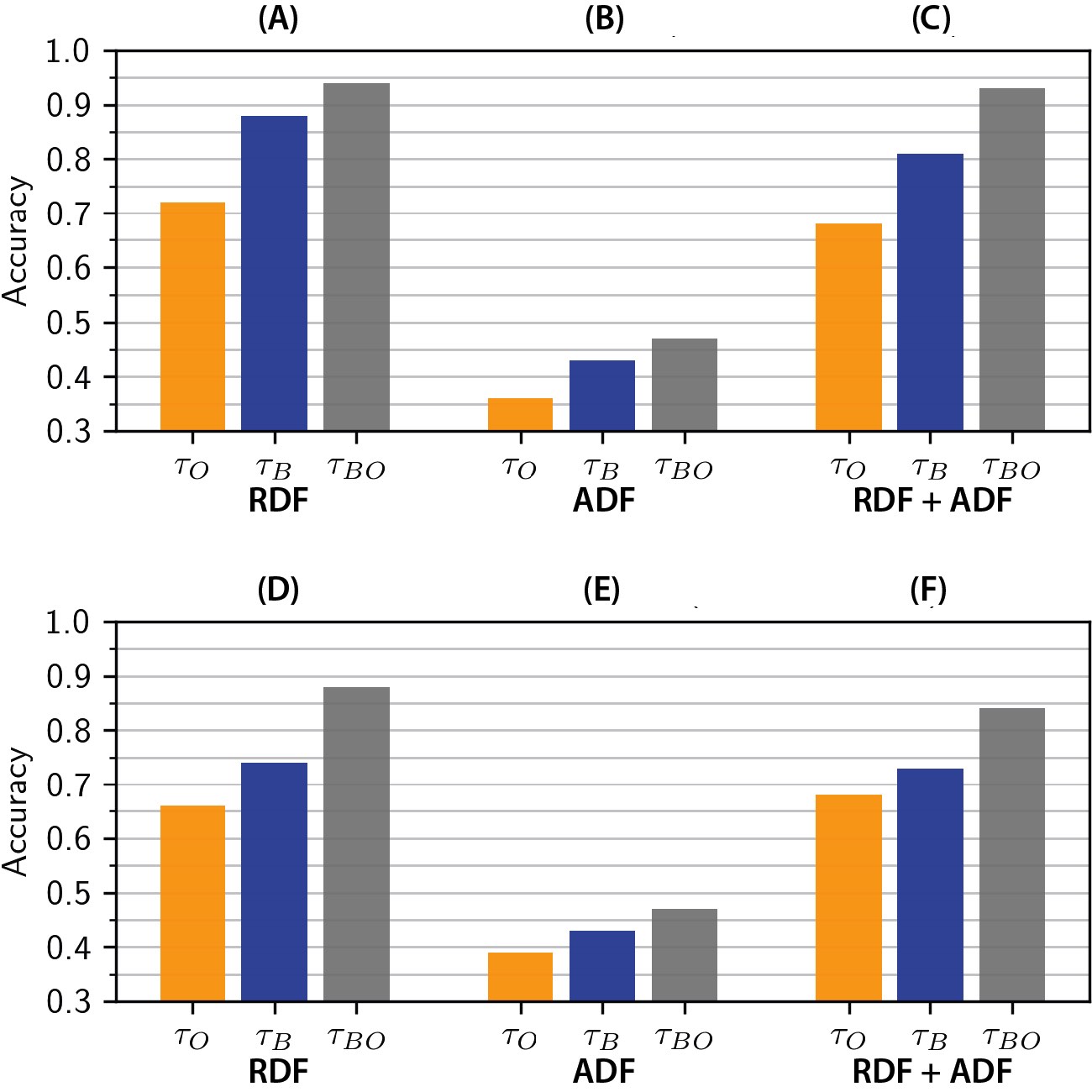}
    \caption{\footnotesize \textbf{Classification accuracy of radial distribution function (RDF) and angular distribution function (ADF) order parameters.} (A-C) at constant population size. (D-F) at varying population size.}
    \label{fig:SI_op_classification}
\end{figure}

\begin{figure}[h]
    \centering
    \includegraphics[scale=0.6]{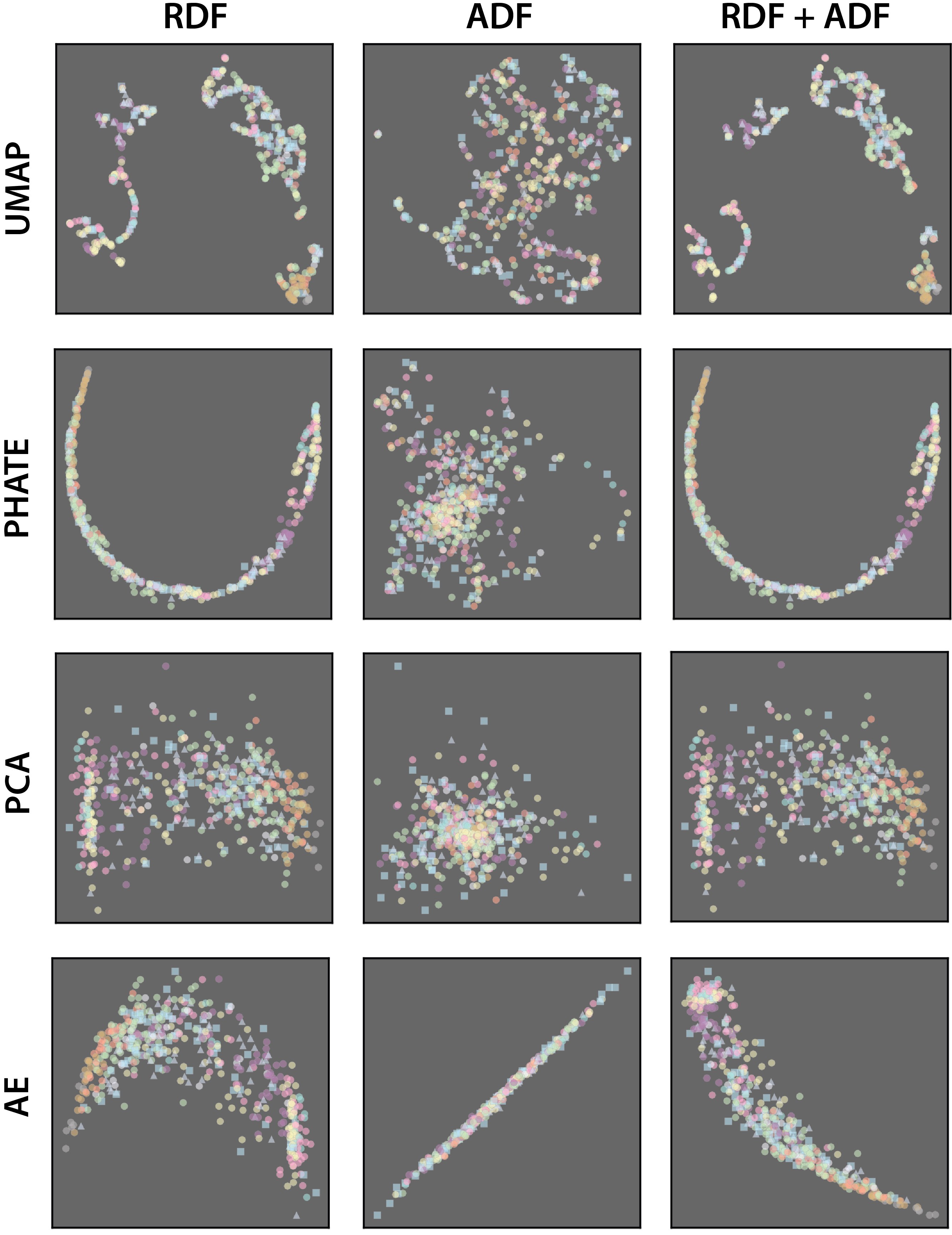}
    \caption{\footnotesize \textbf{2D embeddings of order parameters of non-proliferating particle configurations.} Low dimension representation of simulated particle configurations obtained using UMAP, PHATE, and PCA. Embeddings computed using radial, angular, and concatenated order parameters for simulations with proliferation disabled.}
    \label{fig:SI_op_dim_noprolif}
\end{figure}

\begin{figure}[h]
    \centering
    \includegraphics[scale=0.27]{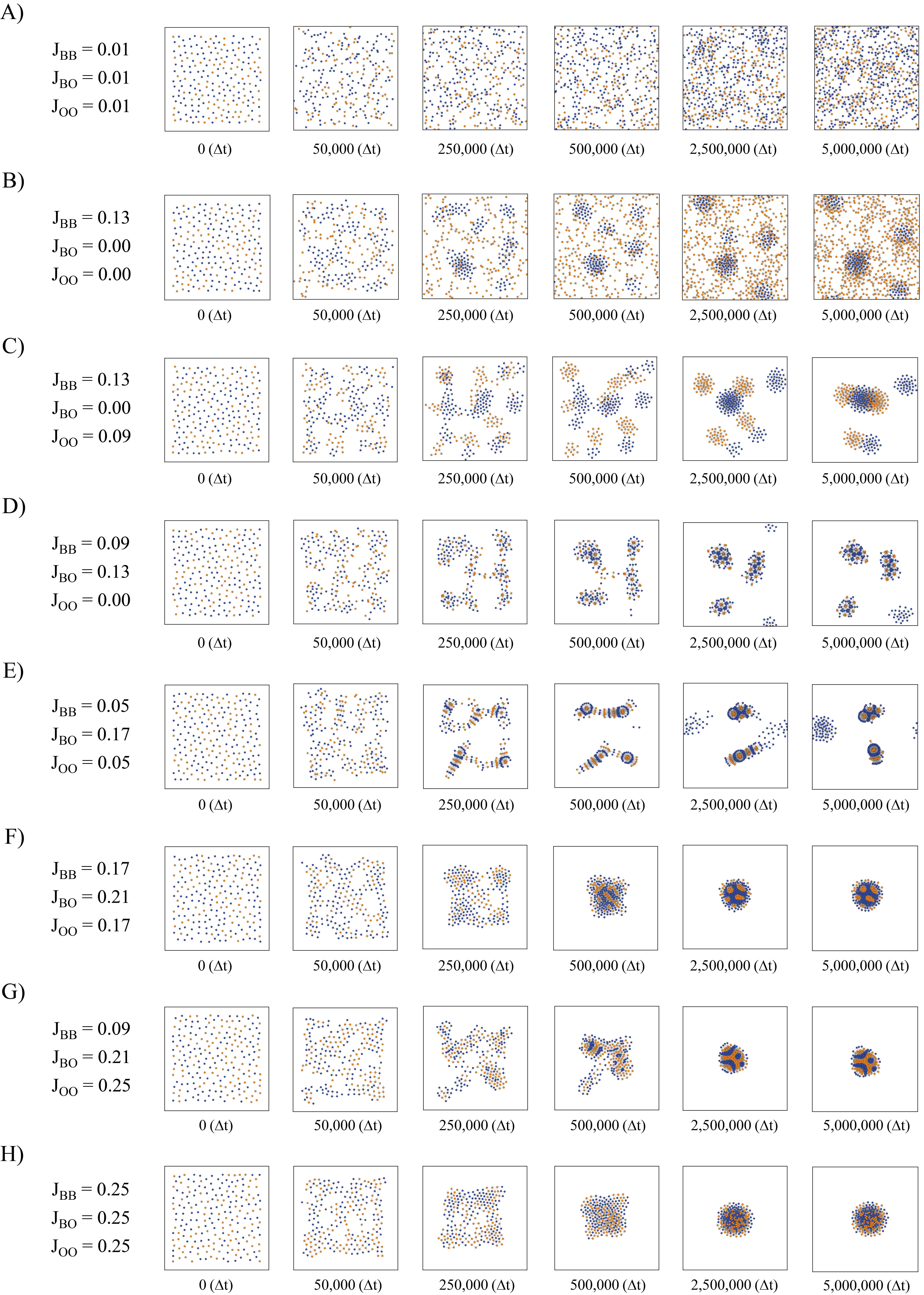}
    \caption{\footnotesize \textbf{Self-organized pattern formation in time-lapse snapshots (log scale) with proliferation turned on at varying differential adhesion values.} (A) All particles remain individually dispersed in the absence of the adhesion force. (B) Blue particles aggregate into clusters due to high blue-blue adhesion. (C) Complete sorting simulation where both blue and orange particles form separate clusters due to high homotypic adhesion. (D-G) High heterotypic adhesion results in configurations that maximize the interaction between the two cell types, forming hexagonal, striped and spotted patterns. (H) Well-mixed clusters are obtained when homotypic and heterotypic adhesion values are greater than zero and (approximately) equal.}
    \label{fig:SI_snaps_prolif_log}
\end{figure}

\begin{figure}[h]
    \centering
    \includegraphics[scale=0.27]{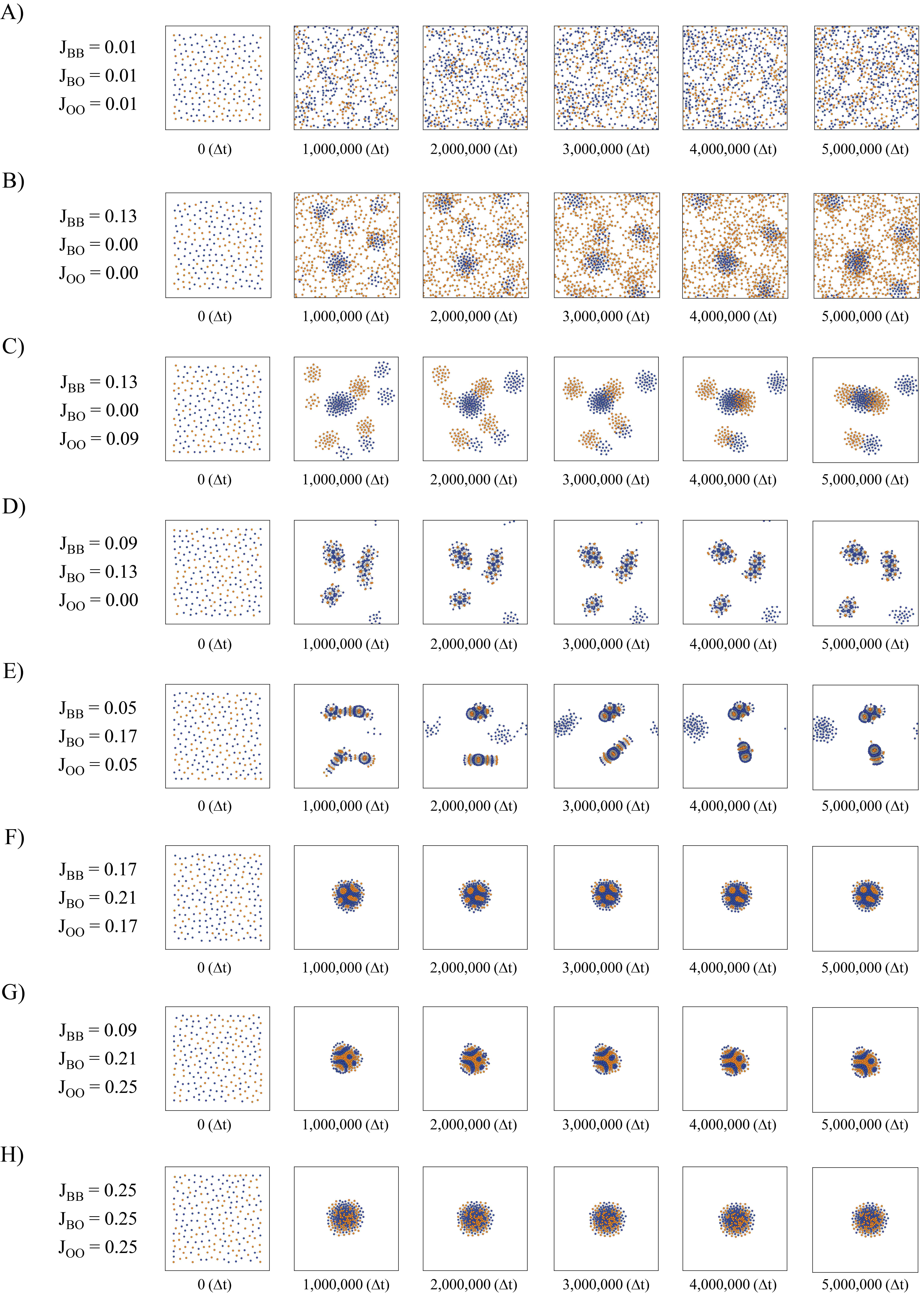}
    \caption{\footnotesize \textbf{Self-organized pattern formation in time-lapse snapshots (linear scale) with proliferation turned on at varying differential adhesion values.} (A) All particles remain individually dispersed in the absence of the adhesion force. (B) Blue particles aggregate into clusters due to high blue-blue adhesion. (C) Complete sorting simulation where both blue and orange particles form separate clusters due to high homotypic adhesion. (D-G) High heterotypic adhesion results in configurations that maximize the interaction between the two cell types, forming hexagonal, striped and spotted patterns. (H) Well-mixed clusters are obtained when homotypic and heterotypic adhesion values are greater than zero and (approximately) equal.}
    \label{fig:SI_snaps_prolif_linear}
\end{figure}

\begin{figure}[h]
    \centering
    \includegraphics[width=\linewidth]{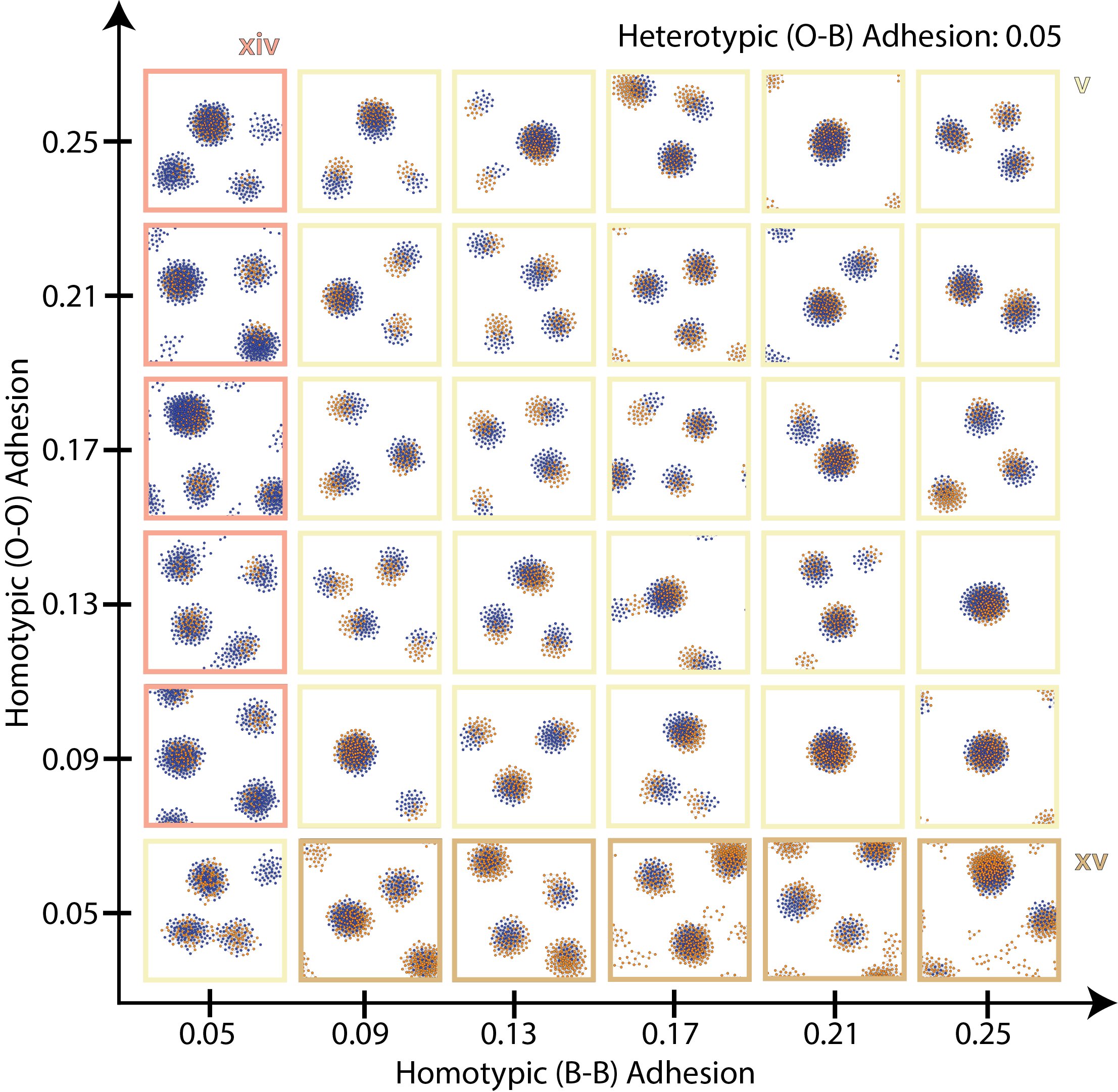}
    \caption{\footnotesize \textbf{Simulation snapshots of multicellular patterning of a heterogeneous, proliferating population at steady state with $\mathbf{J_{BO} = 0.05}$ and varying $\mathbf{J_{BB}, J_{OO}}$.}}
    \label{fig:SI_finalHT1p}
\end{figure}

\begin{figure}[h]
    \centering
    \includegraphics[width=\linewidth]{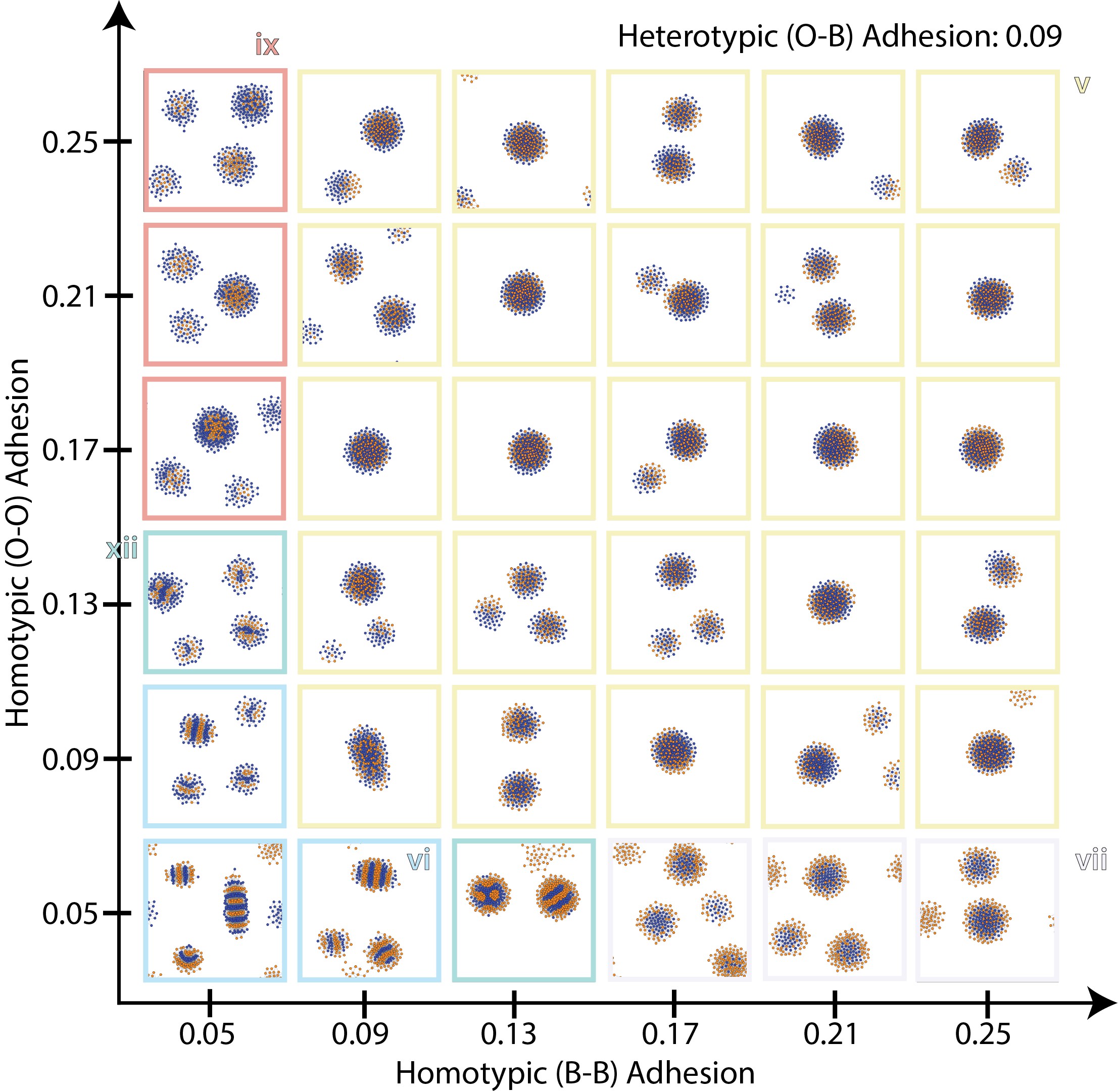}
    \caption{\footnotesize \textbf{Simulation snapshots of multicellular patterning of a heterogeneous, proliferating population at steady state with $\mathbf{J_{BO} = 0.09}$ and varying $\mathbf{J_{BB}, J_{OO}}$.}}
    \label{fig:SI_finalHT2p}
\end{figure}

\begin{figure}[h]
    \centering
    \includegraphics[width=\linewidth]{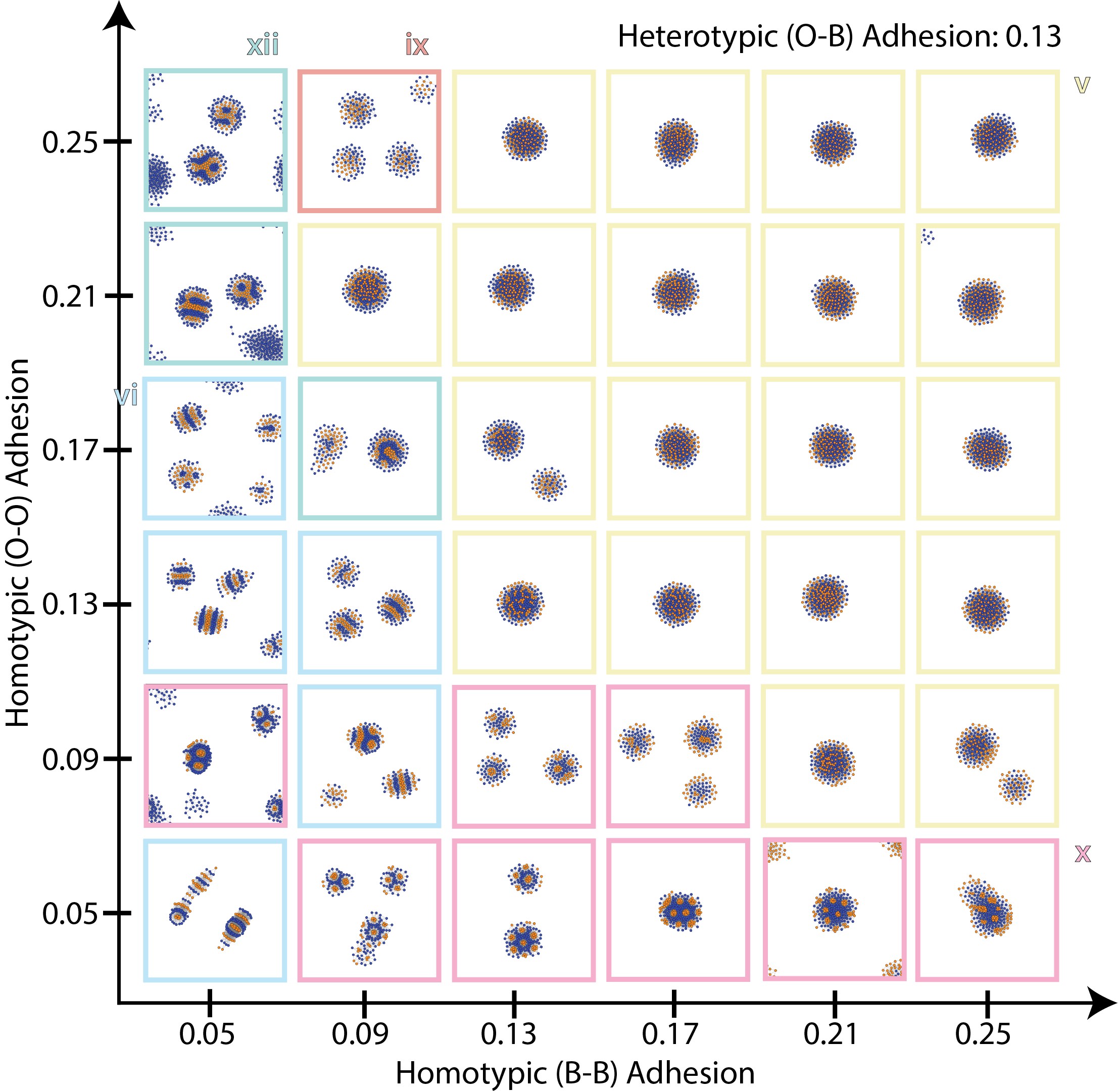}
    \caption{\footnotesize \textbf{Simulation snapshots of multicellular patterning of a heterogeneous, proliferating population at steady state with $\mathbf{J_{BO} = 0.13}$ and varying $\mathbf{J_{BB}, J_{OO}}$.}}
    \label{fig:SI_finalHT3p}
\end{figure}

\begin{figure}[h]
    \centering
    \includegraphics[width=\linewidth]{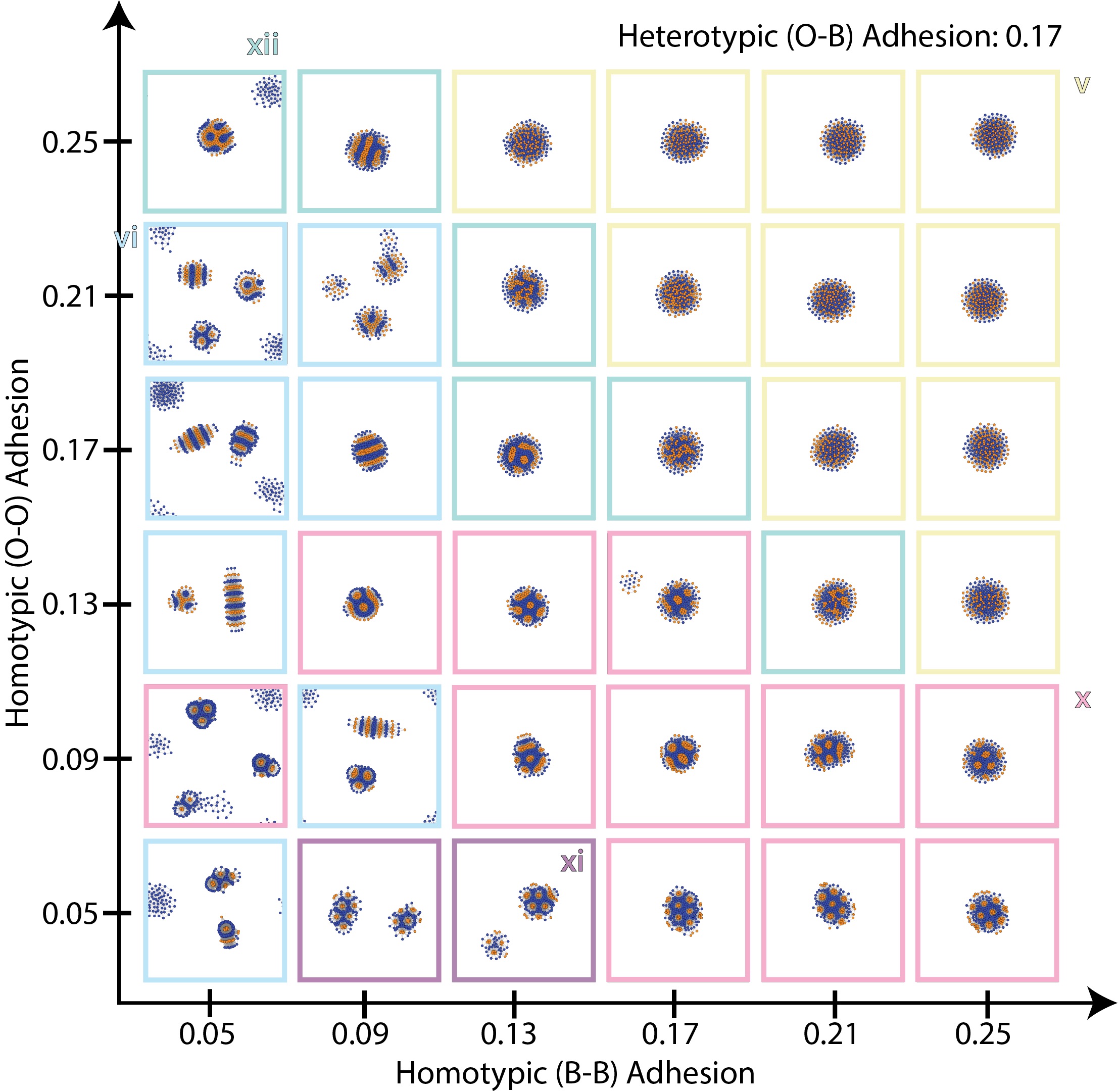}
    \caption{\footnotesize \textbf{Simulation snapshots of multicellular patterning of a heterogeneous, proliferating population at steady state with $\mathbf{J_{BO} = 0.17}$ and varying $\mathbf{J_{BB}, J_{OO}}$.}}
    \label{fig:SI_finalHT4p}
\end{figure}

\begin{figure}[h]
    \centering
    \includegraphics[width=\linewidth]{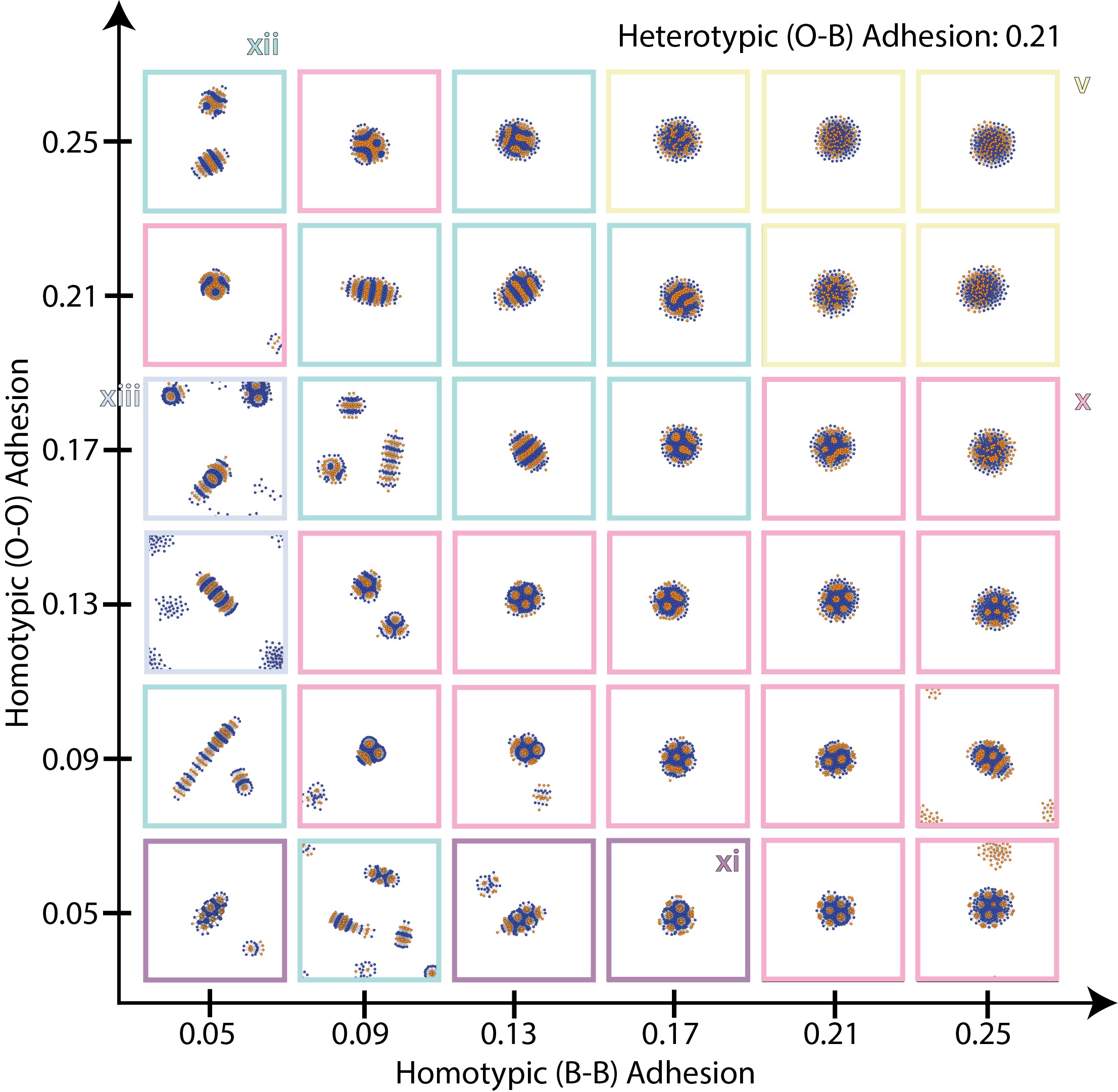}
    \caption{\footnotesize \textbf{Simulation snapshots of multicellular patterning of a heterogeneous, proliferating population at steady state with $\mathbf{J_{BO} = 0.21}$ and varying $\mathbf{J_{BB}, J_{OO}}$.}}
    \label{fig:SI_finalHT5p}
\end{figure}

\begin{figure}[h]
    \centering
    \includegraphics[width=\linewidth]{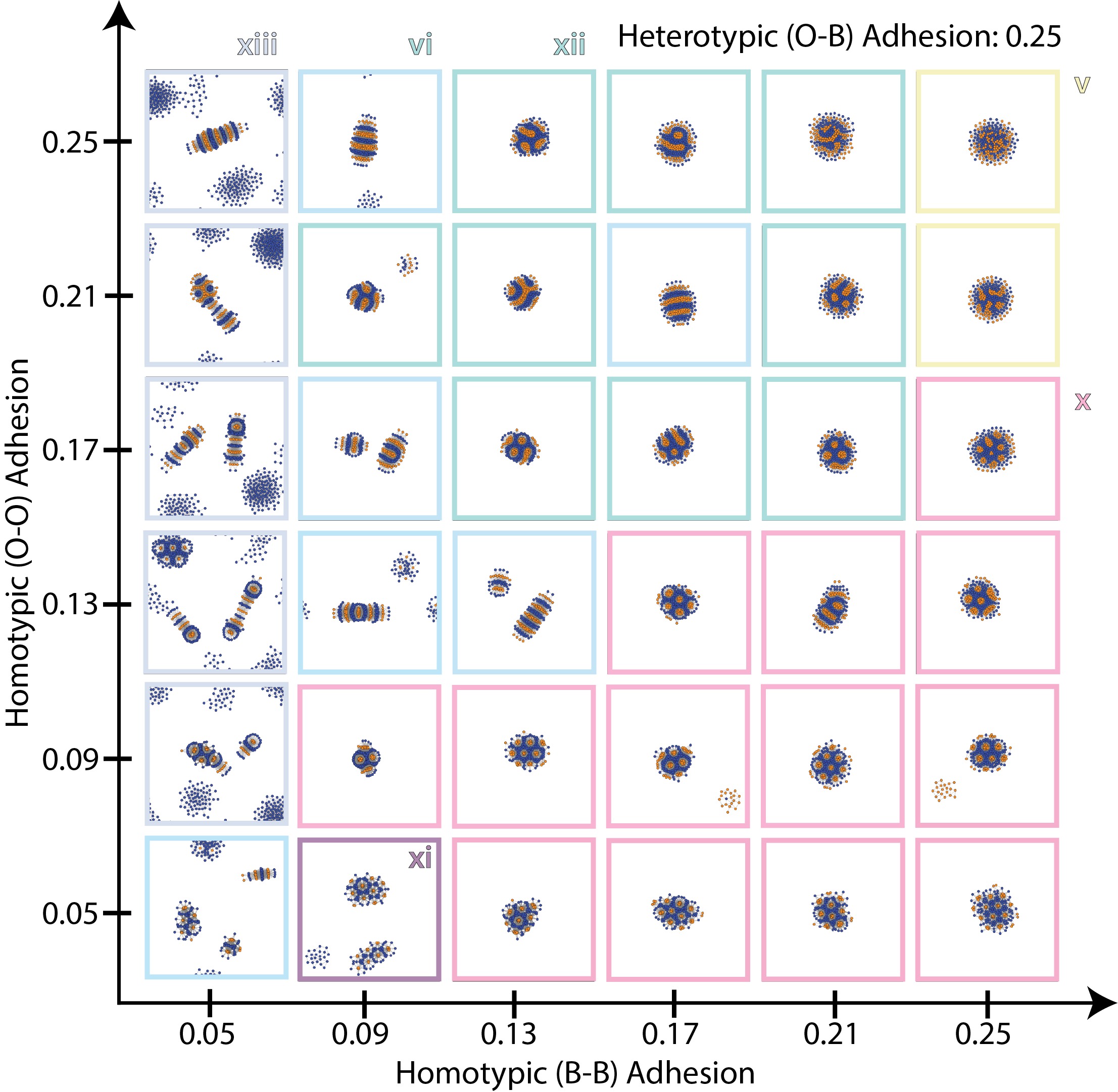}
    \caption{\footnotesize \textbf{Simulation snapshots of multicellular patterning of a heterogeneous, proliferating population at steady state with $\mathbf{J_{BO} = 0.25}$ and varying $\mathbf{J_{BB}, J_{OO}}$.}}
    \label{fig:SI_finalHT6p}
\end{figure}

\begin{figure}[h]
    \centering
    \includegraphics[scale=0.6]{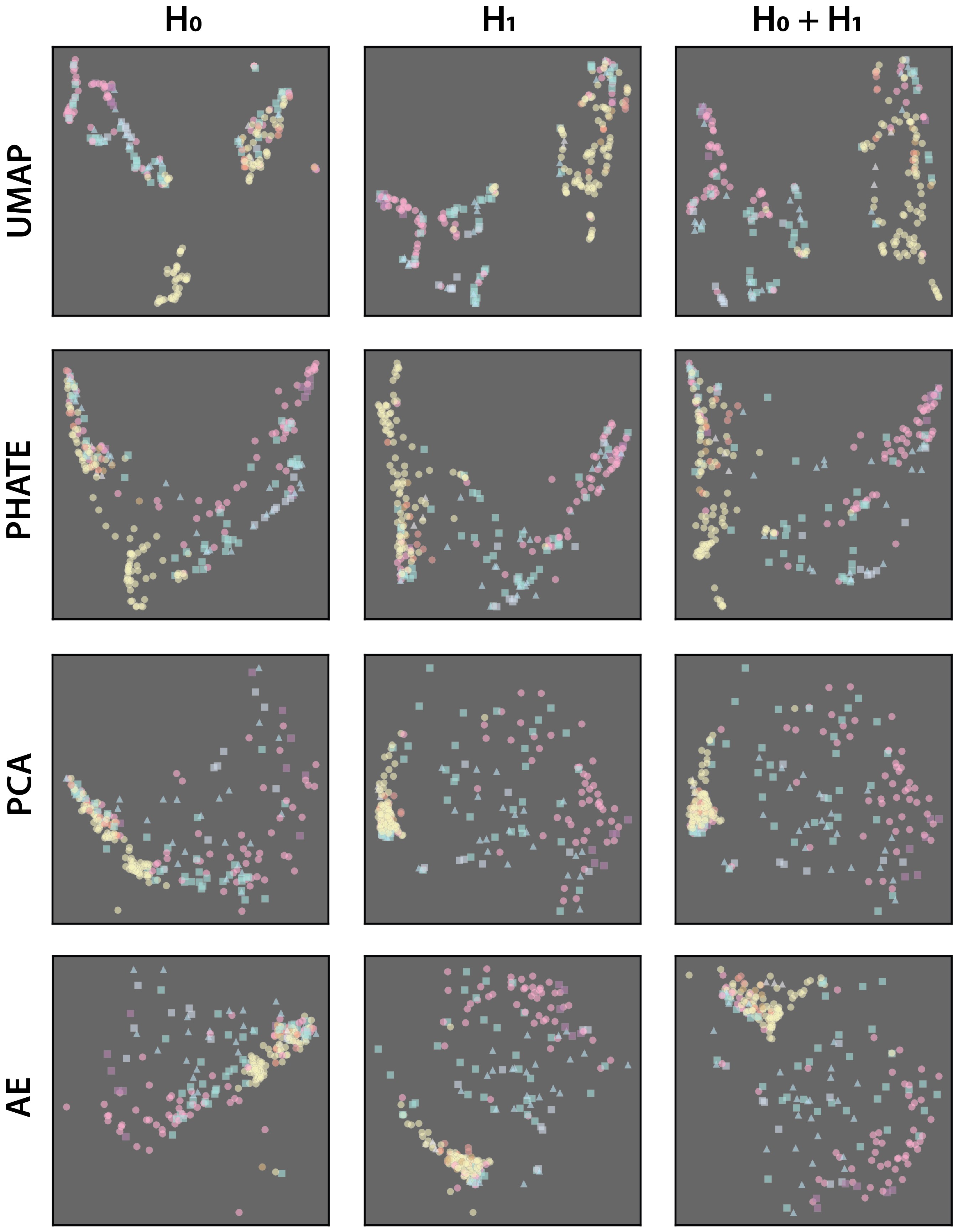}
    \caption{\footnotesize \textbf{2D embeddings of persistence images of proliferating particle configurations.} Low dimension representation of simulated particle configurations obtained using UMAP, PHATE, and PCA. Embeddings computed using dimension 0, dimension 1, and concatenated persistence images for simulations with proliferation enabled.}
    \label{fig:SI_persimg_dim_prolif}
\end{figure}

\begin{figure}[h]
    \centering
    \includegraphics[scale=0.6]{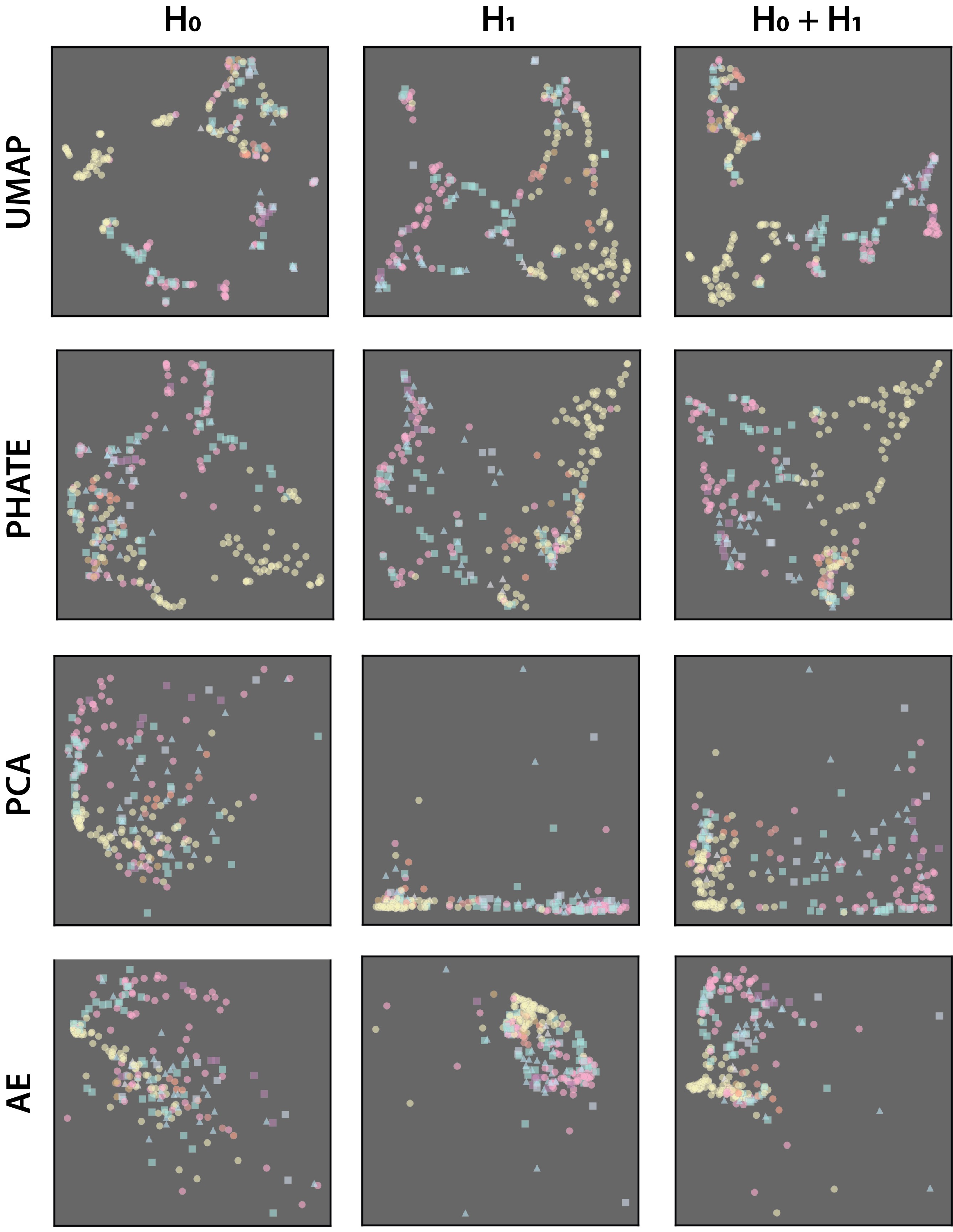}
    \caption{\footnotesize \textbf{2D embeddings of persistence curves of proliferating particle configurations.} Low dimension representation of simulated particle configurations obtained using UMAP, PHATE, and PCA. Embeddings computed using dimension 0, dimension 1, and concatenated persistence curves for simulations with proliferation enabled.}
    \label{fig:SI_perscurve_dim_prolif}
\end{figure}

\begin{figure}[h]
    \centering
    \includegraphics[scale=0.6]{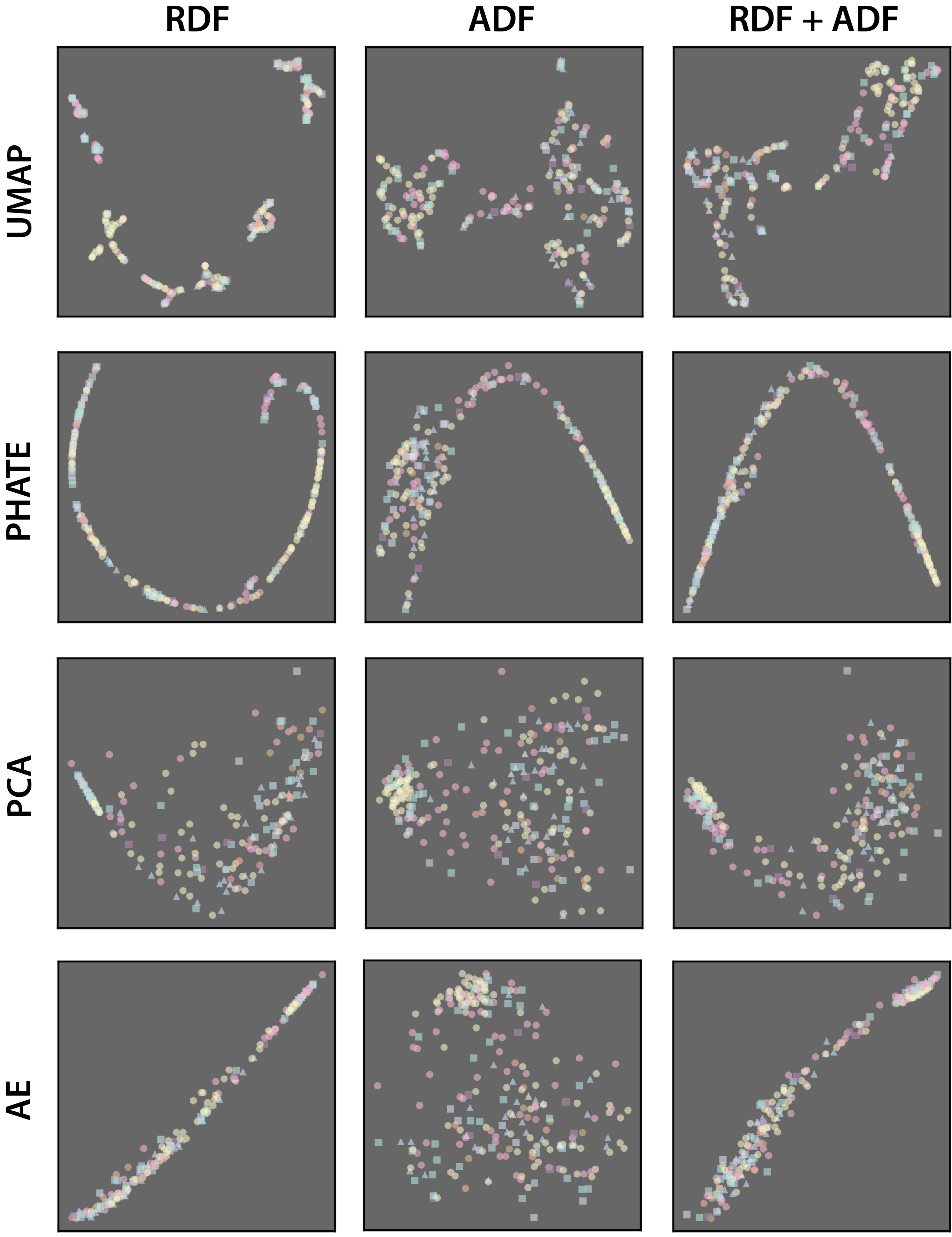}
    \caption{\footnotesize \textbf{2D embeddings of order parameters of proliferating particle configurations.} Low dimension representation of simulated particle configurations obtained using UMAP, PHATE, and PCA. Embeddings computed using radial, angular, and concatenated order parameters for simulations with proliferation enabled.}
    \label{fig:SI_op_dim_prolif}
\end{figure}

\begin{figure}[h]
    \centering
    \includegraphics[scale=0.62]{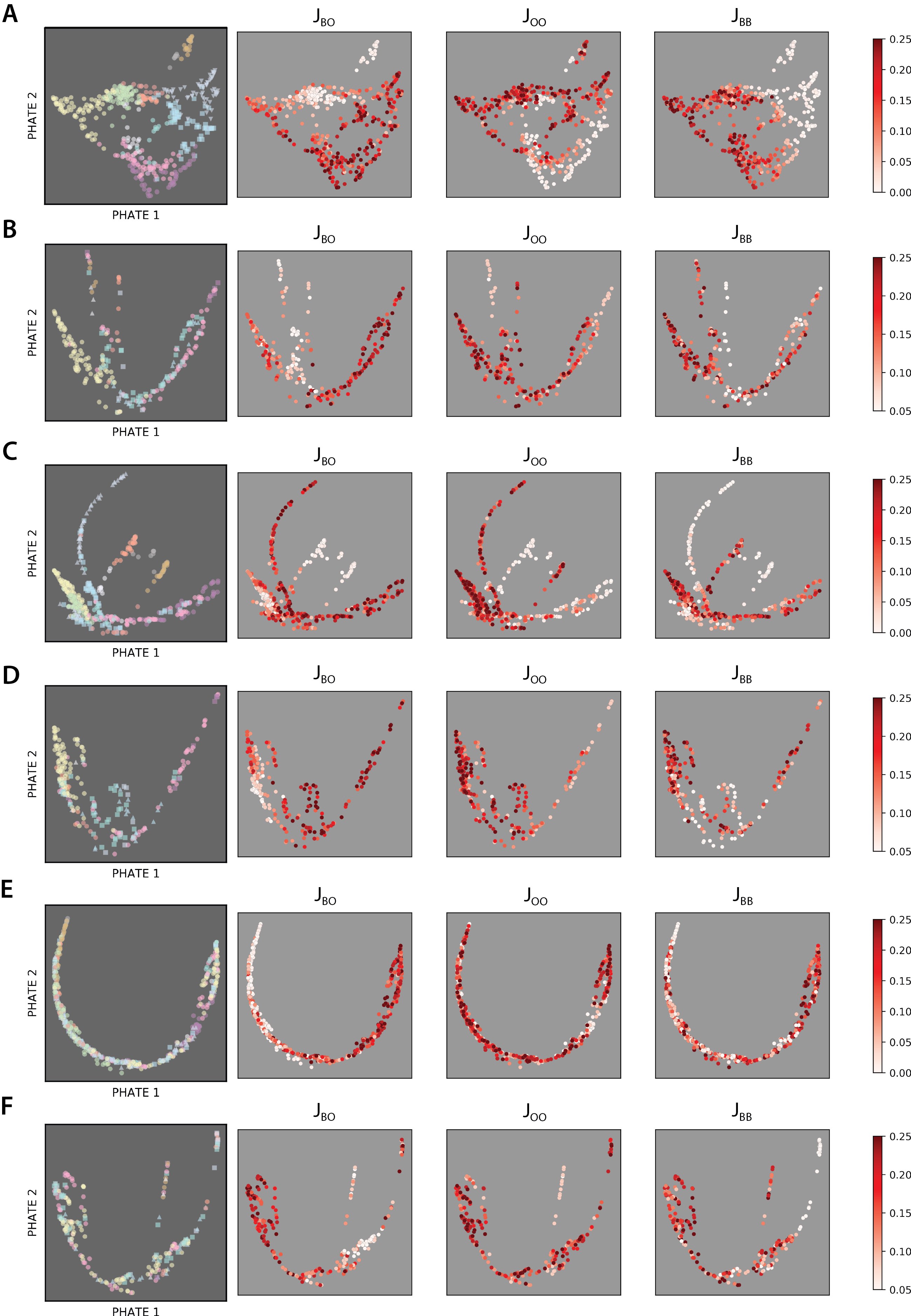}
    \caption{\footnotesize \textbf{Low-dimensional PHATE representations of simulations and corresponding adhesion values.} PHATE plots constructed using persistence images (A, B), persistence curves (C, D), and order parameters (E, F) colored by ground truth classification (left) and adhesion parameter values. Persistence images and curves are generated using dimension 1 topological features and order parameters include both radial and angular distributions. Panels A, C, and E correspond to simulations without proliferation. Panels B, D, and F correspond to simulations with proliferation enabled.}
    \label{fig:SI_phate_repr_adh_vals}
\end{figure}

\begin{figure}[h]
    \centering
    \includegraphics[scale=0.72]{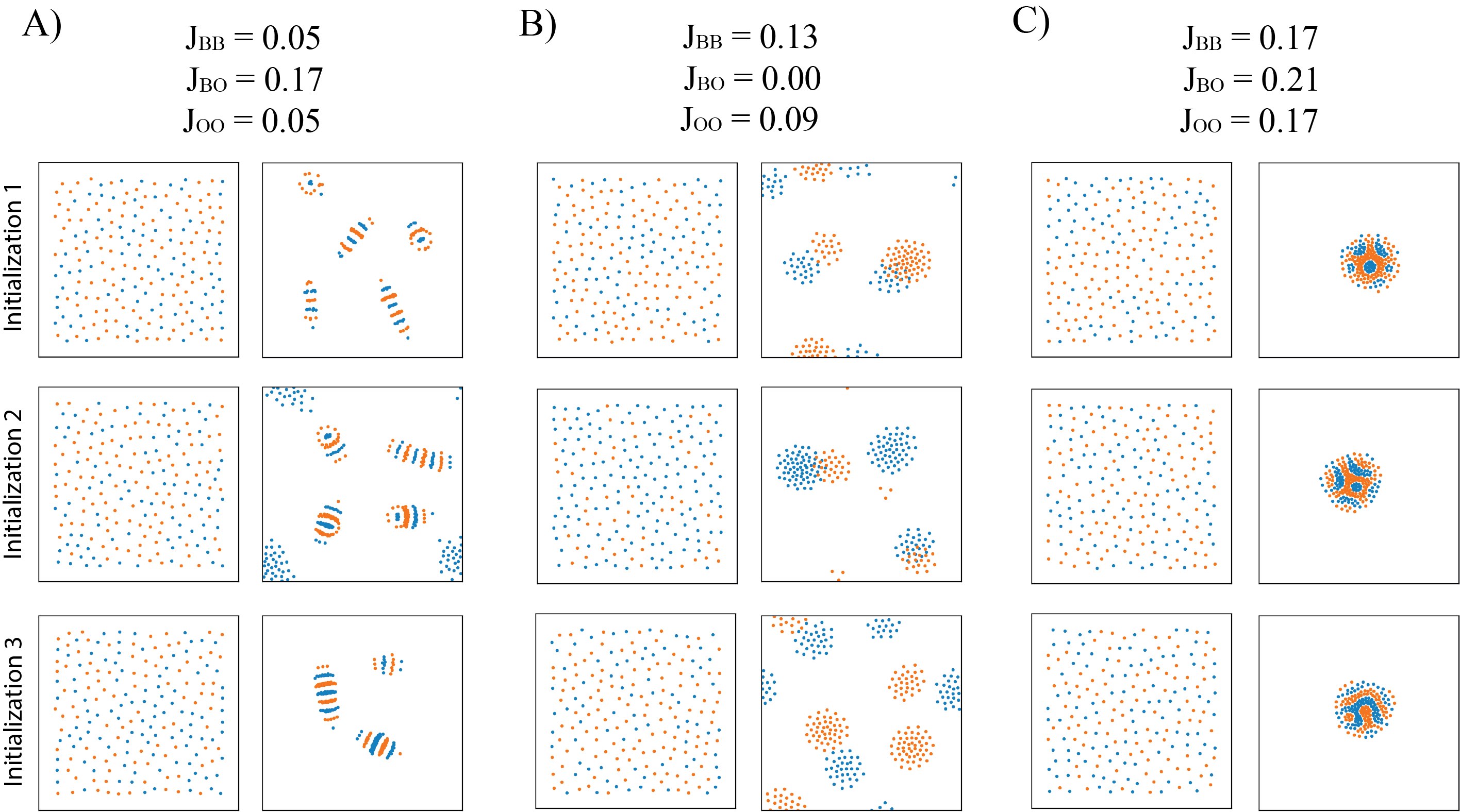}
    \caption{\footnotesize \textbf{Reproducibility of steady-state configurations at varying initial conditions.} (A) High heterotypic adhesion results in the formation of a stripe pattern where blue and orange particles maximize their interaction with each other. (B) Complete sorting simulation where both blue and orange particles form separate clusters due to high homotypic adhesion. (C) Clustering of blue and orange particles at intermediate homotypic and high heterotypic adhesions.}
    \label{fig:SI_init_cond}
\end{figure}

\begin{figure}[h]
    \centering
    \includegraphics[scale=0.7]{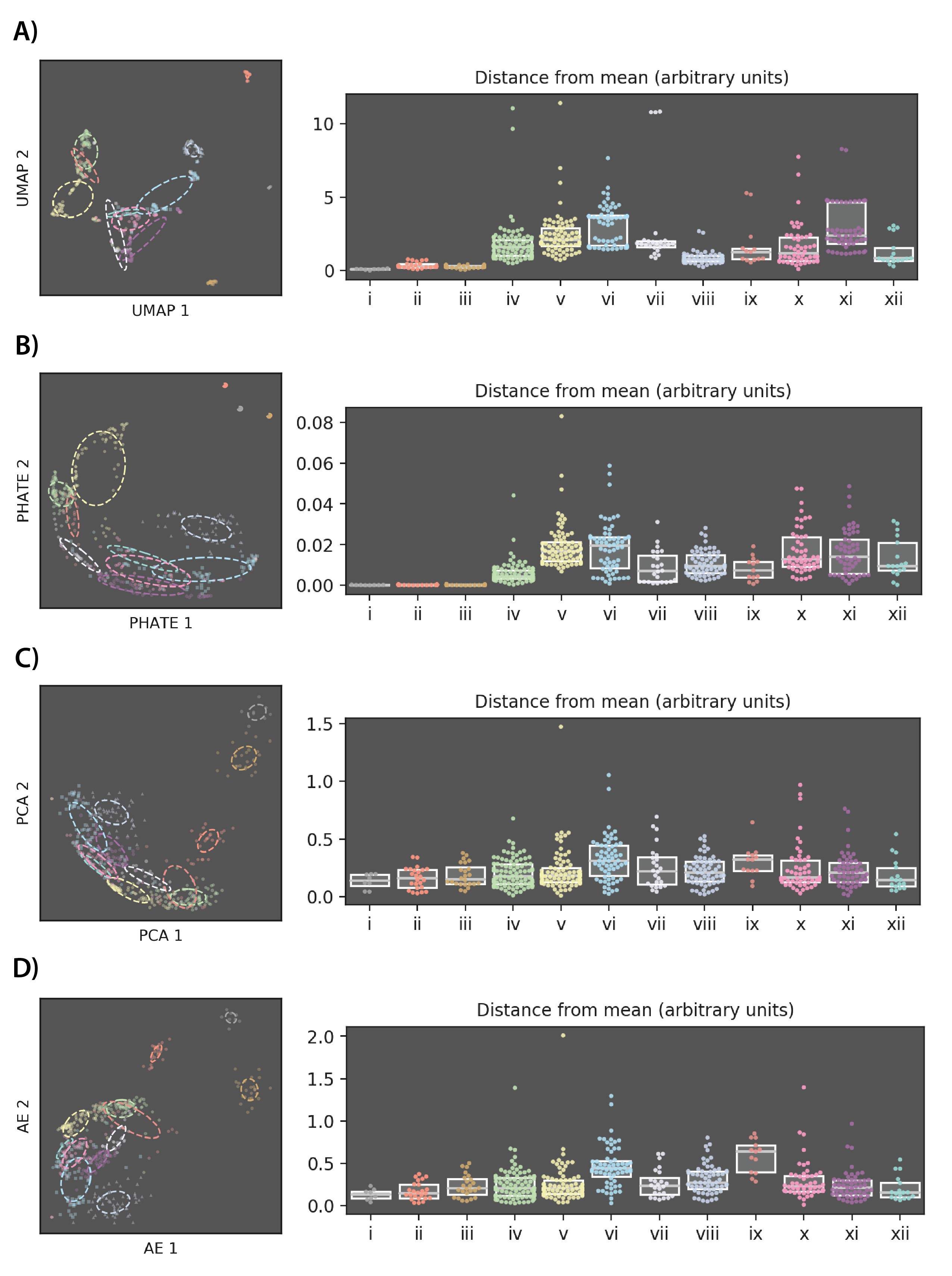}
    \caption{\footnotesize \textbf{GMM fit to low-dimensional embeddings of connected components in persistent images at constant population size.} Dimension reduced topology of connected components extracted using dimension 0 homology of cell configurations at constant population size. Colored dashed ellipses correspond to unit standard deviation in the Gaussian fit for each cell configuration. Beeswarm and boxplot of distance from mean (cluster center) for each configuration using (A) UMAP, (B) PHATE, (C) PCA, and (D) AE embeddings. The boxplot central line denotes the median, the bottom and top of the box correspond to the 25th and 75th percentiles, respectively.}
    \label{fig:SI_GMM_noprolif_H0}
\end{figure}

\begin{figure}[h]
    \centering
    \includegraphics[scale=0.7]{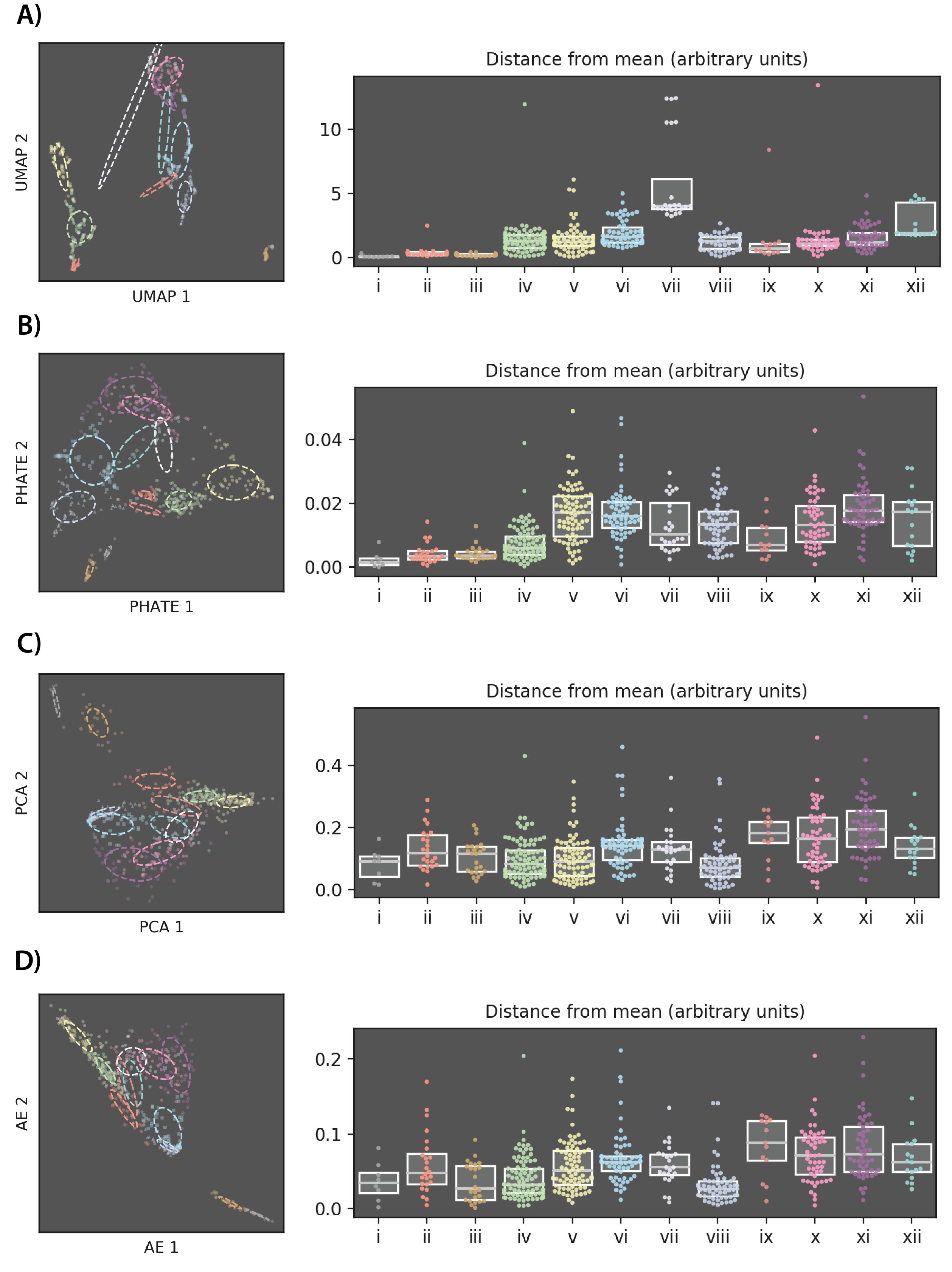}
    \caption{\footnotesize \textbf{GMM fit to low-dimensional embeddings of topological loops in persistent images at constant population size.} Dimension reduced topology of loops extracted using dimension 1 homology of cell configurations at constant population size. Colored dashed ellipses correspond to unit standard deviation in the Gaussian fit for each cell configuration. Beeswarm and boxplot of distance from mean (cluster center) for each configuration using (A) UMAP, (B) PHATE, (C) PCA, and (D) AE embeddings. The boxplot central line denotes the median, the bottom and top of the box correspond to the 25th and 75th percentiles, respectively.}
    \label{fig:SI_GMM_noprolif_H1}
\end{figure}

\begin{figure}[h]
    \centering
    \includegraphics[scale=0.7]{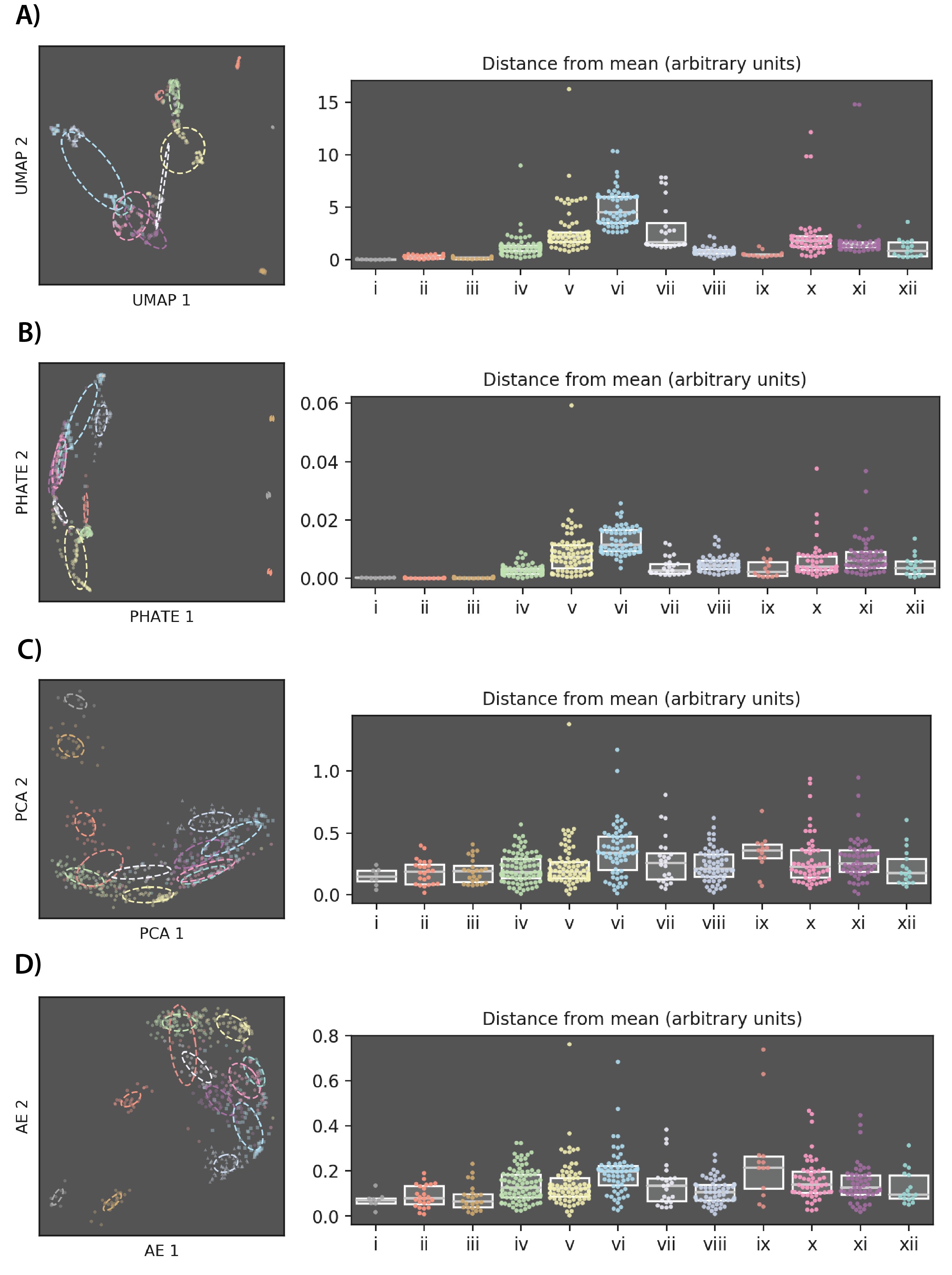}
    \caption{\footnotesize \textbf{GMM fit to low-dimensional embeddings of concatenated topological features in persistent images at constant population size.} Dimension reduced topology of connected components and loops extracted using concatenated dimension 0 and dimension 1 homology of cell configurations at constant population size. Colored dashed ellipses correspond to unit standard deviation in the Gaussian fit for each cell configuration. Beeswarm and boxplot of distance from mean (cluster center) for each configuration using (A) UMAP, (B) PHATE, (C) PCA, and (D) AE embeddings. The boxplot central line denotes the median, the bottom and top of the box correspond to the 25th and 75th percentiles, respectively.}
    \label{fig:SI_GMM_noprolif_H0p1}
\end{figure}

\begin{figure}[h]
    \centering
    \includegraphics[scale=0.7]{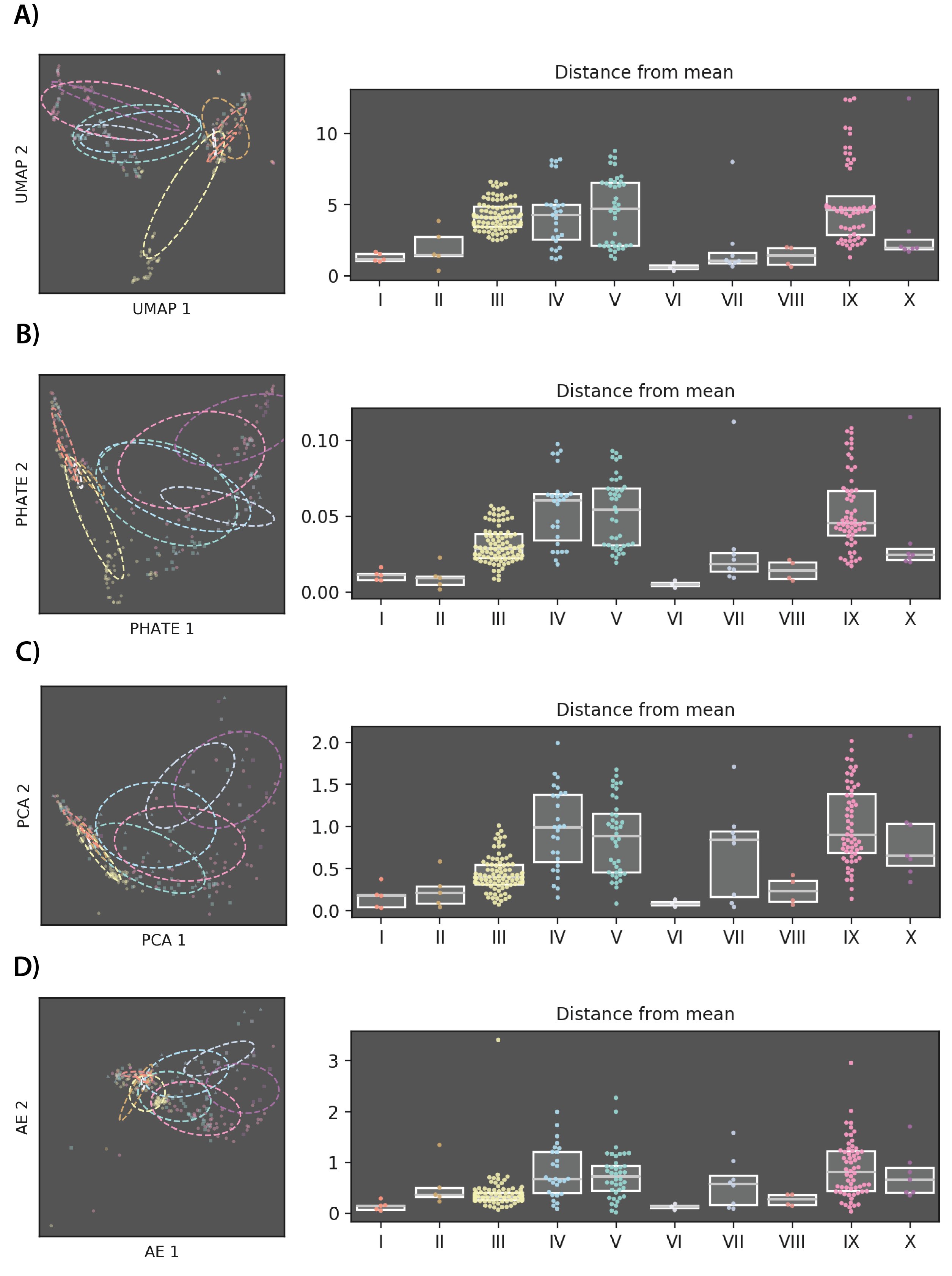}
    \caption{\footnotesize \textbf{GMM fit to low-dimensional embeddings of connected components in persistent images at varying population size.} Dimension reduced topology of connected components extracted using dimension 0 homology of cell configurations at varying population size. Colored dashed ellipses correspond to unit standard deviation in the Gaussian fit for each cell configuration. Beeswarm and boxplot of distance from mean (cluster center) for each configuration using (A) UMAP, (B) PHATE, (C) PCA, and (D) AE embeddings. The boxplot central line denotes the median, the bottom and top of the box correspond to the 25th and 75th percentiles, respectively.}
    \label{fig:SI_GMM_prolif_H0}
\end{figure}

\begin{figure}[h]
    \centering
    \includegraphics[scale=0.7]{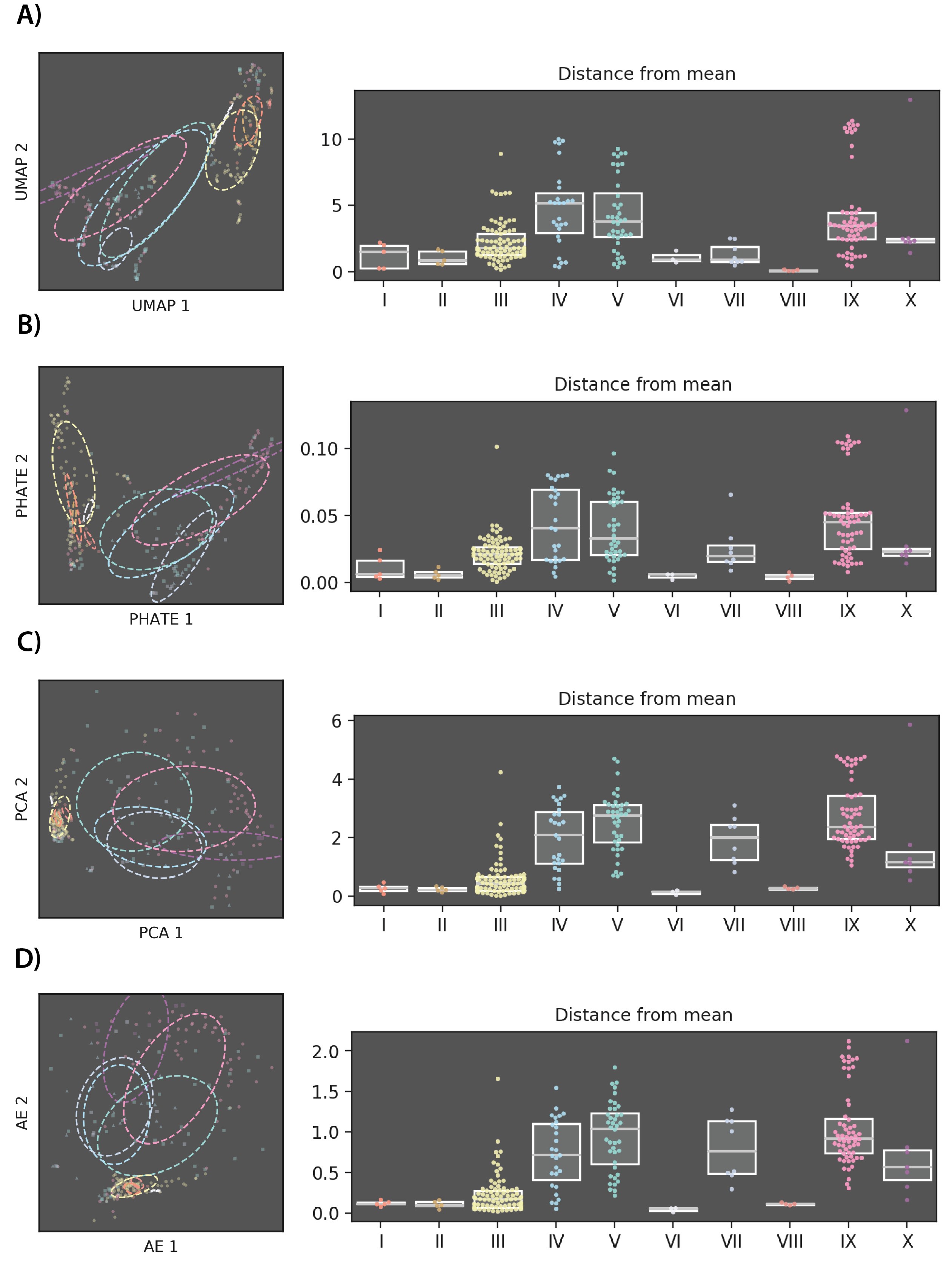}
    \caption{\footnotesize \textbf{GMM fit to low-dimensional embeddings of topological loops in persistent images at varying population size.} Dimension reduced topology of loops extracted using dimension 1 homology of cell configurations at varying population size. Colored dashed ellipses correspond to unit standard deviation in the Gaussian fit for each cell configuration. Beeswarm and boxplot of distance from mean (cluster center) for each configuration using (A) UMAP, (B) PHATE, (C) PCA, and (D) AE embeddings. The boxplot central line denotes the median, the bottom and top of the box correspond to the 25th and 75th percentiles, respectively.}
    \label{fig:SI_GMM_prolif_H1}
\end{figure}

\begin{figure}[h]
    \centering
    \includegraphics[scale=0.7]{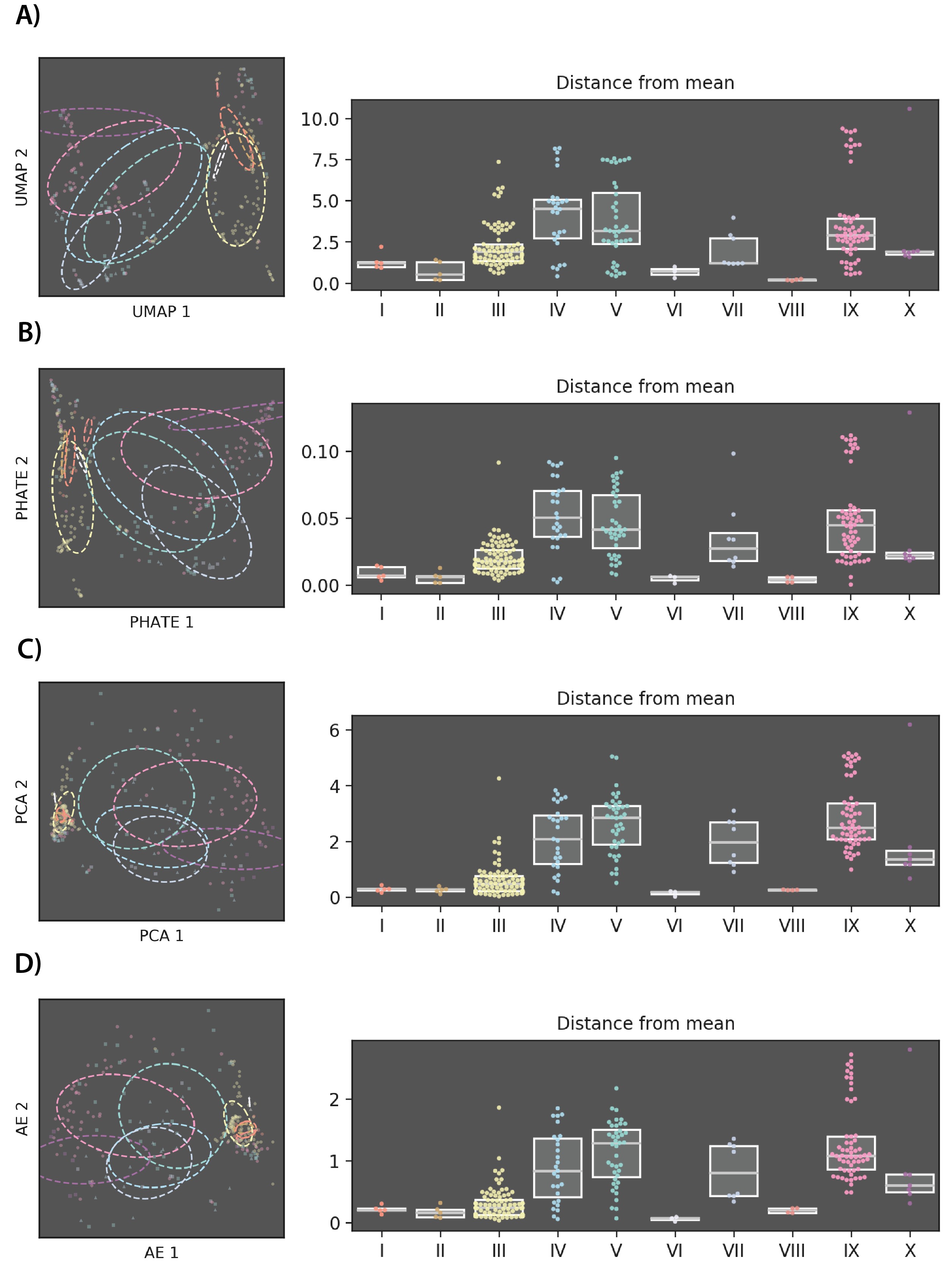}
    \caption{\footnotesize \textbf{GMM fit to low-dimensional embeddings of concatenated topological features in persistent images at varying population size.} Dimension reduced topology of connected components and loops extracted using concatenated dimension 0 and dimension 1 homology of cell configurations at varying population size. Colored dashed ellipses correspond to unit standard deviation in the Gaussian fit for each cell configuration. Beeswarm and boxplot of distance from mean (cluster center) for each configuration using (A) UMAP, (B) PHATE, (C) PCA, and (D) AE embeddings. The boxplot central line denotes the median, the bottom and top of the box correspond to the 25th and 75th percentiles, respectively.}
    \label{fig:SI_GMM_prolif_H0p1}
\end{figure}

\begin{figure}[h]
    \centering
    \includegraphics[scale=0.9]{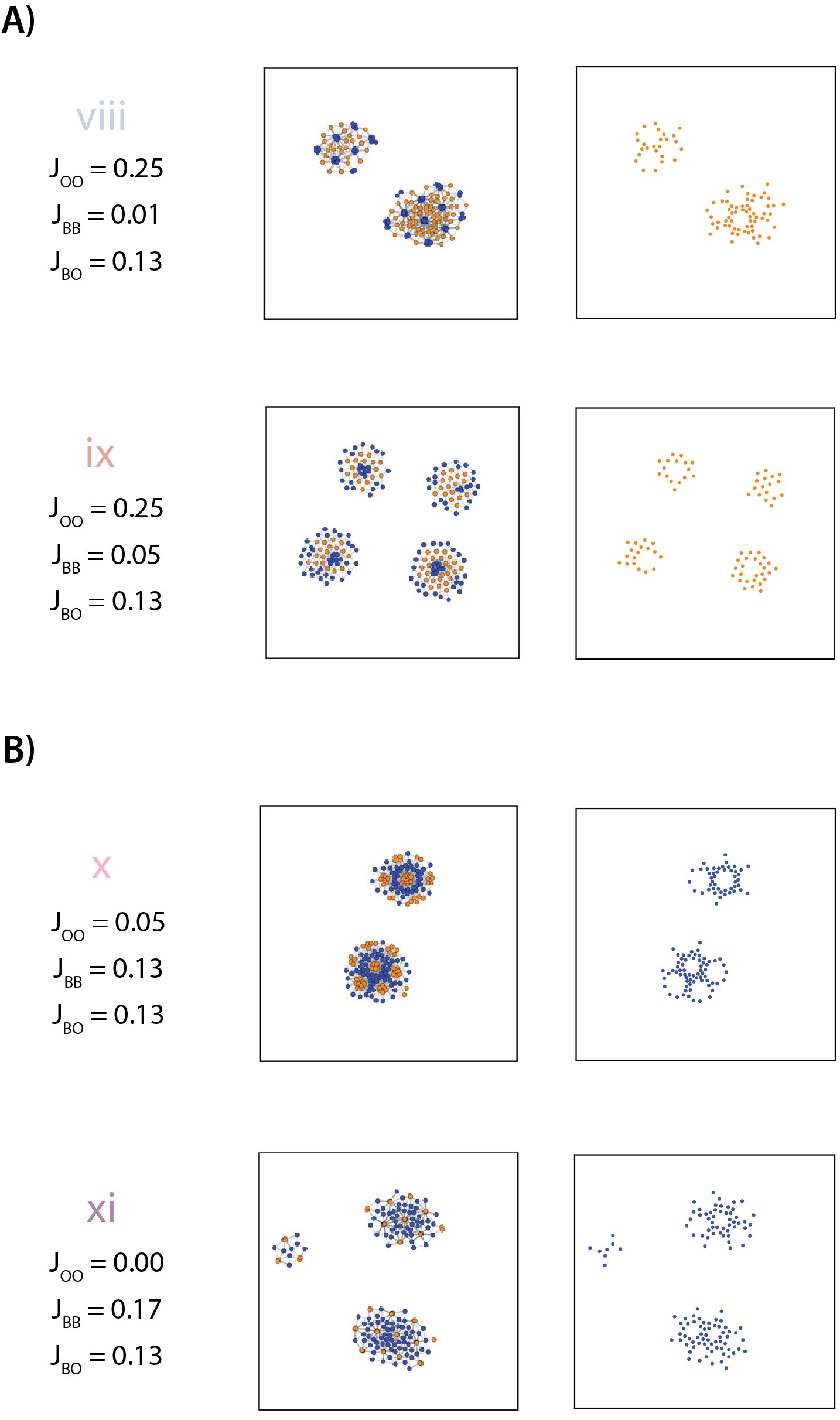}
    \caption{\footnotesize \textbf{Examples of misclassified cell configurations due to similarity of single-cell topological loops.} (A) Simulations are misclassified when only orange cell positions are considered. (B) Simulations are misclassified when only blue cell positions are considered. Note that these simulations are not misclassified in dimension 0 topology, due to differences in cell density.}
    \label{fig:SI_H1_misclassification}
\end{figure}

\begin{figure}[h]
    \centering
    \includegraphics[scale=0.9]{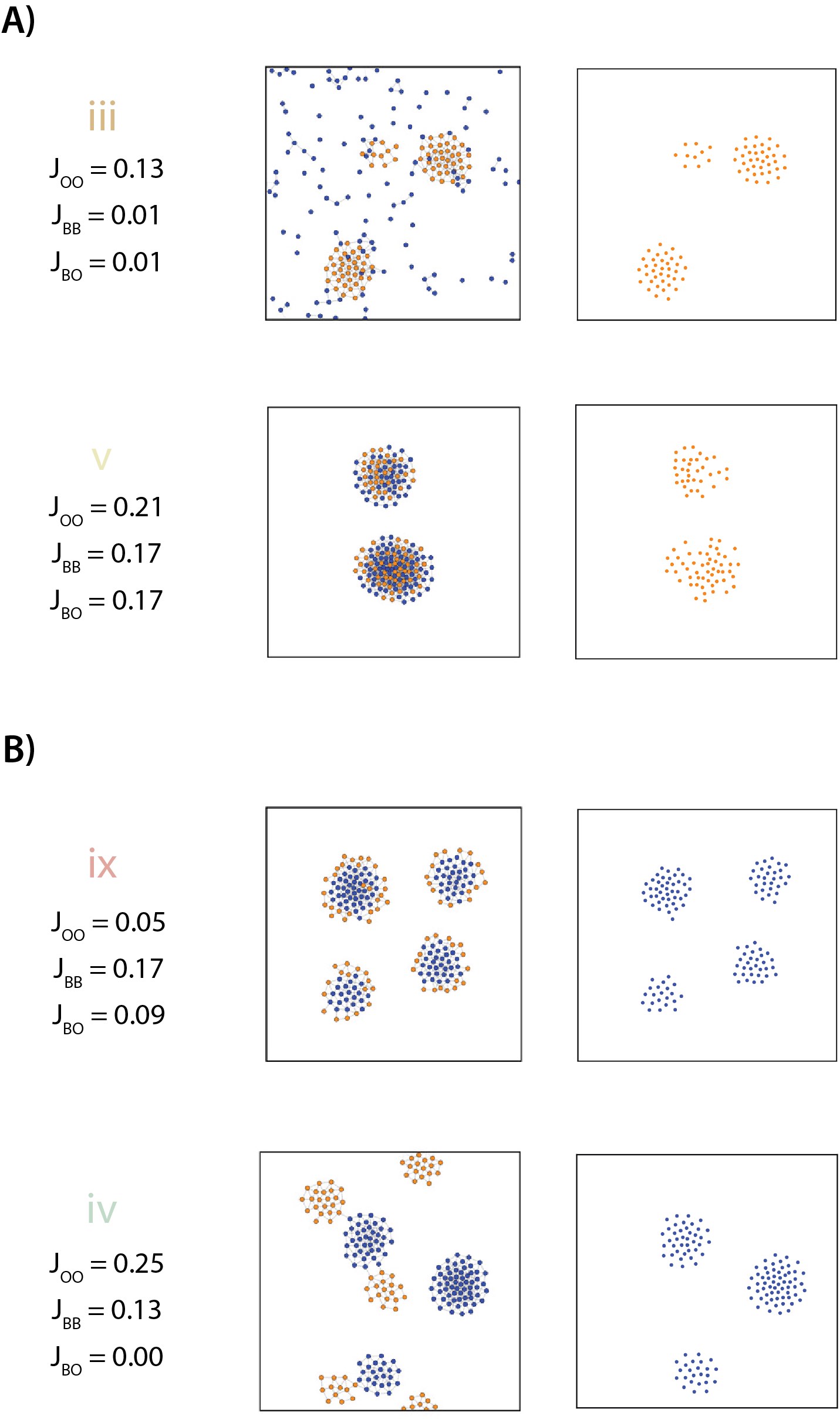}
    \caption{\footnotesize \textbf{Examples of misclassified cell configurations due to similarity of single-cell topological clusters.} (A) Simulations are misclassified when only orange cell positions are considered. (B) Simulations are misclassified when only blue cell positions are considered. Note that these simulations are not misclassified in dimension 1 topology, due to the absence of topological loops.}
    \label{fig:SI_H0_misclassification}
\end{figure}

\begin{figure}[h]
    \centering
    \includegraphics[scale=0.75]{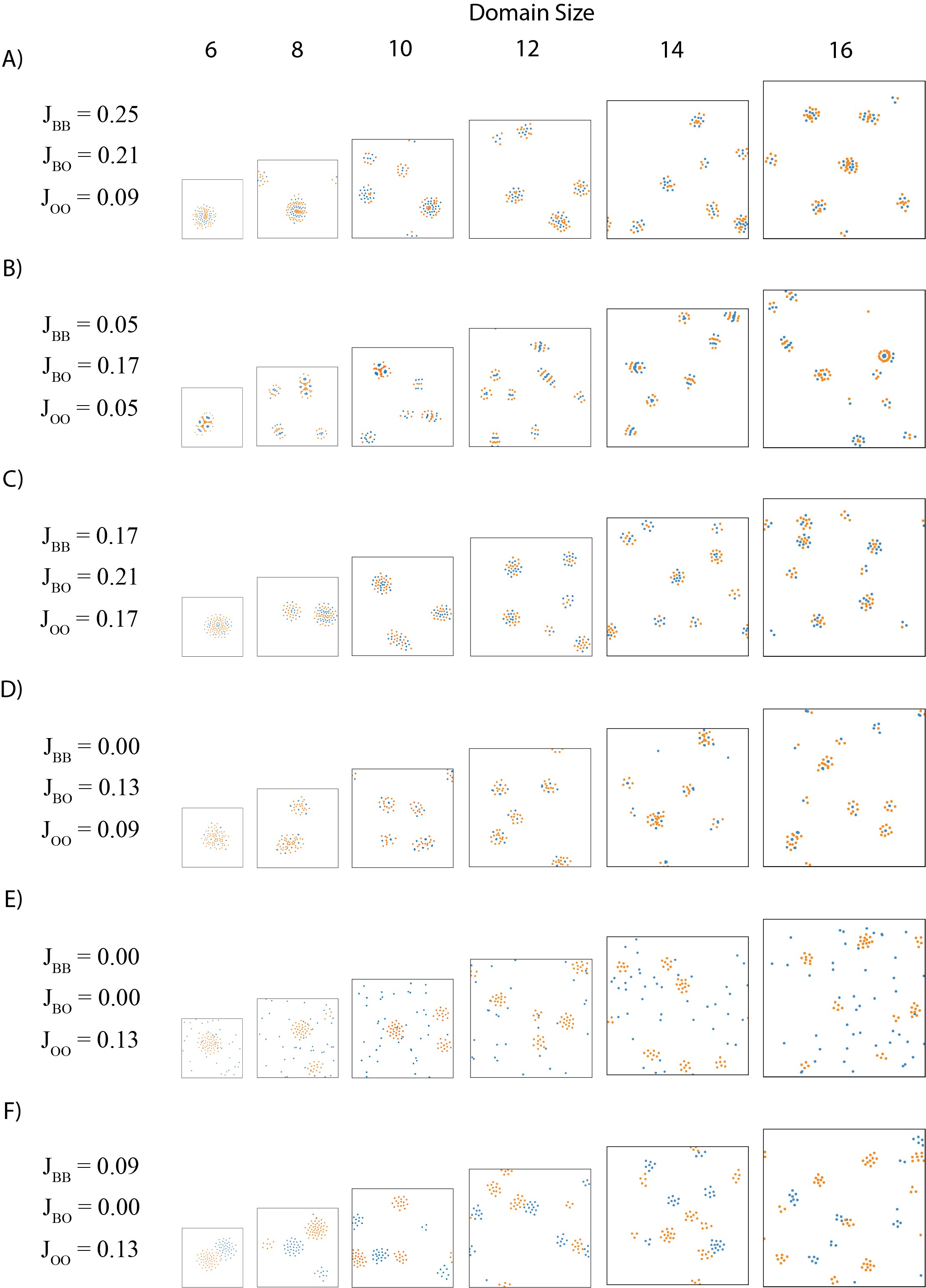}
    \caption{\footnotesize \textbf{Simulations at varying domain sizes with constant population size.} (A) Orange and purple particles aggregate into clusters while blue particles remain individually dispersed. (B) Purple particles are engulfed within orange clusters. (C) Particles aggregate to form clusters due to high adhesion. Clusters contain a mixture of particle types. (D) Blue and purple particles surround the orange particles due to high blue-orange and purple-orange adhesion. (E) Orange and purple particles aggregate into clusters while blue particles remain individually dispersed. (F) Blue and orange particles aggregate into clusters due to high adhesion.}
    \label{fig:SI_domain_size}
\end{figure}

\begin{figure}[h]
    \centering
    \includegraphics[scale=0.72]{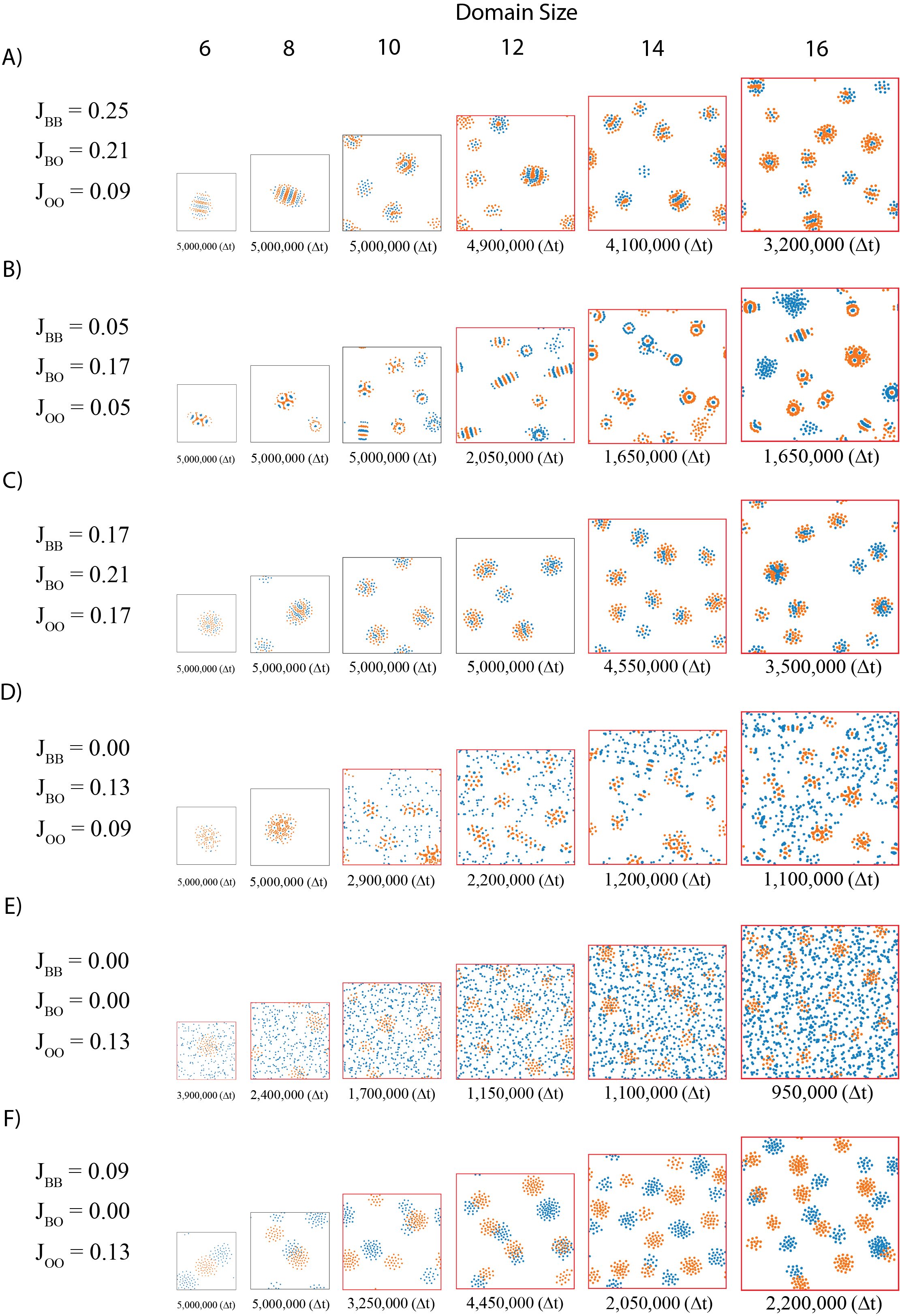}
    \caption{\footnotesize \textbf{Simulations at varying domain sizes with varying population size.} (A) Orange and purple particles aggregate into clusters while blue particles remain individually dispersed. (B) Purple particles are engulfed within orange clusters. (C) Particles aggregate to form clusters due to high adhesion. Clusters contain a mixture of particle types. (D) Blue and purple particles surround the orange particles due to high blue-orange and purple-orange adhesion. (E) Orange and purple particles aggregate into clusters while blue particles remain individually dispersed. (F) Blue and orange particles aggregate into clusters due to high adhesion. Snapshots outlined in red did not run to completion in the allotted time due to the combinatorial complexity of force calculations with exponentially increasing population of orange cells.}
    \label{fig:SI_domain_size_prolif}
\end{figure}

\begin{figure}[h]
    \centering
    \includegraphics[scale=0.8]{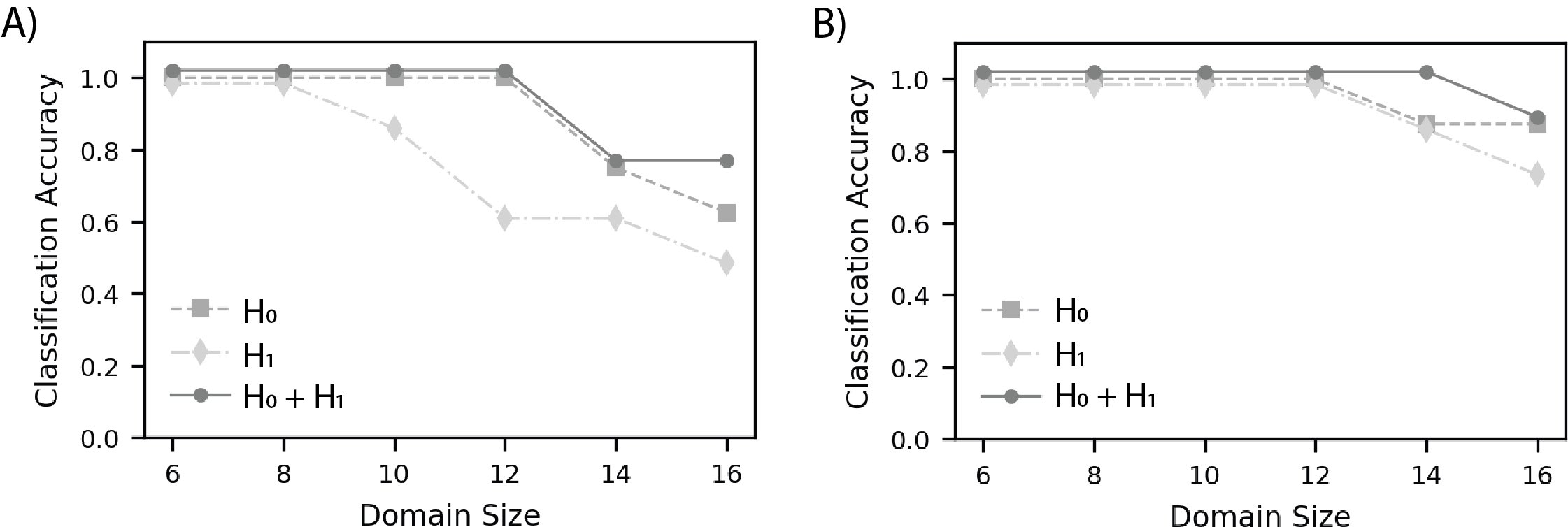}
    \caption{\footnotesize \textbf{Classification accuracy of persistence images at various domain sizes.} (A) at constant population size, and (B) at varying population size.}
    \label{fig:SI_domain_size_classify}
\end{figure}

\begin{figure}[h]
    \centering
    \includegraphics[scale=0.28]{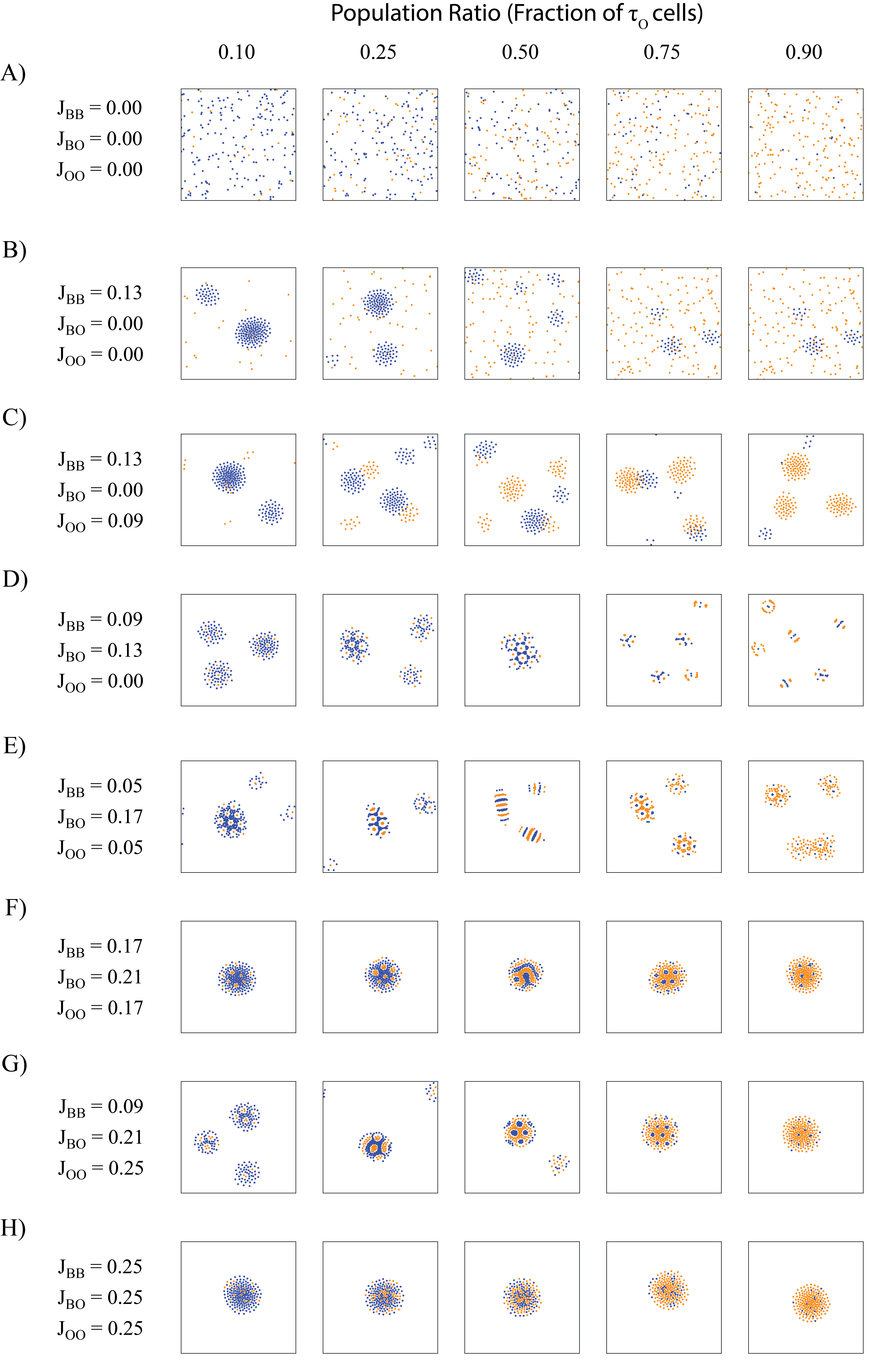}
    \caption{\footnotesize \textbf{Simulation snapshots showing self-organization in a heterogeneous population at varying population ratios.} (A) Particles remain individually dispersed at low adhesion values. (B) Blue particles aggregate into clusters due to high blue-blue adhesion. (C) Complete sorting simulation where both blue and orange particles form separate clusters due to high homotypic adhesion. (D-G) High heterotypic adhesion results in configurations that maximize the interaction between the two cell types, forming hexagonal, striped and spotted patterns. (H) Well-mixed clusters are obtained when homotypic and heterotypic adhesion values are greater than zero and (approximately) equal.}
    \label{fig:SI_snaps_noprolif_ratios}
\end{figure}

\begin{figure}[h]
    \centering
    \includegraphics[scale=0.26]{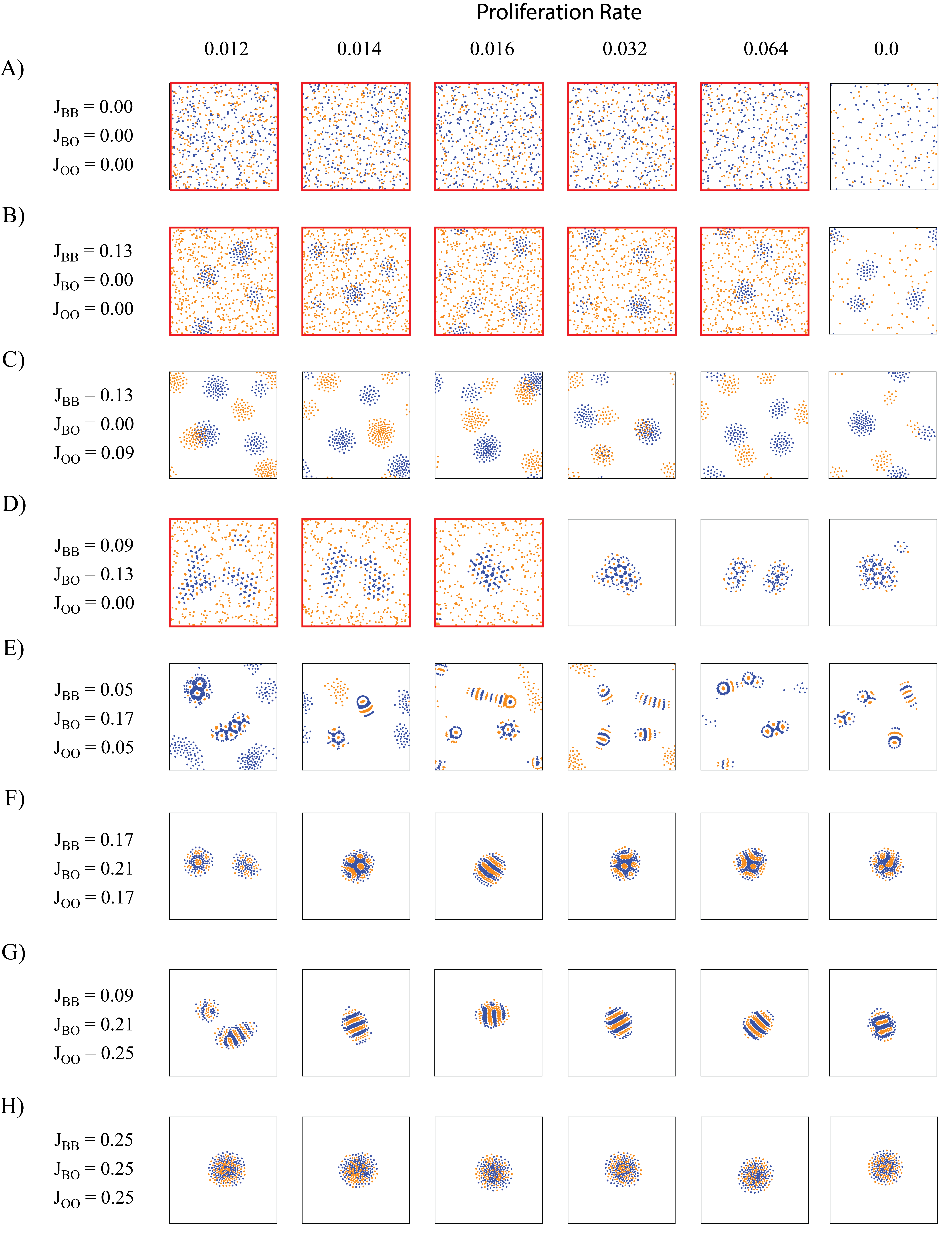}
    \caption{\footnotesize \textbf{Simulation snapshots showing self-organization in a heterogeneous population at varying population size.} (A) Particles remain individually dispersed at low adhesion values. (B) Blue particles aggregate into clusters due to high blue-blue adhesion. (C) Complete sorting simulation where both blue and orange particles form separate clusters due to high homotypic adhesion. (D-G) High heterotypic adhesion results in configurations that maximize the interaction between the two cell types, forming hexagonal, striped and spotted patterns. (H) Well-mixed clusters are obtained when homotypic and heterotypic adhesion values are greater than zero and (approximately) equal. Snapshots outlined in red did not run to completion in the allotted time due to the combinatorial complexity of force calculations with exponentially increasing population of orange cells.}
    \label{fig:SI_snaps_prolif_rates}
\end{figure}

\begin{figure}[h]
    \centering
    \includegraphics[scale=0.72]{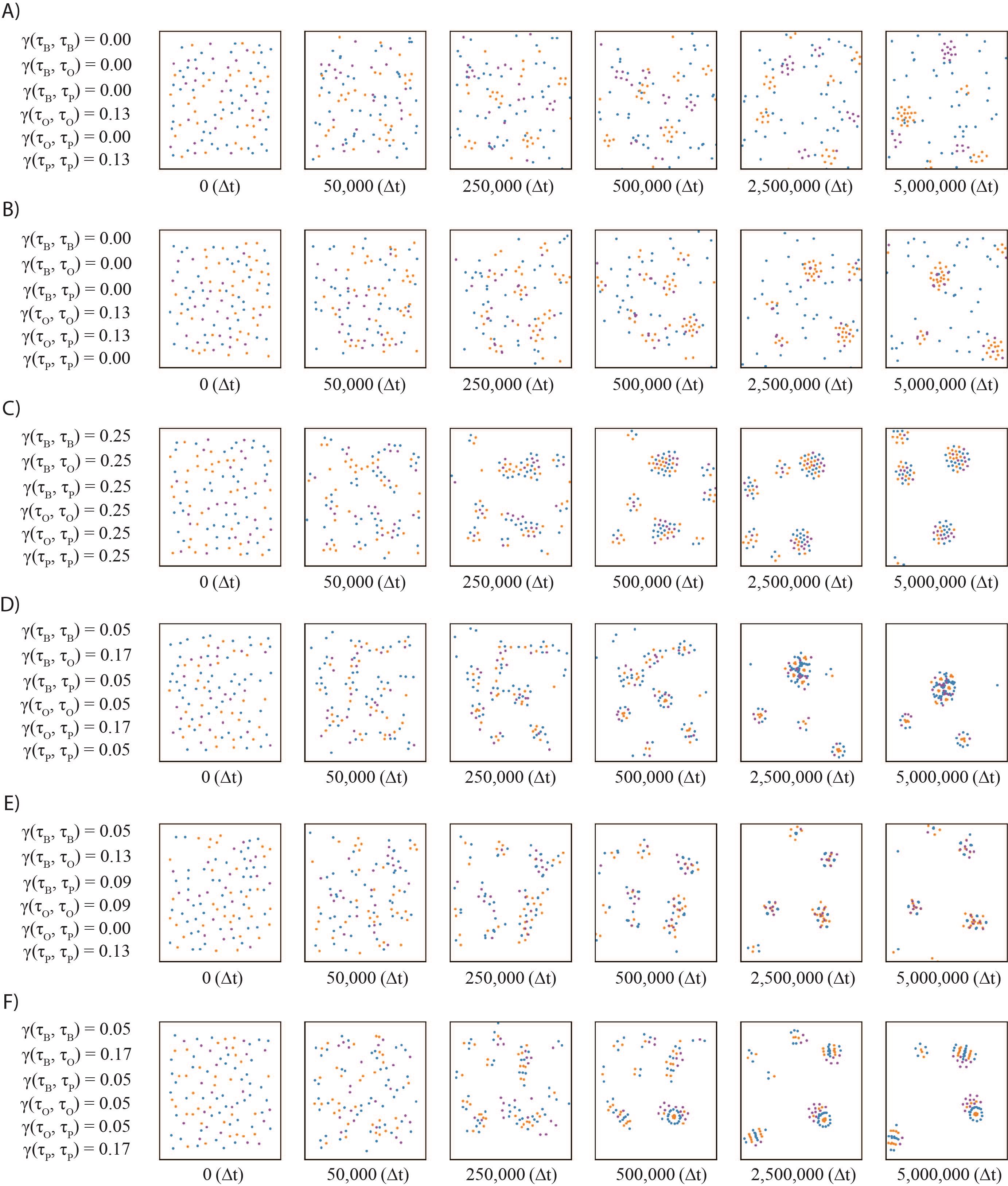}
    \caption{\footnotesize \textbf{Simulation snapshots showing self-organization in a heterogeneous population with three cell types (50\% orange, 30\% blue, 20\% green) at a constant population size of 100.} (A) Orange and green cells aggregate into separate clusters, while blue cells remain individually dispersed. (B) Green cells are engulfed within orange clusters, while blue cells remain individually dispersed. (C) Cells indiscriminately aggregate into clusters due to high adhesion. (D) Hexagonal arrangement of orange clusters surrounded by blue and green cells due to high blue-orange and green-orange adhesion. (E) Hexagonal arrangement of blue cells with offset hexagonal arrangement of orange and green cells (F) Alternating stripes of blue and orange cells with some displaced green cells.}
    \label{fig:SI_multiple_cells_noprolif_100}
\end{figure}

\begin{figure}[h]
    \centering
    \includegraphics[scale=0.72]{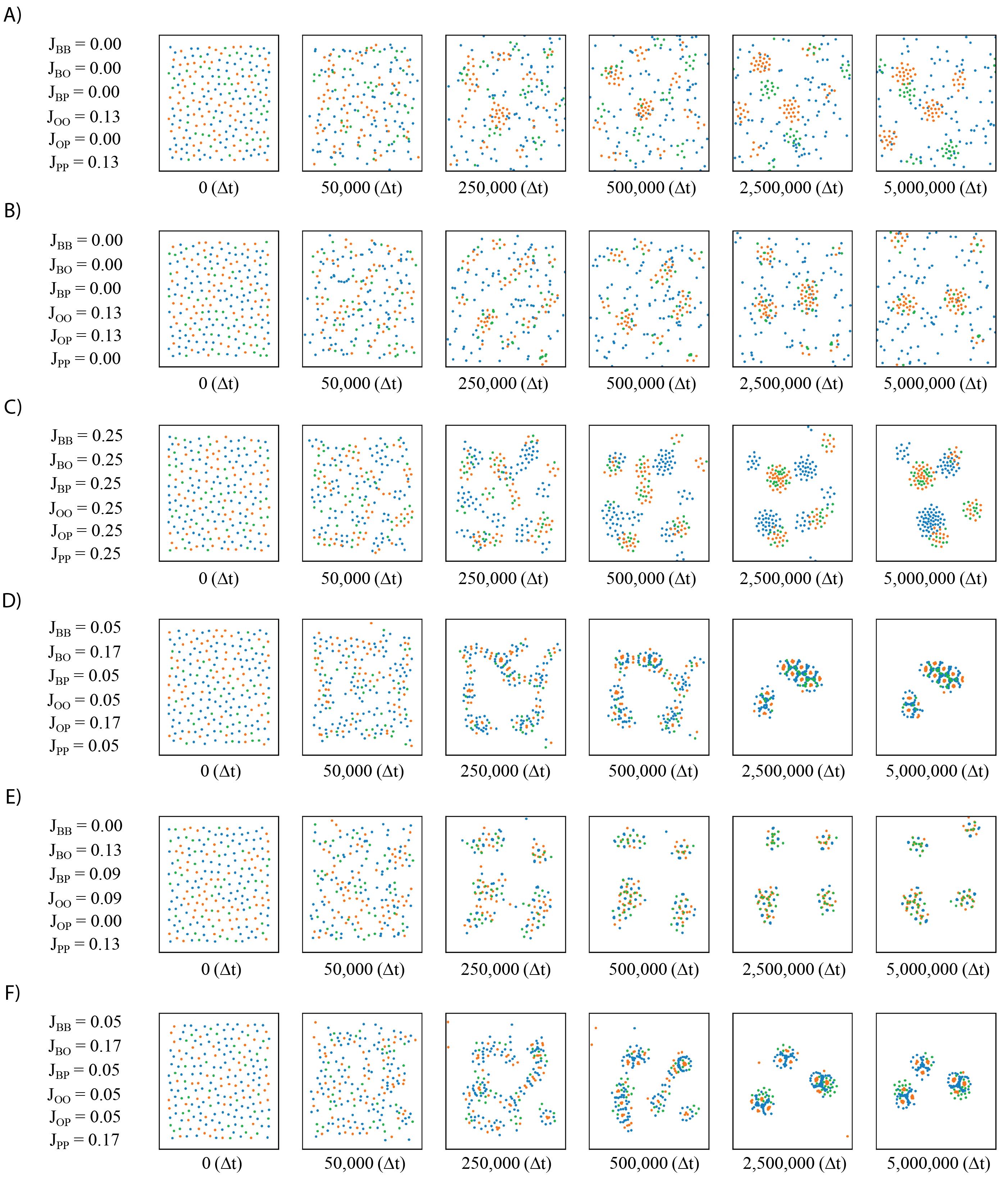}
  \caption{\footnotesize \textbf{Simulation snapshots showing self-organization in a heterogeneous population with three cell types (50\% orange, 30\% blue, 20\% green) at a constant population size of 200.} (A) Orange and green cells aggregate into separate clusters, while blue cells remain individually dispersed. (B) Green cells are engulfed within orange clusters, while blue cells remain individually dispersed. (C) Cells partially sorted within clusters due to high adhesion. (D) Hexagonal arrangement of orange clusters surrounded by blue and green cells due to high blue-orange and green-orange adhesion. (E) Hexagonal arrangement of blue cells with offset hexagonal arrangement of orange and green cells (F) Hexagonal arrangement of orange clusters surrounded by blue and green cells.}
    \label{fig:SI_multiple_cells_noprolif}
\end{figure}

\begin{figure}[h]
    \centering
    \includegraphics[scale=0.72]{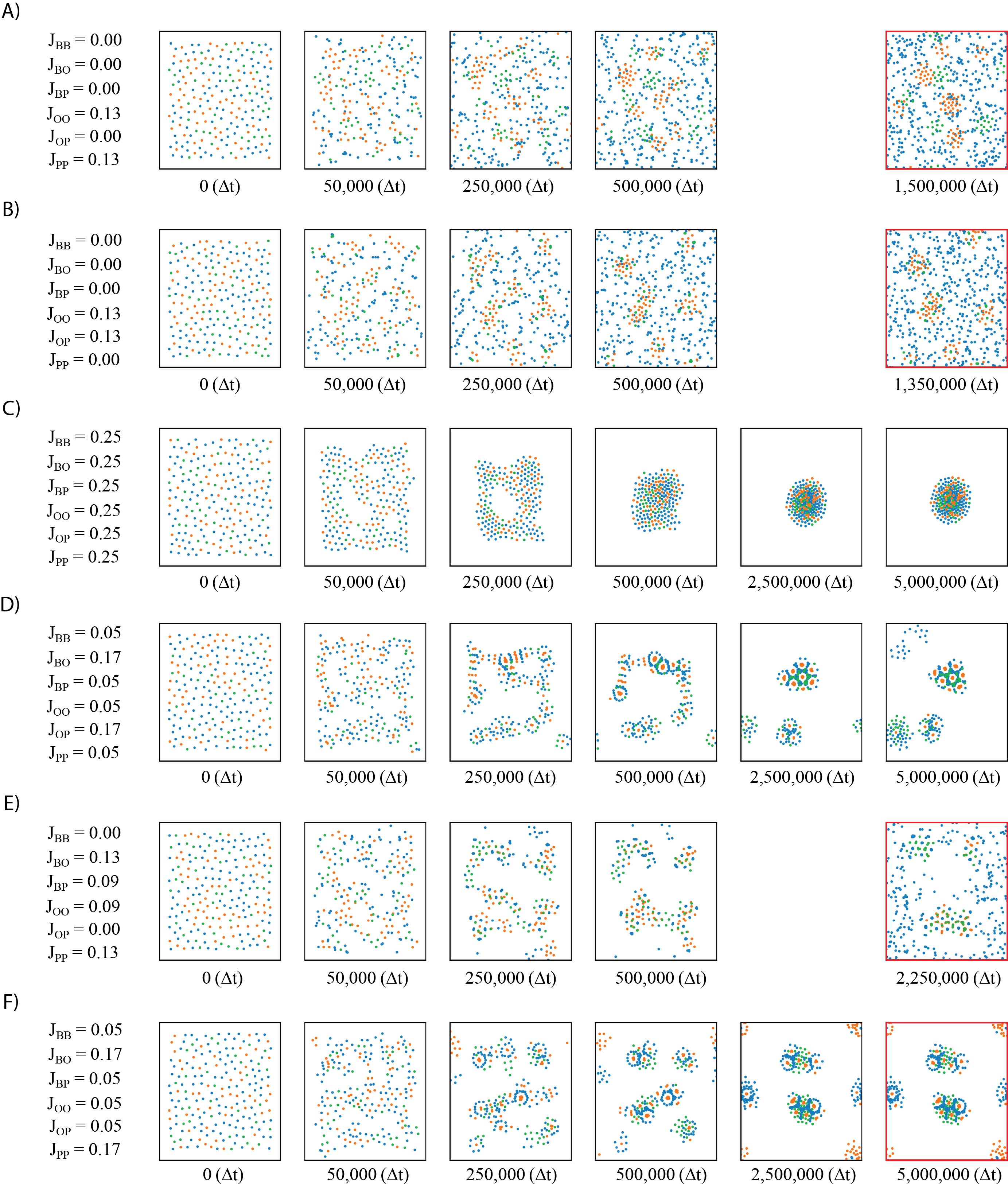}
    \caption{\footnotesize \textbf{Simulation snapshots showing self-organization in a heterogeneous population with three cell types (50\% orange, 30\% blue, 20\% green) with proliferation.} (A) Orange and green cells aggregate into separate clusters while blue cells remain individually dispersed. (B) Green cells are engulfed within orange clusters, while blue cells remain individually dispersed. (C) Cells indiscriminately aggregate into clusters due to high adhesion. Clusters contain a mixture of particle types. (D) Hexagonal arrangement of orange clusters surrounded by blue and green cells due to high blue-orange and green-orange adhesion. (E) Hexagonal arrangement of green cells with offset hexagonal arrangement of orange cells, and dispersed blue cells  (F) Hexagonal arrangement of orange clusters surrounded by blue and green cells.. Snapshots outlined in red did not run to completion in the allotted time due to the combinatorial complexity of force calculations with exponentially increasing population of orange cells.}
    \label{fig:SI_multiple_cells_prolif}
\end{figure}

\begin{figure}[h]
    \centering
    \includegraphics[scale=0.72]{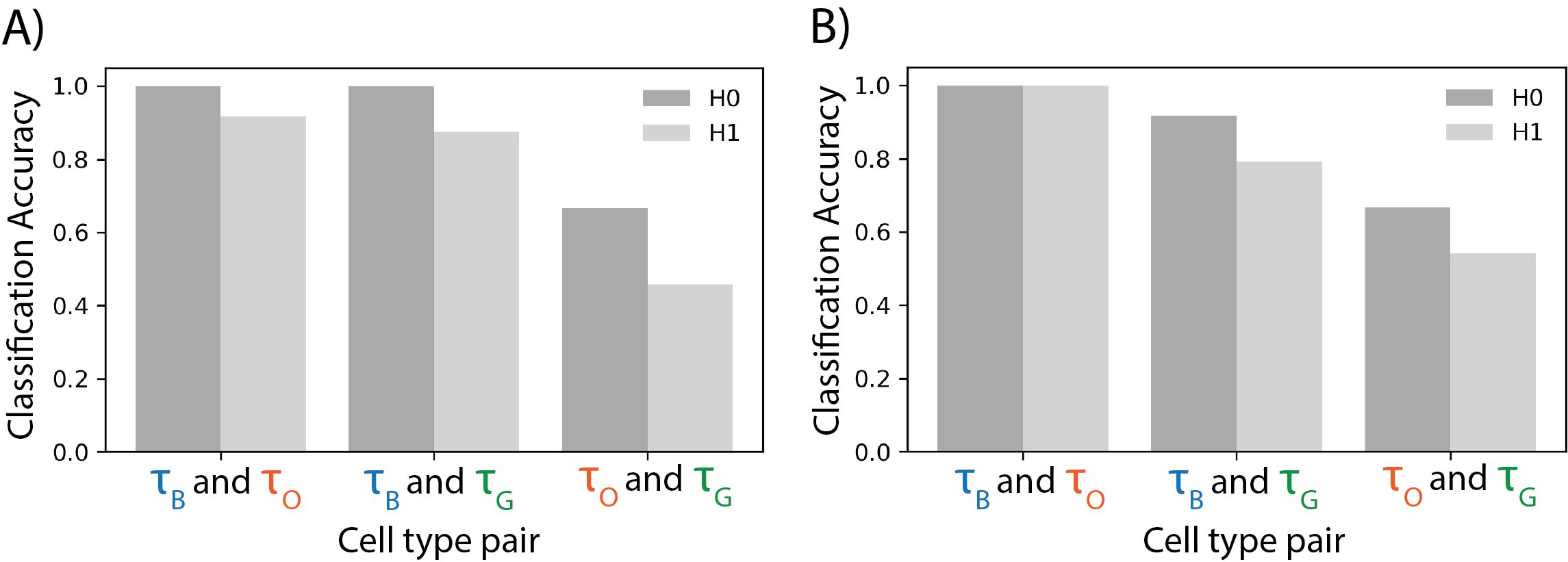}
    \caption{\footnotesize \textbf{Classification accuracy of persistence images computed from different pairs of cell types.} (A) at constant population size, and (B) at varying population size.}
    \label{fig:SI_multicelltype_classify}
\end{figure}

\clearpage

\begin{table}[h]
\begin{center}
{\renewcommand{\arraystretch}{1.2}%
\begin{tabular}{cccccccc}

\toprule

\textbf{Cell} & \textbf{Feature} & \textbf{Dim.} & \textbf{Feature} & \multicolumn{4}{c}{\textbf{Classification Metrics}}\\
\textbf{Types} & & \textbf{Reduction} & \textbf{Size} & \textbf{Accuracy} & \textbf{Precision} & \textbf{Recall} & \textbf{F1}\\

\midrule

$\tau_B, \tau_O$  & $H_0$       & N/A & $3 \times 200$              & $0.99$ & $1.00$ & $0.99$ & $0.99$\\
$\tau_B, \tau_O$  & $H_1$       & N/A & $3 \times 200 \times 100$   & $0.97$ & $1.00$ & $0.97$ & $0.98$\\
$\tau_B, \tau_O$  & $H_0, H_1$  & N/A & $3 \times 200 \times 101$   & $0.99$ & $1.00$ & $0.99$ & $0.99$\\

\hline

$\tau_B, \tau_O$  & $H_0$       & PHATE & $3 \times 20$             & $0.91$ & $1.00$ & $0.93$ & $0.96$\\
$\tau_B, \tau_O$  & $H_1$       & PHATE & $3 \times 20$             & $0.98$ & $0.99$ & $0.97$ & $0.98$\\
$\tau_B, \tau_O$  & $H_0, H_1$  & PHATE & $3 \times 20 \times 2$    & $0.95$ & $1.00$ & $0.97$ & $0.98$\\

\hline

$\tau_B, \tau_O$  & $H_0$       & AE & $3 \times 20$            & $0.93$ & $0.95$ & $0.91$ & $0.93$\\
$\tau_B, \tau_O$  & $H_1$       & AE & $3 \times 20$            & $0.97$ & $0.94$ & $0.89$ & $0.91$\\
$\tau_B, \tau_O$  & $H_0, H_1$  & AE & $3 \times 20 \times 2$   & $0.96$ & $0.96$ & $0.92$ & $0.94$\\

\hline

$\tau_B$  & $H_0$       & N/A & $200$                           & $0.90$ & $0.87$ & $0.82$ & $0.84$\\
$\tau_B$  & $H_1$       & N/A & $200 \times 100$                & $0.83$ & $0.79$ & $0.81$ & $0.80$\\
$\tau_B$  & $H_0, H_1$  & N/A & $200 \times 101$                & $0.92$ & $0.89$ & $0.84$ & $0.86$\\

\hline

$\tau_B$  & $H_0$       & PHATE & $20$              & $0.79$ & $0.80$ & $0.72$ & $0.76$\\
$\tau_B$  & $H_1$       & PHATE & $20$              & $0.72$ & $0.75$ & $0.74$ & $0.74$\\
$\tau_B$  & $H_0, H_1$  & PHATE & $20 \times 2$     & $0.77$ & $0.79$ & $0.76$ & $0.77$\\

\hline

$\tau_B$  & $H_0$       & AE & $20$             & $0.86$ & $0.88$ & $0.83$ & $0.85$\\
$\tau_B$  & $H_1$       & AE & $20$             & $0.74$ & $0.70$ & $0.68$ & $0.69$\\
$\tau_B$  & $H_0, H_1$  & AE & $20 \times 2$    & $0.87$ & $0.88$ & $0.84$ & $0.86$\\

\hline

$\tau_O$  & $H_0$       & N/A & $200$               & $0.73$ & $0.77$ & $0.67$ & $0.72$\\
$\tau_O$  & $H_1$       & N/A & $200 \times 100$    & $0.67$ & $0.69$ & $0.67$ & $0.68$\\
$\tau_O$  & $H_0, H_1$  & N/A & $200 \times 101$    & $0.71$ & $0.76$ & $0.69$ & $0.72$\\

\hline

$\tau_O$  & $H_0$       & PHATE & $20$              & $0.71$ & $0.68$ & $0.73$ & $0.70$\\
$\tau_O$  & $H_1$       & PHATE & $20$              & $0.68$ & $0.64$ & $0.69$ & $0.66$\\
$\tau_O$  & $H_0, H_1$  & PHATE & $20 \times 2$     & $0.67$ & $0.70$ & $0.71$ & $0.70$\\

\hline

$\tau_O$  & $H_0$       & AE & $20$             & $0.69$ & $0.65$ & $0.68$ & $0.66$\\
$\tau_O$  & $H_1$       & AE & $20$             & $0.61$ & $0.62$ & $0.64$ & $0.63$\\
$\tau_O$  & $H_0, H_1$  & AE & $20 \times 2$    & $0.71$ & $0.67$ & $0.68$ & $0.67$\\

\bottomrule

\end{tabular}
}\\
\caption{\label{tab:persimg_noprolif} Unsupervised classification accuracy of persistence images at constant population size.}
\end{center}
\end{table}

\begin{table}[h]
\begin{center}
{\renewcommand{\arraystretch}{1.2}%
\begin{tabular}{cccccccc}

\toprule

\textbf{Cell} & \textbf{Feature} & \textbf{Dim.} & \textbf{Feature} & \multicolumn{4}{c}{\textbf{Classification Metrics}}\\
\textbf{Types} & & \textbf{Reduction} & \textbf{Size} & \textbf{Accuracy} & \textbf{Precision} & \textbf{Recall} & \textbf{F1}\\

\midrule

$\tau_B, \tau_O$  & $H_0$       & N/A & $3 \times 200$              & $0.91$ & $0.85$ & $0.95$ & $0.90$\\
$\tau_B, \tau_O$  & $H_1$       & N/A & $3 \times 200$              & $0.83$ & $0.77$ & $0.79$ & $0.78$\\
$\tau_B, \tau_O$  & $H_0, H_1$  & N/A & $3 \times 200 \times 2$     & $0.93$ & $0.91$ & $0.93$ & $0.92$\\

\hline

$\tau_B, \tau_O$  & $H_0$       & PHATE & $3 \times 20$             & $0.82$ & $0.76$ & $0.79$ & $0.77$\\
$\tau_B, \tau_O$  & $H_1$       & PHATE & $3 \times 20$             & $0.65$ & $0.58$ & $0.61$ & $0.59$\\
$\tau_B, \tau_O$  & $H_0, H_1$  & PHATE & $3 \times 20 \times 2$    & $0.84$ & $0.80$ & $0.79$ & $0.79$\\

\hline

$\tau_B, \tau_O$  & $H_0$       & AE & $3 \times 20$                & $0.86$ & $0.86$ & $0.88$ & $0.87$\\
$\tau_B, \tau_O$  & $H_1$       & AE & $3 \times 20$                & $0.77$ & $0.72$ & $0.74$ & $0.73$\\
$\tau_B, \tau_O$  & $H_0, H_1$  & AE & $3 \times 20 \times 2$       & $0.86$ & $0.84$ & $0.82$ & $0.83$\\

\hline

$\tau_B$  & $H_0$       & N/A & $200$               & $0.77$ & $0.73$ & $0.68$ & $0.70$\\
$\tau_B$  & $H_1$       & N/A & $200$               & $0.74$ & $0.69$ & $0.70$ & $0.69$\\
$\tau_B$  & $H_0, H_1$  & N/A & $200 \times 2$      & $0.79$ & $0.78$ & $0.76$ & $0.77$\\

\hline

$\tau_B$  & $H_0$       & PHATE & $20$              & $0.78$ & $0.71$ & $0.73$ & $0.72$\\
$\tau_B$  & $H_1$       & PHATE & $20$              & $0.64$ & $0.66$ & $0.60$ & $0.63$\\
$\tau_B$  & $H_0, H_1$  & PHATE & $20 \times 2$     & $0.81$ & $0.77$ & $0.76$ & $0.76$\\

\hline

$\tau_B$  & $H_0$       & AE & $20$             & $0.74$ & $0.76$ & $0.73$ & $0.74$\\
$\tau_B$  & $H_1$       & AE & $20$             & $0.70$ & $0.68$ & $0.68$ & $0.68$\\
$\tau_B$  & $H_0, H_1$  & AE & $20 \times 2$    & $0.86$ & $0.81$ & $0.78$ & $0.79$\\

\hline

$\tau_O$  & $H_0$       & N/A & $200$           & $0.66$ & $0.58$ & $0.62$ & $0.60$\\
$\tau_O$  & $H_1$       & N/A & $200$           & $0.58$ & $0.53$ & $0.47$ & $0.50$\\
$\tau_O$  & $H_0, H_1$  & N/A & $200 \times 2$  & $0.65$ & $0.60$ & $0.62$ & $0.61$\\

\hline

$\tau_O$  & $H_0$       & PHATE & $20$              & $0.60$ & $0.57$ & $0.62$ & $0.59$\\
$\tau_O$  & $H_1$       & PHATE & $20$              & $0.53$ & $0.51$ & $0.45$ & $0.48$\\
$\tau_O$  & $H_0, H_1$  & PHATE & $20 \times 2$     & $0.67$ & $0.58$ & $0.61$ & $0.59$\\

\hline

$\tau_O$  & $H_0$       & AE & $20$             & $0.64$ & $0.60$ & $0.59$ & $0.59$\\
$\tau_O$  & $H_1$       & AE & $20$             & $0.57$ & $0.58$ & $0.51$ & $0.54$\\
$\tau_O$  & $H_0, H_1$  & AE & $20 \times 2$    & $0.62$ & $0.61$ & $0.61$ & $0.61$\\

\bottomrule

\end{tabular}
}\\
\caption{\label{tab:perscurve_noprolif} Unsupervised classification accuracy of normalized persistence curves at constant population size.}
\end{center}
\end{table}

\begin{table}[h]
\begin{center}
{\renewcommand{\arraystretch}{1.2}%
\begin{tabular}{cccccccc}

\toprule

\textbf{Cell} & \textbf{Feature} & \textbf{Dim.} & \textbf{Feature} & \multicolumn{4}{c}{\textbf{Classification Metrics}}\\
\textbf{Types} & & \textbf{Reduction} & \textbf{Size} & \textbf{Accuracy} & \textbf{Precision} & \textbf{Recall} & \textbf{F1}\\

\midrule

$\tau_B, \tau_O$  & RDF       & N/A & $3 \times 200$            & $0.98$ & $0.92$ & $0.86$ & $0.89$\\
$\tau_B, \tau_O$  & ADF       & N/A & $3 \times 100$            & $0.52$ & $0.54$ & $0.51$ & $0.52$\\
$\tau_B, \tau_O$  & RDF, ADF  & N/A & $3 \times 300$            & $0.95$ & $0.94$ & $0.91$ & $0.92$\\

\hline

$\tau_B, \tau_O$  & RDF       & PHATE & $3 \times 20$            & $0.97$ & $0.96$ & $0.94$ & $0.95$\\
$\tau_B, \tau_O$  & ADF       & PHATE & $3 \times 20$            & $0.38$ & $0.41$ & $0.35$ & $0.38$\\
$\tau_B, \tau_O$  & RDF, ADF  & PHATE & $3 \times 20 \times 2$   & $0.94$ & $0.91$ & $0.88$ & $0.89$\\

\hline

$\tau_B, \tau_O$  & RDF       & AE & $3 \times 20$               & $0.94$ & $0.97$ & $0.92$ & $0.94$\\
$\tau_B, \tau_O$  & ADF       & AE & $3 \times 20$               & $0.47$ & $0.41$ & $0.36$ & $0.38$\\
$\tau_B, \tau_O$  & RDF, ADF  & AE & $3 \times 20 \times 2$      & $0.93$ & $0.97$ & $0.93$ & $0.95$\\

\hline

$\tau_B$  & RDF       & N/A & $200$     & $0.92$ & $0.89$ & $0.88$ & $0.88$\\
$\tau_B$  & ADF       & N/A & $100$     & $0.44$ & $0.39$ & $0.42$ & $0.40$\\
$\tau_B$  & RDF, ADF  & N/A & $300$     & $0.88$ & $0.93$ & $0.90$ & $0.91$\\

\hline

$\tau_B$  & RDF       & PHATE & $20$            & $0.93$ & $0.88$ & $0.89$ & $0.88$\\
$\tau_B$  & ADF       & PHATE & $20$            & $0.37$ & $0.35$ & $0.29$ & $0.31$\\
$\tau_B$  & RDF, ADF  & PHATE & $20 \times 2$   & $0.90$ & $0.86$ & $0.91$ & $0.88$\\

\hline

$\tau_B$  & RDF       & AE & $20$           & $0.88$ & $0.83$ & $0.84$ & $0.83$\\
$\tau_B$  & ADF       & AE & $20$           & $0.43$ & $0.32$ & $0.39$ & $0.35$\\
$\tau_B$  & RDF, ADF  & AE & $20 \times 2$  & $0.81$ & $0.90$ & $0.84$ & $0.87$\\

\hline

$\tau_O$  & RDF       & N/A & $200$     & $0.74$ & $0.77$ & $0.71$ & $0.74$\\
$\tau_O$  & ADF       & N/A & $100$     & $0.38$ & $0.38$ & $0.34$ & $0.36$\\
$\tau_O$  & RDF, ADF  & N/A & $300$     & $0.74$ & $0.82$ & $0.73$ & $0.77$\\

\hline

$\tau_O$  & RDF       & PHATE & $20$            & $0.75$ & $0.68$ & $0.69$ & $0.68$\\
$\tau_O$  & ADF       & PHATE & $20$            & $0.32$ & $0.28$ & $0.22$ & $0.25$\\
$\tau_O$  & RDF, ADF  & PHATE & $20 \times 2$   & $0.72$ & $0.69$ & $0.69$ & $0.69$\\

\hline

$\tau_O$  & RDF       & AE & $20$           & $0.72$ & $0.64$ & $0.66$ & $0.65$\\
$\tau_O$  & ADF       & AE & $20$           & $0.36$ & $0.31$ & $0.28$ & $0.29$\\
$\tau_O$  & RDF, ADF  & AE & $20 \times 2$  & $0.68$ & $0.66$ & $0.65$ & $0.65$\\

\bottomrule

\end{tabular}
}
\caption{\label{tab:op_noprolif} Unsupervised classification accuracy of order parameters at constant population size.}
\end{center}
\end{table}

\begin{table}[h]
\begin{center}
{\renewcommand{\arraystretch}{1.2}%
\begin{tabular}{cccccccc}

\toprule

\textbf{Cell} & \textbf{Feature} & \textbf{Dim.} & \textbf{Feature} & \multicolumn{4}{c}{\textbf{Classification Metrics}}\\
\textbf{Types} & & \textbf{Reduction} & \textbf{Size} & \textbf{Accuracy} & \textbf{Precision} & \textbf{Recall} & \textbf{F1}\\

\midrule

$\tau_B, \tau_O$  & $H_0$       & N/A & $3 \times 200$             & $0.92$ & $1.00$ & $0.93$ & $0.96$\\
$\tau_B, \tau_O$  & $H_1$       & N/A & $3 \times 200 \times 100$  & $0.85$ & $0.92$ & $0.81$ & $0.86$\\
$\tau_B, \tau_O$  & $H_0, H_1$  & N/A & $3 \times 200 \times 101$  & $0.90$ & $1.00$ & $0.94$ & $0.97$\\

\hline

$\tau_B, \tau_O$  & $H_0$       & PHATE & $3 \times 20$             & $0.74$ & $0.81$ & $0.76$ & $0.78$\\
$\tau_B, \tau_O$  & $H_1$       & PHATE & $3 \times 20$             & $0.52$ & $0.62$ & $0.63$ & $0.62$\\
$\tau_B, \tau_O$  & $H_0, H_1$  & PHATE & $3 \times 20 \times 2$    & $0.79$ & $0.79$ & $0.77$ & $0.78$\\

\hline

$\tau_B, \tau_O$  & $H_0$       & AE & $3 \times 20$               & $0.84$ & $0.89$ & $0.80$ & $0.84$\\
$\tau_B, \tau_O$  & $H_1$       & AE & $3 \times 20$               & $0.81$ & $0.84$ & $0.74$ & $0.79$\\
$\tau_B, \tau_O$  & $H_0, H_1$  & AE & $3 \times 20 \times 2$      & $0.83$ & $0.89$ & $0.81$ & $0.85$\\

\hline

$\tau_B$  & $H_0$       & N/A & $200$                  & $0.78$ & $0.81$ & $0.75$ & $0.78$\\
$\tau_B$  & $H_1$       & N/A & $200 \times 100$       & $0.66$ & $0.62$ & $0.60$ & $0.61$\\
$\tau_B$  & $H_0, H_1$  & N/A & $200 \times 101$       & $0.77$ & $0.79$ & $0.77$ & $0.78$\\

\hline

$\tau_B$  & $H_0$       & PHATE & $20$                 & $0.62$ & $0.59$ & $0.56$ & $0.57$\\
$\tau_B$  & $H_1$       & PHATE & $20$                 & $0.49$ & $0.51$ & $0.45$ & $0.48$\\
$\tau_B$  & $H_0, H_1$  & PHATE & $20 \times 2$        & $0.63$ & $0.66$ & $0.63$ & $0.64$\\

\hline

$\tau_B$  & $H_0$       & AE & $20$                    & $0.71$ & $0.66$ & $0.65$ & $0.65$\\
$\tau_B$  & $H_1$       & AE & $20$                    & $0.68$ & $0.69$ & $0.63$ & $0.66$\\
$\tau_B$  & $H_0, H_1$  & AE & $20 \times 2$           & $0.73$ & $0.70$ & $0.66$ & $0.68$\\

\hline

$\tau_O$  & $H_0$       & N/A & $200$                  & $0.64$ & $0.68$ & $0.61$ & $0.64$\\
$\tau_O$  & $H_1$       & N/A & $200 \times 100$       & $0.66$ & $0.68$ & $0.63$ & $0.65$\\
$\tau_O$  & $H_0, H_1$  & N/A & $200 \times 101$       & $0.63$ & $0.70$ & $0.62$ & $0.66$\\

\hline

$\tau_O$  & $H_0$       & PHATE & $20$                 & $0.61$ & $0.64$ & $0.65$ & $0.64$\\
$\tau_O$  & $H_1$       & PHATE & $20$                 & $0.38$ & $0.44$ & $0.39$ & $0.41$\\
$\tau_O$  & $H_0, H_1$  & PHATE & $20 \times 2$        & $0.59$ & $0.64$ & $0.67$ & $0.65$\\

\hline

$\tau_O$  & $H_0$       & AE & $20$                    & $0.62$ & $0.65$ & $0.59$ & $0.62$\\
$\tau_O$  & $H_1$       & AE & $20$                    & $0.57$ & $0.61$ & $0.62$ & $0.61$\\
$\tau_O$  & $H_0, H_1$  & AE & $20 \times 2$           & $0.63$ & $0.64$ & $0.62$ & $0.63$\\

\bottomrule

\end{tabular}
}
\caption{\label{tab:persimg_prolif} Unsupervised classification accuracy of persistence images at varying population size.}
\end{center}
\end{table}

\begin{table}[h]
\begin{center}
{\renewcommand{\arraystretch}{1.2}%
\begin{tabular}{cccccccc}

\toprule

\textbf{Cell} & \textbf{Feature} & \textbf{Dim.} & \textbf{Feature} & \multicolumn{4}{c}{\textbf{Classification Metrics}}\\
\textbf{Types} & & \textbf{Reduction} & \textbf{Size} & \textbf{Accuracy} & \textbf{Precision} & \textbf{Recall} & \textbf{F1}\\

\midrule

$\tau_B, \tau_O$  & $H_0$       & N/A & $3 \times 200$              & $0.86$ & $0.89$ & $0.81$ & $0.85$\\
$\tau_B, \tau_O$  & $H_1$       & N/A & $3 \times 200$              & $0.55$ & $0.63$ & $0.60$ & $0.61$\\
$\tau_B, \tau_O$  & $H_0, H_1$  & N/A & $3 \times 200 \times 2$     & $0.79$ & $0.82$ & $0.83$ & $0.83$\\

\hline

$\tau_B, \tau_O$  & $H_0$       & PHATE & $3 \times 20$             & $0.79$ & $0.77$ & $0.71$ & $0.74$\\
$\tau_B, \tau_O$  & $H_1$       & PHATE & $3 \times 20$             & $0.52$ & $0.64$ & $0.59$ & $0.61$\\
$\tau_B, \tau_O$  & $H_0, H_1$  & PHATE & $3 \times 20 \times 2$    & $0.80$ & $0.79$ & $0.73$ & $0.76$\\

\hline

$\tau_B, \tau_O$  & $H_0$       & AE & $3 \times 20$                & $0.84$ & $0.87$ & $0.88$ & $0.88$\\
$\tau_B, \tau_O$  & $H_1$       & AE & $3 \times 20$                & $0.49$ & $0.60$ & $0.56$ & $0.58$\\
$\tau_B, \tau_O$  & $H_0, H_1$  & AE & $3 \times 20 \times 2$       & $0.86$ & $0.87$ & $0.89$ & $0.88$\\

\hline

$\tau_B$  & $H_0$       & N/A & $200$           & $0.74$ & $0.78$ & $0.75$ & $0.76$\\
$\tau_B$  & $H_1$       & N/A & $200$           & $0.53$ & $0.49$ & $0.51$ & $0.50$\\
$\tau_B$  & $H_0, H_1$  & N/A & $200 \times 2$  & $0.70$ & $0.76$ & $0.77$ & $0.76$\\

\hline

$\tau_B$  & $H_0$       & PHATE & $20$                  & $0.74$ & $0.73$ & $0.75$ & $0.74$\\
$\tau_B$  & $H_1$       & PHATE & $20$                  & $0.41$ & $0.34$ & $0.39$ & $0.36$\\
$\tau_B$  & $H_0, H_1$  & PHATE & $20 \times 2$         & $0.73$ & $0.74$ & $0.74$ & $0.74$\\

\hline

$\tau_B$  & $H_0$       & AE & $20$             & $0.72$ & $0.76$ & $0.69$ & $0.72$\\
$\tau_B$  & $H_1$       & AE & $20$             & $0.49$ & $0.42$ & $0.51$ & $0.46$\\
$\tau_B$  & $H_0, H_1$  & AE & $20 \times 2$    & $0.76$ & $0.78$ & $0.71$ & $0.74$\\

\hline

$\tau_O$  & $H_0$       & N/A & $200$           & $0.68$ & $0.73$ & $0.70$ & $0.71$\\
$\tau_O$  & $H_1$       & N/A & $200$           & $0.40$ & $0.36$ & $0.39$ & $0.37$\\
$\tau_O$  & $H_0, H_1$  & N/A & $200 \times 2$  & $0.71$ & $0.75$ & $0.69$ & $0.72$\\

\hline

$\tau_O$  & $H_0$       & PHATE & $20$                  & $0.62$ & $0.68$ & $0.56$ & $0.61$\\
$\tau_O$  & $H_1$       & PHATE & $20$                  & $0.44$ & $0.47$ & $0.37$ & $0.41$\\
$\tau_O$  & $H_0, H_1$  & PHATE & $20 \times 2$         & $0.67$ & $0.68$ & $0.60$ & $0.64$\\

\hline

$\tau_O$  & $H_0$       & AE & $20$             & $0.65$ & $0.71$ & $0.63$ & $0.67$\\
$\tau_O$  & $H_1$       & AE & $20$             & $0.38$ & $0.42$ & $0.33$ & $0.37$\\
$\tau_O$  & $H_0, H_1$  & AE & $20 \times 2$    & $0.67$ & $0.68$ & $0.60$ & $0.64$\\

\bottomrule

\end{tabular}
}
\caption{\label{tab:perscurve_prolif} Unsupervised classification accuracy of normalized persistence curves at varying population size.}
\end{center}
\end{table}

\begin{table}[h]
\begin{center}
{\renewcommand{\arraystretch}{1.2}%
\begin{tabular}{cccccccc}

\toprule

\textbf{Cell} & \textbf{Feature} & \textbf{Dim.} & \textbf{Feature} & \multicolumn{4}{c}{\textbf{Classification Metrics}}\\
\textbf{Types} & & \textbf{Reduction} & \textbf{Size} & \textbf{Accuracy} & \textbf{Precision} & \textbf{Recall} & \textbf{F1}\\

\midrule

$\tau_B, \tau_O$  & RDF         & N/A & $3 \times 200$      & $0.94$ & $0.90$ & $0.92$ & $0.91$\\
$\tau_B, \tau_O$  & ADF         & N/A & $3 \times 100$      & $0.45$ & $0.53$ & $0.44$ & $0.48$\\
$\tau_B, \tau_O$  & RDF, ADF    & N/A & $3 \times 300$      & $0.93$ & $0.91$ & $0.91$ & $0.91$\\

\hline

$\tau_B, \tau_O$  & RDF       & PHATE & $3 \times 20$           & $0.92$ & $0.89$ & $0.88$ & $0.89$\\
$\tau_B, \tau_O$  & ADF       & PHATE & $3 \times 20$           & $0.36$ & $0.41$ & $0.39$ & $0.40$\\
$\tau_B, \tau_O$  & RDF, ADF  & PHATE & $3 \times 20 \times 2$  & $0.89$ & $0.91$ & $0.90$ & $0.90$\\

\hline

$\tau_B, \tau_O$  & RDF       & AE & $3 \times 20$              & $0.88$ & $0.89$ & $0.87$ & $0.88$\\
$\tau_B, \tau_O$  & ADF       & AE & $3 \times 20$              & $0.47$ & $0.42$ & $0.40$ & $0.41$\\
$\tau_B, \tau_O$  & RDF, ADF  & AE & $3 \times 20 \times 2$     & $0.84$ & $0.90$ & $0.91$ & $0.90$\\

\hline

$\tau_B$  & RDF       & N/A & $200$     & $0.81$ & $0.79$ & $0.76$ & $0.78$\\
$\tau_B$  & ADF       & N/A & $100$     & $0.46$ & $0.47$ & $0.42$ & $0.44$\\
$\tau_B$  & RDF, ADF  & N/A & $300$     & $0.77$ & $0.77$ & $0.78$ & $0.78$\\

\hline

$\tau_B$  & RDF       & PHATE & $20$            & $0.70$ & $0.67$ & $0.68$ & $0.68$\\
$\tau_B$  & ADF       & PHATE & $20$            & $0.39$ & $0.44$ & $0.41$ & $0.42$\\
$\tau_B$  & RDF, ADF  & PHATE & $20 \times 2$   & $0.64$ & $0.70$ & $0.67$ & $0.68$\\

\hline

$\tau_B$  & RDF       & AE & $20$           & $0.74$ & $0.76$ & $0.70$ & $0.73$\\
$\tau_B$  & ADF       & AE & $20$           & $0.43$ & $0.38$ & $0.41$ & $0.39$\\
$\tau_B$  & RDF, ADF  & AE & $20 \times 2$  & $0.73$ & $0.74$ & $0.72$ & $0.73$\\

\hline

$\tau_O$  & RDF       & N/A & $200$     & $0.69$ & $0.64$ & $0.67$ & $0.65$\\
$\tau_O$  & ADF       & N/A & $100$     & $0.43$ & $0.37$ & $0.39$ & $0.38$\\
$\tau_O$  & RDF, ADF  & N/A & $300$     & $0.76$ & $0.66$ & $0.65$ & $0.65$\\

\hline

$\tau_O$  & RDF       & PHATE & $20$            & $0.72$ & $0.68$ & $0.72$ & $0.70$\\
$\tau_O$  & ADF       & PHATE & $20$            & $0.41$ & $0.38$ & $0.45$ & $0.41$\\
$\tau_O$  & RDF, ADF  & PHATE & $20 \times 2$   & $0.72$ & $0.70$ & $0.71$ & $0.71$\\

\hline

$\tau_O$  & RDF       & AE & $20$           & $0.66$ & $0.68$ & $0.64$ & $0.66$\\
$\tau_O$  & ADF       & AE & $20$           & $0.39$ & $0.41$ & $0.36$ & $0.38$\\
$\tau_O$  & RDF, ADF  & AE & $20 \times 2$  & $0.68$ & $0.70$ & $0.63$ & $0.66$\\

\bottomrule

\end{tabular}
}
\caption{\label{tab:op_prolif} Unsupervised classification accuracy of order parameters at varying population size.}
\end{center}
\end{table}

\FloatBarrier
\textbf{Note S1}: We also considered classification using a softer probabilistic algorithm. Briefly, we used a soft margin support vector machine (SVM) where the soft-margin classifier:
\begin{align*}
\min \frac{1}{2}\|w\|^2 &+ C \sum_{i=1}^n \zeta_i \\
\text{subject to} \quad y_i(w^Tx_i + b) & \ge 1 - \zeta_i \quad \forall \, i = 1,\cdots, n, \quad \zeta_i \ge 0
\end{align*}
trained using the hinge-loss function, $\max \{0, 1-y_i(w^Tx_i +b) \}$, allows flexibility for misclassifications, via slack variables $\zeta_i$. The regularization parameter, $C$, controls the trade-off between maximizing the margin at the decision boundary and minimizing the loss. We trained the soft margin classifier, using the radial basis function for nonlinear transformation of the input, at various values of $C$ and computed the accuracy using $5$-fold cross-validation. Crucially, we modified the scoring function to ignore misclassification between adjacent phases in the accuracy computation. \\

\begin{table}[h]
\begin{center}
{\renewcommand{\arraystretch}{1.2}%
\begin{tabular}{ccccccc}

\toprule

\textbf{Cell} & \textbf{Feature} & \textbf{Dim.} & \textbf{Feature} & \multicolumn{3}{c}{\textbf{Classification}}\\
\textbf{Types} & & \textbf{Reduction} & \textbf{Size} & \multicolumn{3}{c}{\textbf{Accuracy}}\\

\hline

& & & & $C=0.5$ & $C=1.0$ & $C=2.0$ \\

\hline

$\tau_B, \tau_O$  & $H_0$       & N/A & $3 \times 200$              & $1.00$ & $0.98$ & $0.92$ \\
$\tau_B, \tau_O$  & $H_1$       & N/A & $3 \times 200 \times 100$   & $0.96$ & $0.92$ & $0.89$ \\
$\tau_B, \tau_O$  & $H_0, H_1$  & N/A & $3 \times 200 \times 101$   & $1.00$ & $0.97$ & $0.92$ \\

\hline

$\tau_B, \tau_O$  & $H_0$       & PHATE & $3 \times 20$             & $1.00$ & $0.90$ & $0.88$\\
$\tau_B, \tau_O$  & $H_1$       & PHATE & $3 \times 20$             & $0.93$ & $0.87$ & $0.81$\\
$\tau_B, \tau_O$  & $H_0, H_1$  & PHATE & $3 \times 20 \times 2$    & $1.00$ & $0.89$ & $0.87$\\

\hline

$\tau_B, \tau_O$  & $H_0$       & AE & $3 \times 20$                & $1.00$ & $0.90$ & $0.86$ \\
$\tau_B, \tau_O$  & $H_1$       & AE & $3 \times 20$                & $0.95$ & $0.88$ & $0.81$\\
$\tau_B, \tau_O$  & $H_0, H_1$  & AE & $3 \times 20 \times 2$       & $1.00$ & $0.90$ & $0.86$\\

\bottomrule

\end{tabular}
}
\caption{\label{tab:soft_class_noprolif} Soft margin SVM classification accuracy of persistence images at constant population size.}
\end{center}
\end{table}

\begin{table}[h]
\begin{center}
{\renewcommand{\arraystretch}{1.2}%
\begin{tabular}{ccccccc}

\toprule

\textbf{Cell} & \textbf{Feature} & \textbf{Dim.} & \textbf{Feature} & \multicolumn{3}{c}{\textbf{Classification}}\\
\textbf{Types} & & \textbf{Reduction} & \textbf{Size} & \multicolumn{3}{c}{\textbf{Accuracy}}\\

\hline

& & & & $C=0.5$ & $C=1.0$ & $C=2.0$ \\

\hline

$\tau_B, \tau_O$  & $H_0$       & N/A & $3 \times 200$             & $0.97$ & $0.93$ & $0.86$\\
$\tau_B, \tau_O$  & $H_1$       & N/A & $3 \times 200 \times 100$  & $0.85$ & $0.79$ & $0.74$\\
$\tau_B, \tau_O$  & $H_0, H_1$  & N/A & $3 \times 200 \times 101$  & $0.94$ & $0.91$ & $0.87$\\

\hline

$\tau_B, \tau_O$  & $H_0$       & PHATE & $3 \times 20$            & $0.91$ & $0.89$ & $0.88$ \\
$\tau_B, \tau_O$  & $H_1$       & PHATE & $3 \times 20$            & $0.82$ & $0.79$ & $0.76$ \\
$\tau_B, \tau_O$  & $H_0, H_1$  & PHATE & $3 \times 20 \times 2$   & $0.91$ & $0.86$ & $0.87$ \\

\hline

$\tau_B, \tau_O$  & $H_0$       & AE & $3 \times 20$               & $0.79$ & $0.75$ & $0.71$ \\
$\tau_B, \tau_O$  & $H_1$       & AE & $3 \times 20$               & $0.63$ & $0.61$ & $0.58$ \\
$\tau_B, \tau_O$  & $H_0, H_1$  & AE & $3 \times 20 \times 2$      & $0.80$ & $0.73$ & $0.69$ \\

\bottomrule

\end{tabular}
}
\caption{\label{tab:soft_class_prolif} Soft margin SVM classification accuracy of persistence images at varying population size.}
\end{center}
\end{table}

\end{document}